\documentclass[twocolumn]{aastex631}

\usepackage{amsmath,amstext,amssymb}
\usepackage[T1]{fontenc}
\usepackage{apjfonts}
\usepackage[figure,figure*]{hypcap}
\usepackage{xcolor}
\usepackage{soul}
\usepackage{comment}
\usepackage{graphicx}
\usepackage[normalem]{ulem}
\usepackage{multirow}
\usepackage[toc,page]{appendix}

\hypersetup{colorlinks=true}
\graphicspath{{figures/}}
\usepackage{float}

\begin{document}

\title{Detecting prompt and afterglow jet emission of gravitational wave events from LIGO/Virgo/KAGRA and next generation detectors}
\shorttitle{Afterglow Detectability from LIGO/Virgo/KAGRA Events}
\shortauthors{Kaur et al.}

\received{\today}
\revised{--}
\accepted{--}

\submitjournal{ApJ}

\author[0009-0005-1871-7856]{Ravjit Kaur}
\email{ravkaur@ucsc.edu}
\affiliation{Department of Astronomy, University of California Berkeley, \\
501 Campbell Hall 3411, Berkeley, CA 94720-3411, USA}
\affiliation{Department of Astronomy and Astrophysics, University of California Santa Cruz, \\
1156 High St, Santa Cruz, CA 95064, USA}

\author[0000-0002-9700-0036]{Brendan O'Connor}
\email{boconno2@andrew.cmu.edu}
\altaffiliation{McWilliams Fellow}
\affiliation{McWilliams Center for Cosmology and Astrophysics, \\ Department of Physics, Carnegie Mellon University, Pittsburgh, PA 15213, USA}

\author[0000-0002-6011-0530]{Antonella Palmese}
\email{palmese@cmu.edu}
\altaffiliation{NASA Einstein Fellow}
\affiliation{McWilliams Center for Cosmology and Astrophysics, \\ Department of Physics, Carnegie Mellon University, Pittsburgh, PA 15213, USA}
\affiliation{Department of Physics, University of California Berkeley, \\ 366 LeConte Hall MC 7300, Berkeley, CA 94720, USA}

\author[0009-0000-4830-1484]{Keerthi Kunnumkai}
\affiliation{McWilliams Center for Cosmology and Astrophysics, \\ Department of Physics, Carnegie Mellon University, Pittsburgh, PA 15213, USA}

\begin{abstract}

Following the wealth of new results enabled by multimessenger observations of the binary neutron star (BNS) merger GW170817, the next goal is increasing the number of detections of electromagnetic (EM) counterparts to gravitational wave (GW) events. We study the detectability of the prompt emission and afterglows produced by the relativistic jets launched by BNS mergers that will be detected by LIGO-Virgo-KAGRA during their fifth observing run (O5),  and by next generation (XG) GW detectors (Einstein Telescope and Cosmic Explorer). We quantify the impact of various BNS merger and jet afterglow parameters on the likelihood of detection, focusing on the impact of the observer's viewing angle and the jet's core half-opening angle. We explore detectability over a wide range of current state-of-the-art facilities (e.g., the \textit{James Webb Space Telescope}, \textit{Chandra X-ray Observatory}) as well as upcoming next-generation facilities (e.g., \textit{AXIS}, \textit{NewAthena}, ngVLA, SKA). We find that a few GW events ($\sim$\,$0-4$) per year may have a detectable afterglow component in O5, with the largest detection rates expected with SKA in the radio and \textit{JWST} in the near-infrared. In the XG era, hundreds of multimessenger detections of afterglows per year may be possible with a range of instruments, such as \textit{NewAthena} in the X-ray and ngVLA in the radio. While zero to a few GW events per year are expected to be accompanied by a detectable prompt emission in O5, dozens per year may be detectable in XG.

\end{abstract}


\section{Introduction} \label{sec:intro}

Over the past decade, gravitational wave (GW) observations have unlocked a new view of the 
Universe, unleashing a fruitful period of multimessenger astrophysics. Made possible by the Advanced LIGO (Laser Interferometer Gravitational-Wave Observatory; \citealt{2015LIGO}), Virgo \citep{Acernese_2014}, and KAGRA \citep{KAGRA:2020tym}, GW discoveries allow for observations of astrophysical phenomena that cannot often be studied by other means. The first (and thus far, only) multimessenger detection of a GW merger is the binary neutron star (BNS) merger GW170817 \citep{abbott2017gw170817}, detected by LIGO in GWs with a localizing non-detection in Virgo, and followed by multiple EM counterparts. A short duration gamma-ray burst (GRB) was detected by both the \textit{Fermi Gamma-Ray Space Telescope} \citep{Goldstein2017} and \textit{INTEGRAL} \citep{Savchenko2017}, followed by subsequent detections in X-ray, ultraviolet, optical, infrared, and radio wavelengths of both the kilonova \citep{Andreoni2017,Arcavi2017,Chornock2017,Coulter2017,Covino2017,Cowperthwaite2017,Drout2017,Evans2017, Kasliwal2017, Lipunov2017,Nicholl2017,Pian2017,Shappee2017,Smartt2017,Soares-Santos2017,Tanvir2017,Utsumi2017,Valenti2017,Buckley2018} and jet afterglow \citep{Margutti2017,Troja2017,Hallinan2017,Alexander_2017,Haggard2017,Lamb2017jet,Lazzati2018,Resmi2018,Mooley2018,D'Avanzo2018,Alexander2018,Xie2018,Nynka2018,Margutti2018,GG2018,Ghirlanda2019}. This multimessenger event addressed critical open questions regarding jet launching and jet structure, the production of heavy elements, the neutron star equation of state, and the expansion of the Universe \citep[for a review see, e.g.,][]{nakar2020electromagnetic,margutti2021first}. 
The detection of future EM counterparts to GW events, like GW170817, are crucial to continue advancing our knowledge of these topics, and to provide an accurate measurement of the Hubble constant independent of the cosmic distance ladder \citep[e.g.,][]{schutz1986determining,firststandardsiren,Chen2018,feeney,Hotokezaka+19,Dietrich2020,LVK_GWTC3_STS,Palmese_2023,palmese2024standard,GWCosmology}. However, in order to identify a counterpart and maximize the scientific output of a discovery, we require sensitive telescopes at a variety of wavelengths to detect the counterpart across the EM spectrum (radio to X-rays). Given the relatively small GW localization regions ($<$\,$100$ deg$^2$ for a significant fraction of events; e.g., \citealt{Petrov2022,Kiendrebeogo_2023}) expected during the fifth LIGO-Virgo-KAGRA (LVK) observing run (O5, starting $\sim$2028\footnote{\url{https://observing.docs.ligo.org/plan/}}), this observing run will provide a prime opportunity for counterpart discovery. However, the ideal strategy to implement (cadence, sensitivity, filters), and telescopes to use, is still partially an open question for both kilonovae \citep[e.g.,][]{Doctor2017,Andreoni2019,Bianco2019,Almualla2021,McBrien2021,Chase2022,Andreoni2022a,Andreoni2022b,Andreoni2024,2024ApJ...960..122B,Ragosta2024,Kunnumkai230529,kunnumkai} and afterglows \citep[e.g.,][]{Salafia2017,Corsi2019,Duque2020,Ho2020,Yu2021,Perna2021,Boersma2022,Colombo2022,Ronchini2022,Ducoin2023,Bhattacharjee2024,Morsony2024,Pellouin2024,Mondal2024}.  

The ground-breaking detection of GRB 170817A \citep{Savchenko2017,Goldstein2017} following GW170817 \citep{abbott2017gw170817} demonstrated conclusively that BNS mergers are capable of launching highly collimated, highly relativistic outflows. The interaction of the relativistic jet with the surrounding environment causes a forward shock that produces non-thermal synchrotron radiation across the EM spectrum \citep{Meszaros1997,Sari1998,Wijers1999,Granot2002}, called the ``afterglow'' \citep{Costa1997,vanParadijs1997,vanParadijs2000,Frail1997}. In typical observations of cosmological GRBs, we see them on-axis (aligned with our line of sight) where the observer's viewing angle $\theta_\textrm{v}$ is less than the jet's core half-opening angle $\theta_\textrm{c}$ \citep{Matsumoto2019b,OConnor2024}. In this case, both the afterglow and prompt gamma-ray emission is dominated by the jet's core. However, in the case of GW170817, the jet was observed at a large angle ($\sim$\,$15$\,$-$\,$30$ degrees) with respect to the jet's axis \citep{Mooley2018,Ghirlanda2019,Mooley2022,Ghirlanda2022,Hotokezaka+19,Fernandez2022,Makhathini2021ApJ,Hajela2022,Balasubramanian2022,Govreen-Segal2023,Ryan2023,McDowell2023}, revealing new information about the angular structure of the jet and showing that in the nearby Universe the detection of lower luminosity emission from the jet's ``wings'' is possible. Future afterglow observations in conjunction with GWs will reveal the structure and half-opening angle of short GRB jets \citep{BeniaminiNakar2019,Farah2020,Biscoveanu2020,Hayes2020,Sarin2022b,Hayes2023}. 

As the GW amplitude is anisotropic and larger for on-axis events \citep[e.g.,][]{Sathyaprakash2009}, which we define here as $\theta_\textrm{v}$\,$<$\,$\theta_\textrm{c}$ with respect to GRB afterglows and as an inclination angle of $\iota$\,$\approx$\,$0$ for GW mergers (corresponding to a face-on binary merger), this introduces a Malmquist bias that, for an isotropic distribution of viewing angles, leads the events detected in GWs by current advanced detectors to most likely be detected with $\theta_\textrm{v}\approx 35$ deg \citep{Schutz2011}. If we assume a typical jet core half-opening angle of $\theta_\textrm{c}$\,$\approx$\,$6$ deg \citep{Fong2015,RoucoEscorial2022} this naively implies that half of the events will have viewing angles larger than $\theta_\textrm{v}/\theta_\textrm{c}$\,$\gtrsim$\,$6$, similar to the inferred value for GW170817 of $\theta_\textrm{obs}/\theta_\textrm{c}$\,$\approx$\,$5$\,$-$\,$6$ \citep{Mooley2018,Hotokezaka+19,Makhathini2021ApJ,Mooley2022,Ghirlanda2022,Fernandez2022,Govreen-Segal2023,Ryan2023}. As the afterglow of GW170817 was difficult to detect with current instruments at high signal-to-noise over a long period of time, requiring the handful of most sensitive instruments at each wavelength (i.e., \textit{Chandra}, \textit{HST}, VLA), we can expect that GW mergers during LVK O5 (or possible future runs following upgrades to current detectors, such as A$\sharp$; \citealt{Gupta_2024}; McIver et al. in prep.) and next-generation detector runs, when the detection horizon increases substantially, will be even more difficult to detect given they will reach distances beyond $\sim 600$ Mpc in O5 (e.g. \citealt{kunnumkai}), and out to the first mergers for next generation detectors (e.g. \citealt{Gupta_2024}). However, the increase in distance is timed with the launch or commissioning of next generation telescopes at all wavelengths. Here we seek to quantify the ability to detect the off-axis afterglows of BNS mergers using the more sensitive suites of instruments available in the coming decade(s). 

As for far off-axis observers the time of the peak $t_\textrm{peak}$\,$\propto$\,$(\theta_\textrm{v}/\theta_\textrm{c})^2$ and flux at peak $F_\textrm{peak}$\,$\propto$\,$(\theta_\textrm{v}/\theta_\textrm{c})^{-2p}$ are modified significantly compared to the on-axis expectations \citep{Nakar2002,Rossi02,Panaitescu2003,vanEerten2010,BGG2020}, 
we expect a similar Malmquist bias toward viewing angles closer to their jet's core, which are also more likely to be the GW mergers detected at larger distances. At cosmological distances current short GRB observations are dominated by the on-axis sample \citep[$\theta_\textrm{v}$\,$<$\,$\theta_\textrm{c}$;][]{OConnor2024}, but the additional information from the GW merger as a trigger time and localization may allow for this population to be extended to larger viewing angles. However, we note that the lack of clearly off-axis GRBs at cosmological distances may be due to a reduced prompt gamma-ray efficiency outside the narrow core of these jets \citep{BeniaminiNakar2019,OConnor2024}.
In this case an off-axis GRB would appear as an orphan afterglow \citep{Nakar2002,Huang2002,Dalal2002,Totani2002,Levinson2002,Rhoads2003,Piran2013,Cenko2013,Law2018,Ho2020,Andreoni2021,Mooley2022,Sarin2022,Gupta2022,Perley2024,Freeburn2024}. 

The uncertain viewing angle derived for the majority of GW mergers with current generation detectors \citep[e.g.,][]{Chen2019} complicates searches for afterglows as the likely peak time $t_\textrm{p}$\,$\propto$\,$(\theta_\textrm{v}/\theta_\textrm{c})^2$ is not well constrained by the initial information. However, refined GW analysis following the identification of a kilonova and redshift $z$ can allow for accurate determinations of the viewing angle \citep[e.g.,][]{Chen2019}, aiding in afterglow searches. In the absence of a clear narrow-field localization for the merger from either the prompt gamma-ray detection of a GRB as in GW170817, or from a kilonova detection in wide-field searches \citep[e.g.,][]{Chase2022}, the efficient way to search for the afterglow is with wide-field pointed instruments at late times (e.g., \textit{Einstein Probe} \citealt{yuan2022einstein} or the upcoming \textit{ULTRASAT} \citealt{shvartzvald2023ultrasat}) or in upcoming surveys (e.g., the Vera Rubin Observatory's LSST \citealt{ivezic2019lsst} or the Square Kilometer Array \citealt{dewdney2009square}). 

In this paper, we simulate the likelihood of detecting an afterglow counterpart to GW events during LVK O5 and with respect to next-generation (XG) detectors, such as Cosmic Explorer (CE; \citealt{reitze2019cosmic}) and Einstein Telescope (ET; \citealt{maggiore2020science}), expected to reach sensitivities that are 10-100 times better, depending on frequency, than what is expected for O5 \citep{Gupta_2024}, and start operations in the late 2030s. We simulate the likelihood of detection over all wavelengths for a range of telescopes (radio to X-ray). We consider both small and large Field of View (FoV) telescopes, to take into account the case where a KN or sGRB is identified and pointed searches deep-field can occur, as well as the case in which the afterglow is the only EM counterpart to be detected. We use GW distances and viewing angles expected for O5 and next-generation to simulate afterglow lightcurves for comparison to the different telescope sensitivities. Additionally, we look at the impact of jet opening angle on detectability of the afterglow. 

The paper is laid out as follows. In \S \ref{section:simulations}, we discuss the GW simulations, the parameters used for the jet afterglow simulations, and the telescopes and sensitivities considered in this work. In \S \ref{section:Results}, we present our findings and compare the detectability across all telescopes and wavelengths. Lastly, in \S \ref{section:Conclusion}, we discuss implications of these findings and possibilities for future work. Throughout this work we assume the Planck Collaboration results \citep{aghanim2020planck} under a flat $\Lambda$CDM cosmology.

\section{Simulations}\label{section:simulations}

\begin{figure}
    \includegraphics[width=\columnwidth]{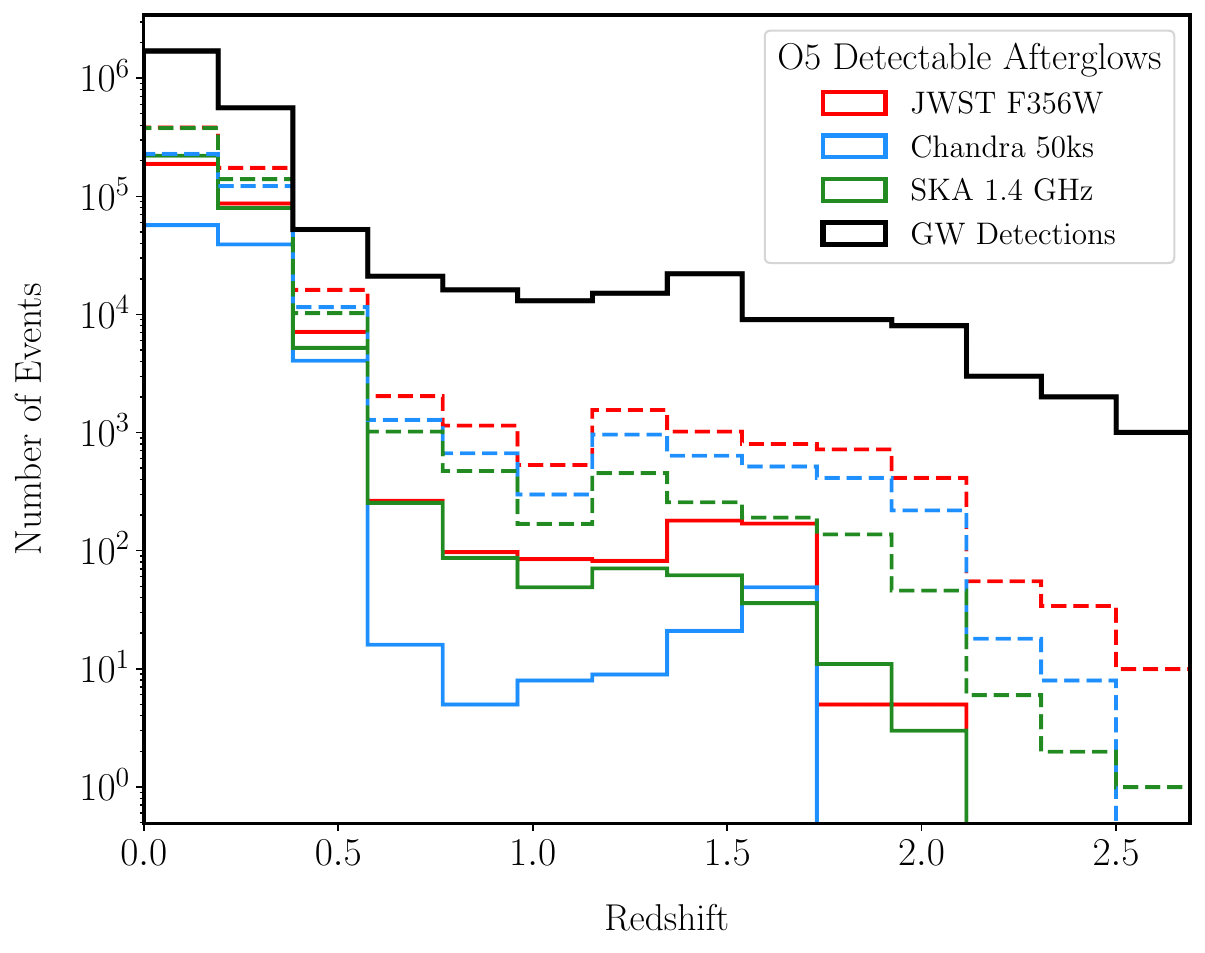}
    \includegraphics[width=\columnwidth]{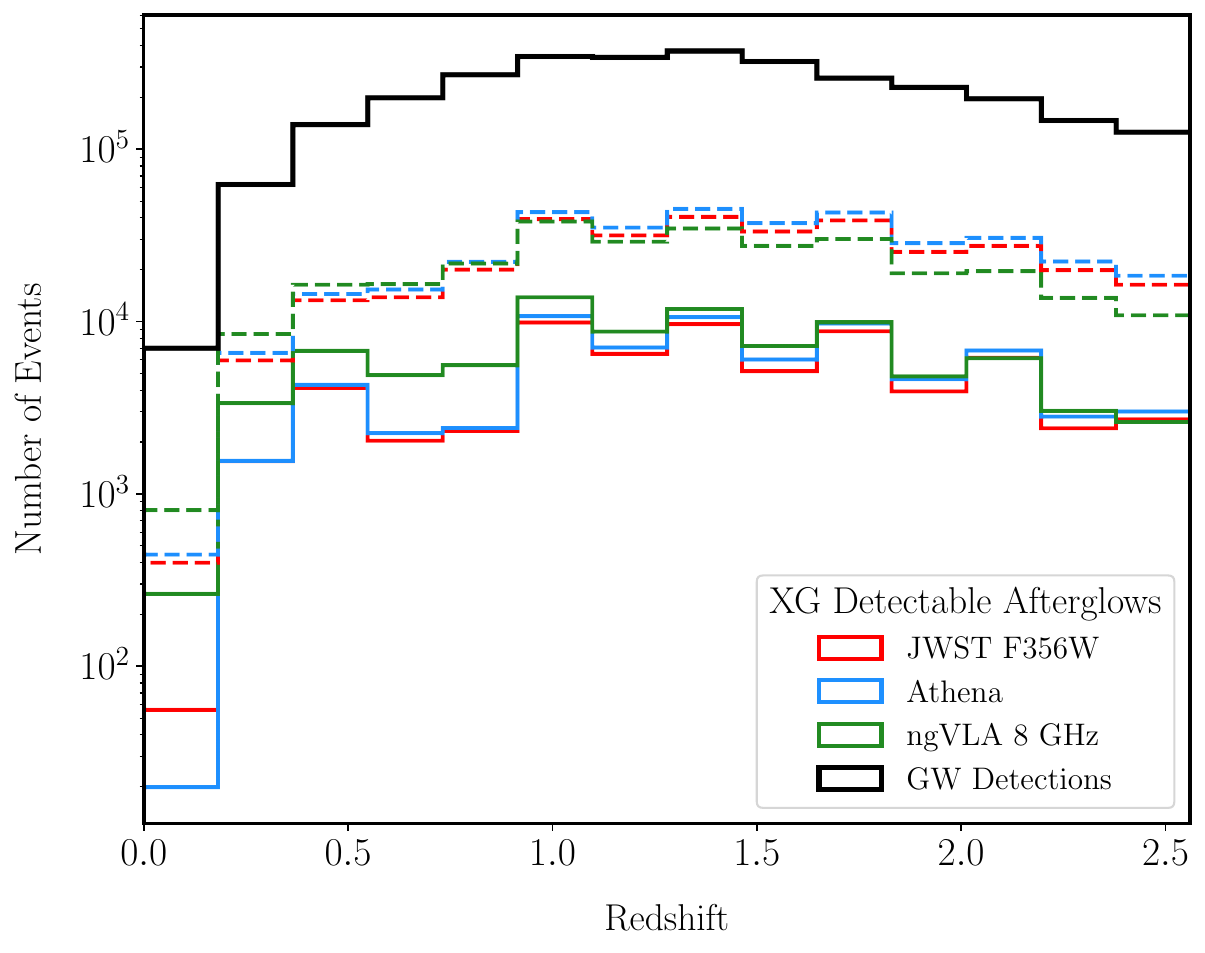}
    \caption{Redshift distribution of all simulated BNS mergers, where for each GW event we simulated 1,000 afterglows. Events with GW detections are shown in black, and the distribution of those events that had an afterglow detectable in our Run 7 by instruments in X-ray (blue), UVOIR (red), and Radio (green) are shown in the solid lines. Run 1 results are shown in dashed lines. The top figure displays these distributions for O5, and the bottom for XG. 
    }
    \label{fig:redshift}
\end{figure}

\subsection{GW simulations}
\label{sec:GWsims}

We produce simulations of GW events with LVK O5 sensitivities, as well as next-generation CE \citep{reitze2019cosmic} and ET \citep{maggiore2020science} detector sensitivities. We use \texttt{BAYESTAR}, a rapid sky localization code to localize the sky area \citep{singer2016rapid}, tools from LALSuite \citep{lalsuite} to make simulations of GW events, and the public software \texttt{ligo.skymap} to generate and visualize the GW skymaps. The mass and spin distributions for the O5 simulations are chosen as described in \citet{abbott2023population,fishbach2020does,farah2022bridging}. We use the population posterior samples from \citet{farah2022bridging} to draw masses. We uniformly distribute the mass and spin samples in comoving volume and isotropically distribute them in the orbital orientation and sky location. We draw $10^5$ samples from the distribution, which include binary neutron star (BNS) mergers, binary black hole (BBH) mergers, and neutron star-black hole (NSBH) mergers. We filter these events to satisfy our detection criteria for LVK O5 (network SNR $>$\,$8$, following \citealt{Petrov2022}), assuming the detector sensitivities from \url{https://dcc.ligo.org/LIGO-T2000012-v1/public} (A+ for the two LIGO detectors, avirgo\_O5low\_NEW.txt for Virgo, kagra\_128Mpc.txt for KAGRA), and a 70\% duty cycle for each detector. We then filter by mass to extract the BNS mergers: mergers with a primary and secondary mass below $3 M_\odot$ are considered to be BNSs. Through this process, we obtain 2,403 BNS GW events for O5. For more details about the O5 simulations see \citet{Kunnumkai230529,kunnumkai}.

For next generation detectors, we do not use the \citet{farah2022bridging} population posterior samples as we would need to extrapolate those to significantly larger redshifts compared to O5, which may not be realistic to assume. Instead, we assume a Gaussian distribution centered at $1.5~M_\odot$ with standard deviation $1.1~M_\odot$, consistent with the \citet{abbott2023population} parametric fits results. The redshift distribution follows the cosmic star formation density \citep{mandau}. This time, we inject $10^6$ events to ensure that we have enough samples at all redshifts, given that the observable volume is larger. We assume 1 CE and 1 ET detector, using the sensitivity curves from \citet{Srivastava_2022}\footnote{\url{https://dcc.cosmicexplorer.org/cgi-bin/DocDB/ShowDocument?docid=T2000017}} for a baseline 40 km CE facility and the ET-D design from \cite{Hild_ET}. We require an SNR threshold of 4 for single detectors, and a minimum network SNR of 12 for detection (following \citealt{KAGRA:2013rdx}). We require an event to be detected by 2 detectors and assume a duty cycle of 70\% for both ET and CE. We have 490,835 detected BNS GW events for this next-generation network. We select a random sub-sample of 3,000 events, comparable to the number for O5, in order to use within this work.  We show the redshift distribution of our GW detections in Figure \ref{fig:redshift}. The distance and viewing angle of all simulated events for O5 and XG are shown in Appendix \ref{sec:appendixsimulatedevents} (see Figures \ref{fig:O5detectionsJWST} and \ref{fig:XGdetectionsJWST}).

We compute the expected GW BNS merger detection rates per year for both O5 and XG simulations using the volumetric merger rate of 170$^{+270}_{-120}$ yr$^{-1}$Gpc$^{-3}$ following Table II of \citet{abbott2023population}. We appropriately rescale the rate to account for the non-detection of BNS mergers in O4 until the end of O4b\footnote{We follow the LVK calculation from \url{https://dcc.ligo.org/P2400022/public}},  which is due to the intrinsically low volumetric rates of megers. For O5 we find a 90\% credible interval rate of detections of 29$^{+47}_{-20}$ yr$^{-1}$. For XG simulations, we use the cosmic star formation rate density from \citet{mandau} as a proxy for redshift evolution of merger rate, folded in with a $\propto$\,$t^{-1}$ time delay distribution, and extrapolate the merger rate in the local universe \citep{abbott2023population} to higher redshifts. We then consider detectability of the injected events as a function of redshift to derive the detection rates. We find the BNS merger detection rate in a 90\% credible interval with XG detectors to be 50$^{+79}_{-35} \times 10^3$ yr$^{-1}$.

\subsubsection{Jet Launching Fraction}
\label{sec:jetlaunch}

We determine whether each event launches a successful relativistic jet based on the disk mass and fate of the central remnant. We assume that a black hole should be formed as a result of the merger and that an accretion disk must be present in order for the jet to be launched. The disk mass surrounding the post-merger remnant is a function of both the total mass of the merging binary as well as the threshold mass $M_{\rm th}$, the limiting total mass of the binary system beyond which the BNS merger would result in prompt collapse to a black hole~\citep{Bauswein:2013jpa,Agathos:2019sah}. The disk mass is computed as follows \citep{Dietrich_2020}:
\begin{equation}
    \label{eq:mdisk_bns}
    \log_{\rm 10}\left(M_{\rm disk}^{\rm BNS}\right) = \max \left (-3, a\left(1 + b \tanh \left [\frac{c-{M_{\rm tot}}/{M_{\rm th}}}{d} \right] \right) \right)\, ,
\end{equation}
where $a$ and $b$ are defined as:
\begin{equation}
  \begin{split}
    &a =a_{0} + \delta_{a}x_{i} \\
    &b =  b_{0} + \delta_{b}x_{i}\\
    &x_i = 0.5 \tanh(\beta(q - q_{t})) \\
  \end{split}
\end{equation}
with $a_{0} = -1.581$, $b_{0} = -0.538$, $c = 0.953$, $d = 0.0417$, $\delta_{a}=-2.439$, $\delta_{b} = -0.406$, $\beta=3.910$, $q_{t}=0.900$ and $q$ is the mass ratio of the merging binaries. For successful launching of the jet, we require $M_{\rm disk}^{\rm BNS}$ > $10^{-3}$ $M_\odot$ \citep{PhysRevD.73.064027,Salafia_22} and $M_{\rm tot} > 1.2~M_{\rm TOV}$ \citep{Salafia_22,PhysRevD.97.021501,Ruiz_2019}. $M_{\rm TOV}$ is the Tolman-Oppenheimer-Volkoff limit, which is defined as the maximum mass of a non-rotating neutron star. According to the maximum posterior equation of state from \citet{Huth:2021bsp} used in this study, $M_{\rm TOV}$ is 2.44 $M_{\odot}$. The requirement that $M_{\rm tot} > 1.2~M_{\rm TOV}$ implies that either a hypermassive neutron star that soon collapses into a black hole or a prompt collapse to a black hole are needed to launch the jet.

Using the above criteria, we find a fraction of $\sim\,92\%$ and $\sim\,57\%$ of events launch an event for O5 and XG, respectively. This difference in jet launching fraction is due to the different mass distributions assumed for O5 versus XG, since the O5 population is based on the GWTC-3 population results which could not be properly extrapolated for XG due to the significant difference in distance reach. As a result, the GWTC-3 population allows for a larger population of the high mass population than XG, since that is a truncated power law as opposed to a broad Gaussian distribution. Thus, a larger fraction of the population in O5 satisfied the criterion that $M_{\rm tot} > 1.2~M_{\rm TOV}$. We apply this jet-launch criteria to our simulations, and calculate afterglow and gamma ray detectability only if a jet was launched for that GW event.

\begin{table*}
\centering
\caption{
The X-ray, ultraviolet, optical, infrared, and radio wavelength instruments and their compiled parameters used throughout this work. The run column depicts whether the instrument was considered for our O5 runs or XG runs, according to whether it will be operational during the correct time frame. The detectable afterglows columns refer to the results for simulation Run 7. We count an afterglow event as detectable by a specific instrument if it launches a jet, and the simulated lightcurve flux values for that event are at any point greater than the sensitivity for that instrument. X-ray sensitivities are reported at 1 keV assuming a photon index of $\Gamma$\,$=$\,$1.7$. 
}

    \begin{tabular}{| c| c| c | c | c| c | c | c | c| c|c|} 
    \hline
     Telescope & Frequency & Exposure & Sensitivity & Field of & Run & \multicolumn{4}{|c|}{Detectable Afterglows (Run 7)} & Ref. \\
     \cline{7-10}
     && Time &(mJy) & View && \% O5 & \% XG & \# O5  & \# XG  &\\
     \hline

\textbf{X-ray} & (keV) &&&&&&&($\textrm{yr}^{-1}$)&($\textrm{yr}^{-1}$)& \\

\hline

\textit{Chandra} & 0.5-8 & 50 ks & $4.4 \times 10^{-7}$ & 9.7 arcmin$^2$ & O5 & 4.1 & -- & 0.4 -- 3.1 & -- & [1] \\  

eROSITA & 0.2-2.3 & 1.6 ks & $2.2 \times 10^{-6}$ & 0.833 deg$^2$ & O5 & 2.9 & -- & 0.3 -- 2.2 & -- & [2] \\

\textit{Einstein Probe} & 0.5-2 & 1.5 ks & $1.4 \times 10^{-5}$ & 3600 deg$^2$ & O5 & 2.0 & -- & 0.2 -- 1.5 & -- & [3, 4]  \\

\textit{Swift}/XRT & 0.3-10 & 5 ks & $4.8 \times 10^{-5}$ & $23.6\arcmin \times 23.6\arcmin$ & O5 & 1.5 & -- & 0.1 -- 1.2 & -- &  [5]\\
 
\textit{SVOM}/MXT & 0.2-10 & 10 ks & $9.5 \times 10^{-5}$ & $1.1\degr \times 1.1\degr$ & O5 & 1.3 & -- & 0.1 -- 1.0 & -- & [6] \\

\textit{NewAthena} & 0.2-10 & 50 ks & $5.9 \times 10^{-9}$ & $40\arcmin \times 40\arcmin$ & XG & -- & 2.4 & -- & 355 -- 3090 & [7, 8]\\

\textit{AXIS} & 0.2-10 & 50 ks & $9.6 \times 10^{-9}$ & 0.125 deg$^2$ & XG & -- & 2.2 & -- & 321 -- 2790 & [9]  \\

\textit{Lynx} & 0.5-2 & 100 ks & $1.9 \times 10^{-9}$ & 0.105 deg$^2$ & XG & -- & 3.1 & -- & 464 -- 4038 & [10]\\

 \hline

\textbf{UVOIR} &  (Hz) &&&&&&&& \\

\hline

\textit{ULTRASAT} (NUV) & $1.3 \times 10^{15}$ & 900 s & $4.4 \times 10^{-3}$ & 204 deg$^2$ & O5 & 1.3 & -- & 0.1 -- 1.0 & -- & [11]\\

\textit{UVEX} (NUV) & $1.3 \times 10^{15}$ & 900 s & $5.8 \times 10^{-4}$ & 12 deg$^2$ & O5, XG & 2.1 & 0.5 & 0.2 -- 1.6 & 72 -- 624 & [12] \\

Rubin (g-band) & $6.2 \times 10^{14}$ & 30 s & $4.8 \times 10^{-4}$ & 9.6 deg$^2$ & O5, XG & 2.5 & 0.6 & 0.2 -- 1.9 & 86 -- 752 & [13] \\

Rubin (i-band) & $4.0 \times 10^{14}$ & 30 s & $1.1 \times 10^{-3}$ & 9.6 deg$^2$ &  O5, XG & 2.2 & 0.5 & 0.2 -- 1.7 & 71 -- 621 & [13] \\

Rubin (g-band) & $6.2 \times 10^{14}$ & 180 s & $1.4 \times 10^{-4}$ & 9.6 deg$^2$ & O5, XG & 3.3 & 0.8 & 0.3 -- 2.5 & 119 -- 1032 & [13] \\

Rubin (i-band) & $4.0 \times 10^{14}$ & 180 s & $3.2 \times 10^{-4}$ & 9.6 deg$^2$ & O5, XG & 2.9 & 0.7 & 0.3 -- 2.2 & 103 -- 900 & [13] \\

\textit{Euclid} (H-band) & $1.7 \times 10^{14}$ & 360 s & $9.1 \times 10^{-4}$ & 0.57 deg$^2$ &O5, XG & 2.7 & 0.6 & 0.2 -- 2.0 & 358 -- 1953 & [14] \\

\textit{Roman} (R-band) & $4.9 \times 10^{14}$ & 67 s & $1.2 \times 10^{-4}$ & 0.28 deg$^2$ & O5, XG & 3.6 & 0.9 & 0.3 -- 2.7 & 128 -- 1114 & [15, 16] \\

\textit{Roman} (J-band) & $2.3 \times 10^{14}$ & 67 s & $2.3 \times 10^{-4}$ & 0.28 deg$^2$ & O5, XG & 3.5 & 0.8 & 0.3 -- 2.6 & 123 -- 1066 & [15, 16] \\

\textit{JWST} (F150W) & $2.0 \times 10^{14}$ & 7000 s & $6.9 \times 10^{-6}$ & 9.7 arcmin$^2$ & O5, XG & 9.0 & 1.8 & 0.8 -- 6.8 & 262 -- 2281 & [17]\\

\textit{JWST} (F356W) & $8.5 \times 10^{13}$ & 7000 s & $5.3 \times 10^{-6}$ & 9.7 arcmin$^2$ & O5, XG & 11.6 & 2.2 & 1.0 -- 8.8 & 322 -- 2801 & [17] \\

ELT (J-band) & $2.3 \times 10^{14}$ & 5 hr & $1.3 \times 10^{-5}$  & 78.5 arcmin$^2$ & O5, XG & 7.3 & 1.5 & 0.6 -- 5.5 & 225 -- 1957 & [18] \\

ELT (H-band) & $1.9 \times 10^{14}$ & 5 hr & $5.8 \times 10^{-6}$ & 78.5 arcmin$^2$ & O5, XG & 9.7 & 1.9 & 0.8 -- 7.3 & 275 -- 2397 & [18] \\

\hline

\textbf{Radio} &  (GHz) &&&&&&&& \\

\hline

ALMA  & 97.5 & 1 hr & $3.0 \times 10^{-2}$ & -- & O5 & 3.0 & -- & 0.3 -- 2.3 & -- & [19]\\

ALMA  & 187 & 1 hr & $6.5 \times 10^{-2}$ & -- & O5 & 1.9 & -- & 0.2 -- 1.4 & -- & [19]\\

ALMA  & 324 & 1 hr & $2.2 \times 10^{-1}$ & -- & O5 & 1.0 & -- & 0.1 -- 0.7 & -- & [19]\\

ATCA & 2.1 & 2 hr & $3.6 \times 10^{-2}$ & -- & O5 & 5.5 & -- & 0.5 -- 4.2 & -- &  [20]\\

ATCA & 5.5 & 2 hr & $2.4 \times 10^{-2}$ & -- & O5 & 5.5 & -- & 0.5 -- 4.2 & -- & [20]\\

ATCA & 9 & 2 hr & $2.4 \times 10^{-2}$ & -- & O5 & 5.1 & -- & 0.4 -- 3.9 & -- & [20]\\

VLA  & 6 & 2 hr & $7.0 \times 10^{-3}$ & -- & O5 & 9.1 & -- & 0.8 -- 6.9 & -- & [21] \\

VLA  & 10 & 2 hr & $7.5 \times 10^{-3}$ & -- & O5 & 8.1 & -- & 0.7 -- 6.1 & -- & [21]\\

VLA  & 22.25 & 2 hr & $1.3 \times 10^{-2}$ & -- & O5 & 5.6 & -- & 0.5 -- 4.2 & -- & [21] \\

SKA & 1.4 & 1 hr & $6.0 \times 10^{-3}$ & -- & O5, XG & 12.6 & 0.8 & 1.1 -- 9.5 & 112 -- 978 & [22, 23] \\

SKA & 9 & 1 hr & $3.6 \times 10^{-3}$ & -- & O5, XG & 10.7 & 0.9 & 0.9 -- 8.1 & 129 -- 1120 & [22, 23] \\

SKA & 12.5 & 1 hr & $3.6 \times 10^{-3}$ & -- & O5, XG & 10.0 & 0.8 & 0.9 -- 7.6 & 123 -- 1067 & [22, 23] \\

ngVLA & 2.4 & 1 hr & $7.2 \times 10^{-4}$ & -- & XG & -- & 2.8 & -- & 423 -- 3680 & [24, 25] \\

ngVLA & 8 & 1 hr & $4.2 \times 10^{-4}$ & -- & XG & -- & 3.0 & -- & 439 -- 3819 & [24, 25] \\

ngVLA & 16 & 1 hr & $4.8 \times 10^{-4}$ & -- & XG & -- & 2.4 & -- & 359 -- 3120 & [24, 25] \\

\hline
    \end{tabular}
\label{tab:telescopes}
\tablecomments{References to the telescope information are as follows. For X-ray: [1] \citet{weisskopf2000chandra}, [2] \citet{predehl2021erosita}, [3] \citet{zhang2022estimate}, [4] \citet{yuan2022einstein}, [5] \citet{Evans2020}, [6] \citet{Schneider2023}, [7] \citet{piro2021multi}, [8] \citet{nandra2013hot}, [9] \citet{mushotzky2019advanced}, [10] \citet{gaskin2019lynx}. For UVOIR: [11] \citet{shvartzvald2023ultrasat}, [12] \citet{kulkarni2021science}, [13] \citet{ivezic2019lsst}, [14] \citet{laureijs2011euclid}, [15] \citet{Spergel2013}, [16] \citet{Spergel2015}, [17] \citet{rigby2023science}, [18] \citet{Davies2021}. For Radio: [19] \citet{wootten2009atacama}, [20] \citet{wilson2011australia}, [21] \citet{condon1998nrao}, [22] \citet{braun2017anticipated}, [23] \citet{dewdney2009square}, [24] \citet{butler2018sensitivity}, [25] \citet{butler2019ngvla}.}
\end{table*}

\subsection{Multi-wavelength telescopes and instrument sensitivities}
\label{sec: telescopes}

The afterglow of a binary neutron star merger can be discovered through three routes: 
\textit{i}) the rapid identification ($<$\,$1$ day) of either an X-ray, optical, or radio afterglow following a joint GW-GRB trigger, \textit{ii}) localized late-time ($\sim$\,$10-100+$ days) observations following the detection of a kilonova in wide-field searches after the initial localization of either a GW or joint GW-GRB with a large sky localization, 
\textit{iii}) late-time observations covering the localization of either a GW or GRB in the event a kilonova was not discovered (i.e., no arcsecond localization is available). 
The first method depends largely on the viewing angle $\theta_\textrm{v}$ with respect to the jet's core half-opening angle $\theta_\textrm{c}$, as their ratio $\theta_\textrm{v}/\theta_\textrm{c}$ dictates the  behavior of the afterglow lightcurve \citep[e.g.,][]{Ryan2020}. 
The second method depends largely on source visibility to telescopes in either the Northern or Southern hemispheres or to space-based observatories (e.g., \textit{Roman}, \textit{ULTRASAT}) to discover the kilonova and allow for well-localized and sensitive late-time observations (e.g., VLA, \textit{Chandra}, \textit{JWST}) to detect the afterglow peak (for an off-axis merger). The third method requires wide-field survey instruments to perform the follow-up using a similar tiling strategy as carried out at earlier times for kilonovae. In this case the exact strategy and timing of the observations to best catch the peak is unclear initially, and largely depends on the unknown viewing angle. 

As there are multiple methods, strategies, and telescopes to localize and detect the EM counterpart of a BNS merger, we focus simply on the ability of a variety of telescopes to detect the afterglow, as opposed to their likelihood to localize and then characterize the afterglow behavior. We consider a range of telescopes that will be available during LVK O5 and mission concepts that may exist at the same time as XG detectors. We consider both sensitive narrow-field instruments and wide-field survey telescopes covering a range of wavelengths (X-ray, ultraviolet,  optical, infrared, and radio) in our analysis. 
The telescopes and their assumed sensitivities are listed in Table \ref{tab:telescopes}.

\begin{table*}
    \centering
    \caption{
    The gamma-ray satellites and their compiled parameters used throughout this work. For simplicity we consider each of these satellites for both LVK O5 and XG detector runs as representative gamma-ray sensitivities in these eras. Both the detectable fraction of GRBs and the rate of events is corrected for the instrument field of view and duty cycle. The results are shown for our fiducial run (Run 7).}
    \begin{tabular}{|c|c|c|c|c|c|c|c|c|}

    \hline
    Telescope & Energy (keV) & Flux Limit & Run & \multicolumn{4}{|c|}{Detectable GRBs} & References \\
    \cline{5-8}
     &&(erg/cm$^2$/s)&& \% O5 & \% XG & \# O5 ($\textrm{yr}^{-1}$) & \# XG ($\textrm{yr}^{-1}$) & \\

     \hline

\textit{Swift}-BAT & 15-150 & $2 \times 10^{-8}$ & O5, XG & 0.5 & 0.1 & 0.04 -- 0.4 & 17 -- 150 & [1]\\
\textit{Fermi} & 50-300 & $2 \times 10^{-7}$ & O5, XG & 1.7 & 0.2 & 0.2 -- 1.3 & 24 -- 208 & [2] \\
\textit{SVOM} & 4-150 & $2 \times 10^{-8}$ & O5, XG & 0.5 & 0.1 & 0.05 -- 0.4 & 19 -- 167 & [3] \\
\textit{Daksha} & 20-200 & $4 \times 10^{-8}$ & O5, XG & 3.7 & 0.8 & 0.3 -- 2.8 & 113 -- 982 & [4, 5] \\
\textit{THESEUS} & 2-10,000 & $3 \times 10^{-8}$ & O5, XG & 0.3 & 0.1  & 0.02 -- 0.2 & 11 -- 97 & [6, 7] \\
\textit{COSI} & 200-1,000 & $6 \times 10^{-7}$ & O5, XG & 0.5 & 0.02 & 0.04 -- 0.4 & 2.9 -- 25 & [8] \\

    \hline
    \end{tabular}
    \label{tab:gammasens}
    \tablecomments{The references are as follows: [1] \citet{lien2014probing}, [2] \citet{vonKienlin2020}, [3] \citet{Arcier2022}, [4] \citet{Bhalerao2024a}, [5] \citet{Bhalerao2024b}, [6] \citet{Amati2018}, [7] \citet{Amati2021}, [8] \textit{COSI} Level 2 requirements document, J. Tomsick private communication.}
\end{table*}

\subsubsection{LVK O5}

For LVK O5 we consider a majority of telescopes that are already currently active and likely to still be operational in $\sim$2028, as well as new facilities that will begin operating in that time frame. At X-ray wavelengths these include the \textit{Chandra X-ray Observatory} \citep{weisskopf2000chandra}, the eROSITA instrument onboard the \textit{Spectrum-Roentgen-Gamma} mission \citep{predehl2021erosita}, \textit{Neil Gehrels Swift Observatory} \citep{Gehrels2004} X-ray telescope (XRT; \citealt{Burrows2005}), \textit{Space-based multi-band astronomical Variable Objects Monitor} (\textit{SVOM};  \citealt{Paul2011,Cordier2015}) Microchannel X-ray Telescope (MXT; \citealt{Schneider2023}), and the Wide-field X-ray Telescope (WXT) on the \textit{Einstein Probe} \citep{yuan2022einstein}. 
At ultraviolet wavelengths we consider the \textit{Ultraviolet Transient Astronomy Satellite} (\textit{ULTRASAT}, $\sim$2026; \citealt{shvartzvald2023ultrasat}) and the \textit{Ultraviolet Explorer} (\textit{UVEX}, $\sim$2030; \citealt{kulkarni2021science}), 
while at optical and near-infrared wavelengths we have included the 
\textit{James Webb Space Telescope}  \citep[\textit{JWST};][]{rigby2023science}, \textit{Euclid} \citep{2024arXiv240513491E}, the \textit{Nancy Grace Roman Space Telescope} (\textit{Roman}, $\sim$2026; \citealt{mosby2020properties}), the Vera Rubin Observatory's Legacy Survey of Space and Time (LSST, $\sim$2025; \citealt{ivezic2019lsst}), and the Extremely Large Telescope (ELT, $\sim$2028; \citealt{Davies2021}). 
At radio wavelengths we include the Very Large Array \citep[VLA;][]{condon1998nrao}, the Australia Telescope Compact Array  \citep[ATCA;][]{wilson2011australia}, the Atacama Large Millimeter Array  \citep[ALMA;][]{wootten2009atacama}, and the Square Kilometer Array (SKA, $\sim$2029; \citealt{dewdney2009square}). 

The typical exposure times and sensitivities for each instrument are compiled in Table \ref{tab:telescopes} and are tabulated from the literature for mission concepts or using the instrument exposure time calculators for active missions (e.g., \textit{Chandra}, \textit{Swift},  \textit{JWST}, VLA, and ATCA). Field of views (FoVs) for the instruments are also tabulated in Table \ref{tab:telescopes}.

This study is not a comprehensive list of all available instruments that will be available for use during LVK O5. We aim to provide a range of facilities across all wavelengths that can serve as a reference for the expected sensitivities. We note that there are other proposed missions that are likely to be operational in the O5 time frame, such as the \textit{Chinese Space Station Survey Telescope (CSST)} \citep{CSST2025} and the \textit{Lazuli Space Observatory} \citep{Roy2026}, but we do not consider them at this time.
Moreover, a handful of these missions, which cannot be predicted, may still be operational in the mid- to late-2030s when XG detectors are expected to come online. However, the expectation is that current mission concepts will replace them in sensitivity, and thus we consider the majority of these instruments only for LVK O5. 

\subsubsection{XG detectors}

For XG detector runs, the majority of instruments we consider are currently only mission concepts (though some are actively being developed) that may be active in the mid- to late-2030s as a representative case for the available instruments in the XG era. 
The only currently operational missions we consider for the XG era are \textit{Euclid} and \textit{JWST}, though we tentatively include Rubin in this category as well given its recent start date. In the XG era the LSST is likely complete (after its 10 year run) and more Rubin Observatory time may be available to dedicate to GW and GRB follow-up. 
We therefore consider also the potential for longer exposures (e.g., 180 s) in each filter \citep[e.g.,][]{Andreoni2022a}. 

For potential future missions we consider \textit{NewAthena} \citep[$\sim$2037;][]{nandra2013hot,piro2021multi,Macculi2023,Cruise2025}, \textit{AXIS} \citep[$\sim$2032;][]{mushotzky2019advanced}, and \textit{Lynx} \citep[$\sim$2036;][]{gaskin2019lynx} at X-ray wavelengths and the next-generation VLA (ngVLA, $\sim$2035; \citealt{butler2018sensitivity,butler2019ngvla,Corsi2019}) and SKA \citep{dewdney2009square,braun2017anticipated}  at radio wavelengths. 
At ultraviolet and near-infrared wavelengths we also consider \textit{UVEX}, \textit{Roman}, and thirty meter class telescopes such as the ELT. 

While the available missions during XG detector runs are not clear and can change with time, these instruments offer a range of possible sensitivities during this timeframe across the EM spectrum. 

\subsubsection{Gamma-ray satellites}

We consider a range of available gamma-ray satellites that currently exist and which are likely to exist during LVK O5. The available missions at the time XG detectors will be operational are not known, and we consider the current available detectors and some likely mission concepts as characteristic gamma-ray detector sensitivities for missions to launch in the mid- to late-2030s. Sensitive and large field of view gamma-ray telescopes are required to increase the chance for joint GRB and GW detections, which in turn greatly increase the chances of identifying an additional EM counterpart, namely a kilonova or jet afterglow.

For current instruments we consider the \textit{Swift} Burst Alert Telescope \citep[BAT;][]{Barthelmy2005}, the \textit{Fermi Gamma-ray Burst Monitor} \citep{Meegan2009}, and \textit{SVOM}/ECLAIRs \citep{Arcier2022,Llamas2024}. 
As proposed mission concepts we consider \textit{Daksha} \citep{Bhalerao2024a,Bhalerao2024b,Bhattacharjee2024} and the \textit{Transient High-Energy Sky and Early Universe Surveyor}  \citep[\textit{THESEUS};][]{Amati2018,Amati2021}. We also consider the Compton Spectrometer and Imager \citep[\textit{COSI};][]{Tomsick2019,Tomsick2023,Gulick2024} which has  planned launch in $\sim$2027. We note that there are a number of other currently proposed mission concepts that may be selected or that already have been selected, and that our study is not a comprehensive analysis of all available or potential instruments, see, e.g., \textit{AMEGO-X} \citep{Caputo2022}, \textit{StarBurst}\footnote{\url{https://science.nasa.gov/mission/starburst/}}, \textit{MoonBEAM} \citep{Fletcher2023}, and many others.

\subsection{Afterglow simulations}
\label{sec:AGsims}

\subsubsection{Afterglowpy}

We use the publicly available \texttt{afterglowpy} package\footnote{\url{https://github.com/geoffryan/afterglowpy}} \citep[\texttt{v0.8.0;}][]{Ryan2020,Ryan2023} to simulate GRB jet afterglow lightcurves over a range of afterglow parameters and jet structures. 
We apply \texttt{specType}\,$=$\,$0$, corresponding to synchroton radiation from a forward shock \citep{Sari1998,Wijers1999,Granot2002} without the inclusion of inverse Compton effects \citep{Sari2001,Zou2009,Beniamini2015,Jacovich2020,McCarthy2024}. We consider only jets propagating into a uniform density environment, typical of short GRBs \citep{Fong2015,OConnor2020}. 
The generated afterglow depends on the following parameters: the observer's viewing angle, $\theta_\textrm{v}$;  the isotropic equivalent kinetic energy of the jet's core, $E_\textrm{kin,0}$; the jet's core half-opening angle, $\theta_\textrm{c}$;  the jet truncation angle outside of which the energy is 0, $\theta_\textrm{w}$; the density of the surrounding medium (henceforth called the circumburst density), $n$;  the electron energy distribution index, $p$;  the fraction of energy that goes into relativistic electrons, $\epsilon_e$;  the fraction of energy that goes into magnetic energy, $\epsilon_B$;  
the luminosity distance to the source, $d_\textrm{L}$; and the corresponding redshift $z$.

We consider GRB jets with both a power-law (PLJ; \texttt{jetType}\,$=$\,$4$) and Gaussian (GJ; \texttt{jetType}\,$=$\,$0$) angular kinetic energy structure $E_\textrm{kin}(\theta)$. These are defined by \citep{Ryan2020}:
\begin{align}
\label{eqn: GJ}
    \textrm{GJ:} & \:\:\:\:\:\:\:\:\:\: E_\textrm{kin}(\theta)=E_\textrm{kin,0}\, \exp\Bigg(-\frac{1}{2}\frac{\theta^2}{\theta_\textrm{c}^2}\Bigg) \:\:\:\:\:\:\:\:\:\:\:\: \textrm{for} \:\:\: \theta\leq\theta_\textrm{w}, \\
\label{eqn: PLJ}
    \textrm{PLJ:} & \:\:\:\:\:\:\:\:\:\: E_\textrm{kin}(\theta)=E_\textrm{kin,0}\,\Bigg(1+\frac{1}{b}\frac{\theta^2}{\theta_\textrm{c}^2}\Bigg)^{-b/2} \:\:\:\:\:\:\:\:\:\: \textrm{for} \:\:\: \theta\leq \pi/2,
\end{align}
where $\theta_\textrm{w}$\,$\leq$\,$\pi/2$ is the truncation angle of the GJ and $b$ is the power-law slope of the PLJ outside of the core. We apply a typical truncation angle of $\theta_\textrm{w}$\,$=$\,$4.9\theta_\textrm{c}$ \citep{Ryan2020}, but also consider $\theta_\textrm{w}$\,$=$\,$3\theta_\textrm{c}$ as a test of this assumption's impact on our results. The truncation angle is an arbitrary parameterization of a sharp cutoff to the energy profile, observed in some numerical simulations \citep[e.g.,][]{Aloy2005}, which also serves to decrease the calculation time in \texttt{afterglowpy} for material that does not influence the observed lightcurve. In general $\theta_\textrm{w}$ is unconstrained in afterglow modeling, though we adopt values similar to those derived for GW170817 \citep[e.g.,][]{Ryan2023}. 
We note that the PLJ model, for essentially all values of $b$, allows for more energetic material outside the jet's core than the GJ model. Both the PLJ and GJ models in \texttt{afterglowpy} have been used to model the full lightcurve of GW170817 \citep[e.g.,][]{Ryan2020,Troja2020,Troja2022,Ryan2023}. For the PLJ model, \citet{Ryan2023} found a power-law slope of $b$\,$\approx$\,$9$\,$-$\,$10$ was able to match the data and we adopt similar values here.

By default \texttt{afterglowpy} does not include an angular profile for the bulk Lorentz factor $\Gamma$ (see Appendix \ref{sec: afterglowpyassumptions} for a discussion), and generally adopts an infinite bulk Lorentz factor at all angles, leading to a deceleration radius of zero; i.e., material at all angles is in the deceleration regime (where $\Gamma(\theta)$\,$\propto$\,$E(\theta)^{1/2}$; \citealt{Ryan2020}) and has skipped the initial coasting phase (where $\Gamma$\,$=$\,constant). 
For on-axis observers there is no change to the lightcurve if observations only occur after the (on-axis) deceleration time $t_\textrm{dec,0}$  \citep{Liang2010,Lu2012,Ghirlanda2012,Ghirlanda2018}. This is the case for cosmological short GRBs, even those with rapid follow-up ($<$\,$100$ s) by \textit{Swift}, which are almost always observed in the deceleration regime, see discussions in \citet{OConnor2020,OConnor2024}.

However, the angular Lorentz factor profile also dictates when material at different angles becomes no longer beamed to the observer, and, therefore, can significantly impact the observed lightcurve for off-axis observers (Appendix \ref{sec: afterglowpyassumptions}) and the inferred afterglow detectability. Therefore, we adopt an angular profile for the Lorentz factor that is the same as the energy profile (Equations \ref{eqn: GJ} and \ref{eqn: PLJ}) such that  
\begin{equation}
\label{eqn:GammaEvenMass}
    \frac{\Gamma(\theta)-1}{\Gamma_0-1} = \frac{E_\gamma(\theta)}{E_{\gamma,0}},
\end{equation}
where $\Gamma_0$ is the initial bulk Lorentz factor at the jet's core. This setup, defined within \texttt{afterglowpy} \citep{Ryan2020} through the \texttt{GammaEvenMass} flag (Appendix \ref{sec: afterglowpyassumptions}),  provides identical mass loading throughout the jet. We adopt an initial bulk Lorentz factor at the jet's core of $\Gamma_0$\,$=$\,$300$ throughout this work \citep{Ghirlanda2018}. 

\subsubsection{Connecting to GW simulations}

From our GW simulations outlined in \S \ref{sec:GWsims} we have produced a population of GW events for both O5 and XG detector runs containing 2,403 and 3,000 BNS mergers for O5 and XG, respectively. For each of these events we obtain the inclination angle $\iota$ and luminosity distance $d_\textrm{L}$, which we convert to a redshift $z$ assuming a flat $\Lambda$CDM \textit{Planck} cosmology \citep{aghanim2020planck}. 
Given the inclination angle, $\iota$, we can determine the viewing angle with respect to the jet's axis, which we assume is aligned with the binary's orbital angular momentum axis (which may not always be the case, see, e.g., \citealt{Muller2024} for a discussion). The viewing angle is then: 
\begin{equation}
    \theta_\textrm{v} = \min(\iota, \pi - \iota). 
\end{equation}
We determine whether each event launched a successful relativistic jet according to the criteria outlined in \S \ref{sec:jetlaunch}. If an event did not launch a jet then we automatically consider both the GRB and afterglow not detectable in our simulations. We also consider the afterglow and jet structure to be independent of the compact object masses as with only a single multimessenger event these properties are not suitably connected to observations. 

\begin{table*}
\centering
\caption{The afterglow parameter cases assumed throughout this work. Here $U[x_1,x_2]$ refers to a uniform distribution between $x_1$ and $x_2$, and $N[y_1,y_2]$ refers to a normal distribution with median $y_1$ and standard deviation $y_2$. 
}
\begin{tabular}{|c|c|c|c|c|c|}
\hline
 Label & p & $\log(n/\textrm{cm}^{-3})$  & $\log\varepsilon_e $& $\log\varepsilon_B$  & $\log\varepsilon_\gamma$ \\
\hline
Case 1 & $U[2.1,2.8]$ & $N[-3,1]$ & $N[-0.82,0.15]$ & $N[-3,1]$ & $N[-0.82,0.15]$  \\
Case 2 & $U[2.1,2.8]$ & $U[-4,0]$ & $U[-2,-0.5]$ & $U[-4,-0.5]$ &   -- \\
\hline
\end{tabular}
\label{tab: afterglowparameters}
\end{table*}

\begin{table*}
\centering
\caption{The simulation setups performed in this work. The same setups were used for both LVK O5 and XG detectors. GJ refers to the Gaussian GRB jet, and PLJ refers to the power-law GRB jet. G16 refers to the energy distribution derived from \citet{Ghirlanda2016} while F15 refers to the inferred kinetic energy distribution from \citet{Fong2015}. The ``Narrow'' distribution of core angles refers to the log-normal distribution centered at 6 degrees, the "Very Narrow" refers to the log-normal distribution centered at 3.4 degrees, the "Fixed" distribution corresponds to a fixed core angle value of 3.4 degrees, while "RE23" refers to the inferred core angle distribution from \citet{RoucoEscorial2022}. We take Run 7 to be our fiducial run. 
}
\begin{tabular}{|c|c|c|c|c|c|}
\hline
 Label & Structure & Afterglow & $E_\textrm{kin,0}$ & $\theta_\textrm{c}$ & $\theta_\textrm{w}$   \\
\hline
Run 1 & GJ & Case 1 & G16 & RE23 & $\min(4.9\theta_\textrm{c},\pi/2)$ \\
Run 2 & GJ & Case 1 & G16 & RE23 & $\min(3\theta_\textrm{c},\pi/2)$\\
Run 3 & GJ & Case 1 & G16 &  Narrow & $\min(4.9\theta_\textrm{c},\pi/2)$\\
Run 4 & PLJ & Case 1 & G16 & Narrow & -- \\ 
Run 5 & GJ & Case 2 & F15 & RE23 & $\min(4.9\theta_\textrm{c},\pi/2)$ \\
Run 6 & GJ & Case 1 & G16 & Fixed & $\min(4.9\theta_\textrm{c},\pi/2)$\\
Run 7 & GJ & Case 1 & G16 & Very Narrow & $\min(4.9\theta_\textrm{c},\pi/2)$\\
\hline
\end{tabular}
\label{tab: simulationruns}
\end{table*}

\begin{figure*}
    \includegraphics[width=\columnwidth]{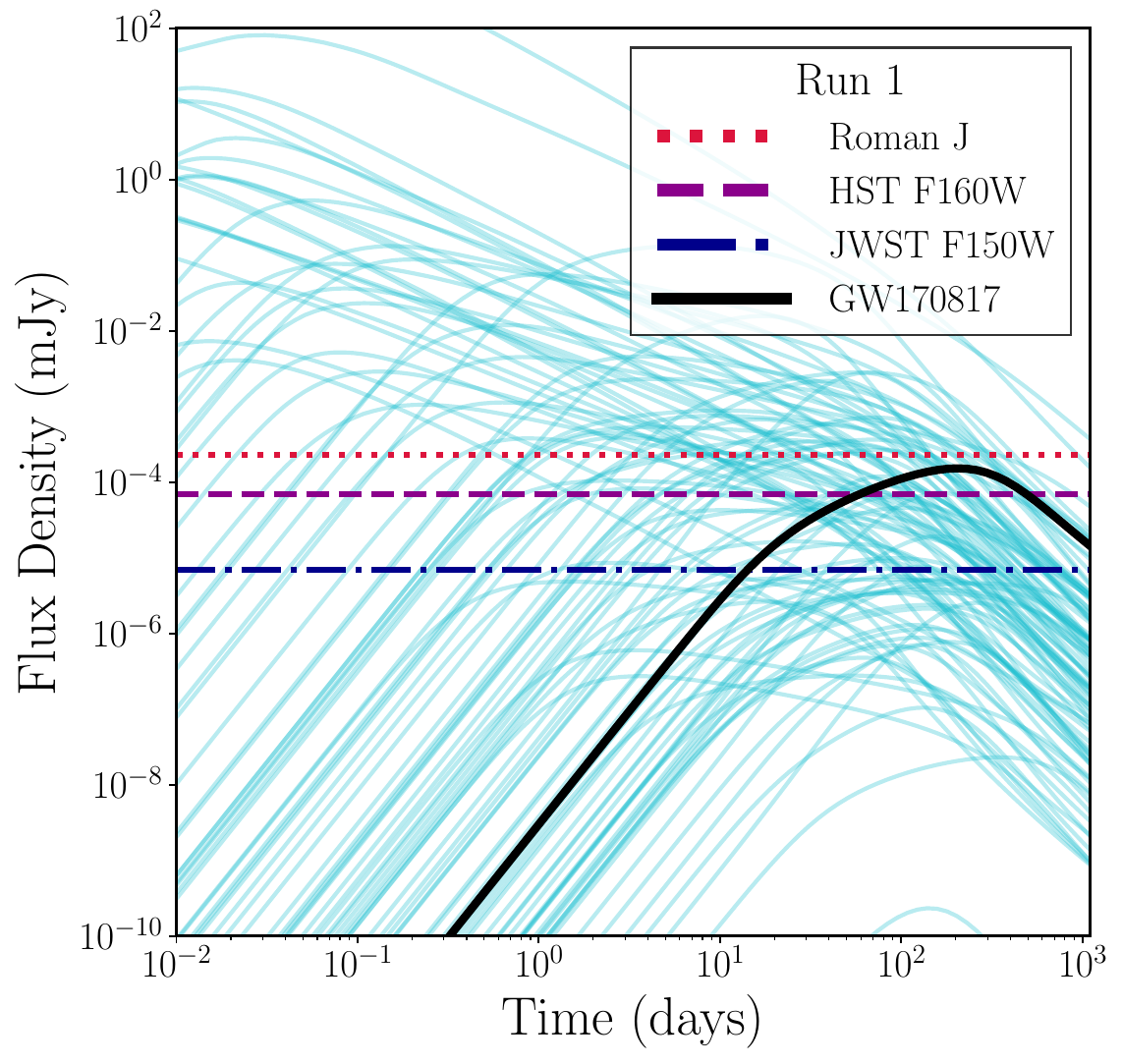}
    \includegraphics[width=\columnwidth]{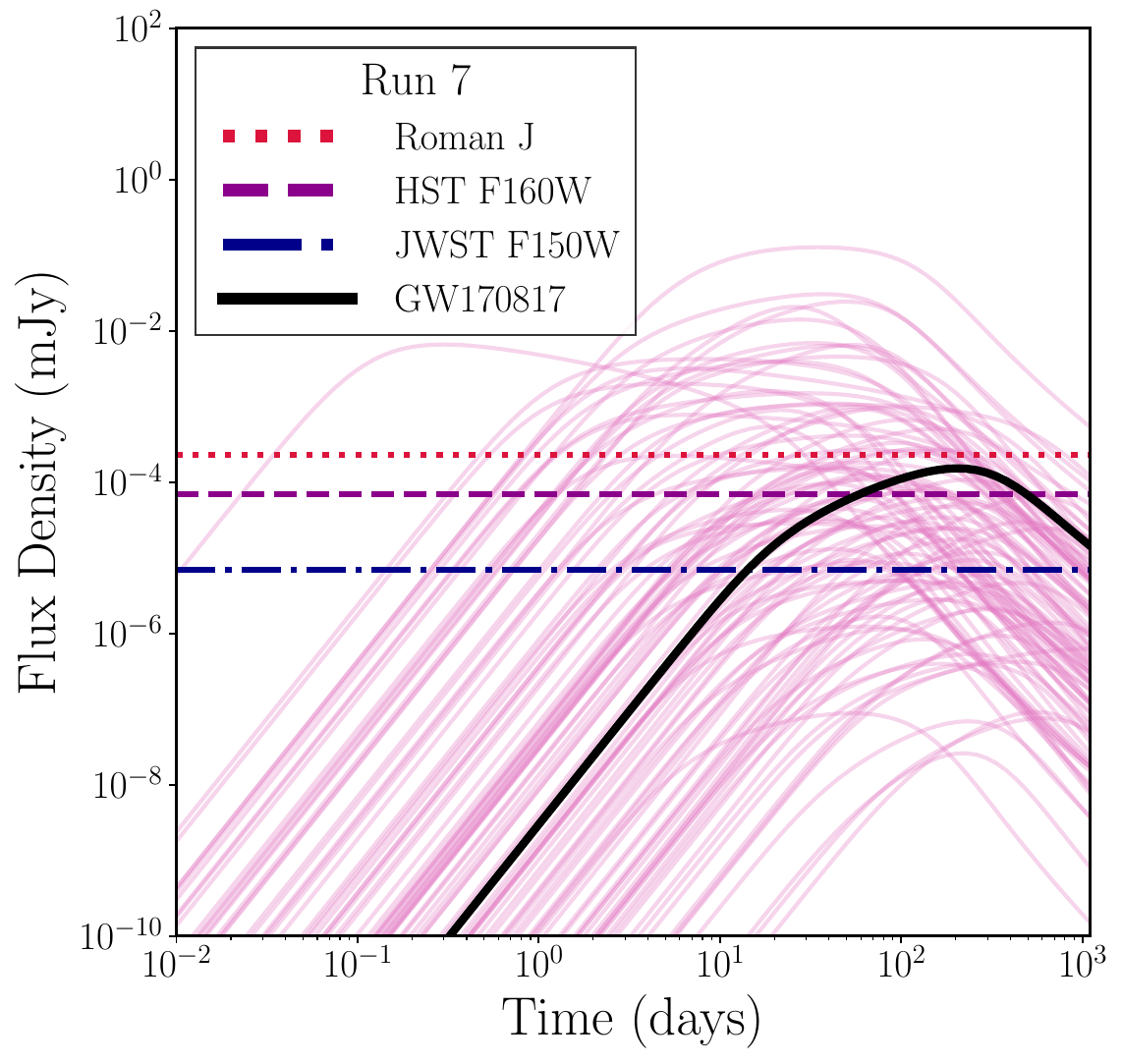}
    \caption{100 simulated \texttt{afterglowpy} lightcurves at a frequency of $2.017 \times 10^{14}$ Hz in the near-infrared showing the impact of afterglow parameter assumptions for a simulated GW event with fixed $d_\textrm{L}$\,$=$\,$130$ Mpc, 
    $\theta_\textrm{v}$\,$=$\,$0.36$ rad (20 deg). The afterglow parameters are sampled from Case 1. Run 1 is shown on the left in cyan, and Run 7, using the very narrow distribution for $\theta_\textrm{C}$, is shown on the right in pink. A lightcurve simulated from the best-fit GW170817 afterglow parameters is shown by a thick black line \citep{Ryan2023}. The sensitivities of a few instruments, \textit{Roman}, \textit{HST}, and \textit{JWST}, are shown by the dashed horizontal lines. 
    }
    \label{fig:170817lightcurves}
\end{figure*}

\subsubsection{Afterglow parameter distributions}
\label{sec:AGparams}

Here we outline our choices of afterglow parameter distributions that are typically found from afterglow modeling of short GRBs. The two main cases of afterglow parameters we consider are shown in Table \ref{tab: afterglowparameters}. 
For the slope $p$ of the electron's power-law energy distribution $N(E)$\,$\propto$\,$E^{-p}$ we adopt a uniform distribution between $U[2.1, 2.8]$. The density of the surrounding uniform environment is taken to be a normal distribution in log$_{10}$, with mean and standard deviation $\log(n/\textrm{cm}^{-3})$\,$=$\,$-3\pm1$ and is based on observations of short GRB afterglows \citep{Fong2015,OConnor2020}. The low density matches the expected distribution of interstellar medium (ISM) densities \citep{OConnor2020,Mandhai2021} at their typically large host galaxy offsets \citep{Fong2013,Fong2022,Nugent2022,OConnor2022}. We also adopt a lognormal distribution for the microphysical parameters of $\log\varepsilon_e$\,$=$\,$-0.82\pm0.15$ \citep{Nava2014,BeniaminiVanDerHorst,Duncan2023} for the energy fraction that goes to electrons and $\log\varepsilon_B$\,$=$\,$-3\pm1$ \citep{Barniol2014,Santana2014,Zhang2015} for the magnetic fields energy fraction. In order to test our base assumptions on these parameters (Table \ref{tab: afterglowparameters}) we also consider uniform distributions in $\log(n/\textrm{cm}^{-3})$\,$=$\,$U[-4,0]$, $\log\varepsilon_e$\,$=$\,$U[-2,-0.5]$, and $\log\varepsilon_B$\,$=$\,$U[-4,-0.5]$. This is discussed further in \S \ref{sec: AGparamimpact}.

For the jet's core half-opening angle we consider four different distributions, as listed in Table \ref{tab: simulationruns}. In our "RE23" distribution, we adopt the probability distribution of opening angles derived in \citet{RoucoEscorial2022} based on late-time X-ray observations of short GRBs and the corresponding constraints to the time of their jet break \citep{SariPiranHalpern1999,Rhoads1999,Frail2001,Nakar2002,Lamb2021rev}. As we have selected a truncation angle $\theta_\textrm{w}$ for the GJ structure that depends on the core angle, e.g., $\theta_\textrm{w}$\,$=$\,$4.9\theta_\textrm{c}$, we must set $\theta_\textrm{w}$\,$\leq$\,$\pi/2$ as an upper limit for the small sample of wide jets ($\theta_\textrm{c}$\,$>$\,$15$ deg; 0.26 rad) derived in \citep{RoucoEscorial2022}. In order to test the impact of the tail of wide core angles we also utilize a distribution of narrow opening angles described by a normal distribution with mean and standard deviation of $\log(\theta_\textrm{c}/\textrm{rad})$\,$=$\,$-1\pm0.15$. This distribution, listed as "Narrow" in Table \ref{tab: simulationruns}, is centered at around 6 deg (0.1 rad) and chosen to match the approximately Gaussian distribution of core angles when excluding those with wide jets \citep{Fong2015,RoucoEscorial2022}. A major reason for this choice is that extremely wide jets (with opening angles as wide as 30 deg; 0.5 rad) will be detectable at extreme viewing angles covering all the way out to $\theta_\textrm{v}$\,$\approx$\,$\pi/2$\footnote{We performed tests including a counter-jet within \texttt{afterglowpy} \citep[e.g.,][]{Ryan2023,Li2024counterjet}, but found the impact on afterglow detection to be negligible.}, and we wish to test how this small sample of events (which are quite different from the inferred core angle for GW170817 of $\sim$\,$3$\,$-$\,$5$ deg; 0.06 rad) impacts our results. We do find a significant impact, and discuss this further in \S \ref{sec: AGparamimpact}. To also test events similar to GW170817, we simulate Run 6 with $\theta_\textrm{c}$ fixed to a value of 3.4 degrees, labeled as "Fixed" in Table \ref{tab: simulationruns}. Lastly, for a more realistic case than Run 6, we apply our "Very Narrow" distribution: a normal distribution with mean and standard deviation of $\log(\theta_\textrm{c}/\textrm{rad})$\,$=$\,$-1.23\pm0.15$. This distribution centers at around 3.4 degrees, thus still testing events similar to GW170817 but allowing for variation. We choose Run 7 with our Very Narrow distribution as our fiducial and most conservative run that still allows us to sample from a distribution of core opening angles.

In order to determine the isotropic equivalent core kinetic energy $E_\textrm{kin,0}$ of the jet as input for \texttt{afterglowpy} we employ two methodologies. The first is to adopt the observed distribution of short GRB isotropic equivalent core gamma-ray luminosities $L_{\gamma,0}$ from \citet{Ghirlanda2016} and convert them to energies $E_{\gamma,0}$\,$=$\,$L_{\gamma,0} T_{90}$ assuming a short GRB rest-frame duration of $T_{90}$\,$=$\,$0.2$ s. The kinetic energy $E_\textrm{kin,0}$\,$=$\,$E_{\gamma,0}(1-\varepsilon_\gamma)/\varepsilon_\gamma$ is then computed assuming a gamma-ray efficiency $\varepsilon_\gamma$ that is sampled from a normal distribution with mean and standard deviation  $-0.82$ and $0.15$ in $\log\varepsilon_\gamma$  \citep{Nava2014,Beniamini2016corr,BeniaminiVanDerHorst}. We set a lower bound to the gamma-ray efficiency of 1\%. This method is listed as "G16" in Table \ref{tab: simulationruns}. The second method, listed as "F15" in Table \ref{tab: simulationruns}, is to utilize the inferred core kinetic energies from short GRB afterglow modeling. In this case the kinetic energies are taken to be a normal distribution in $\log(E_\textrm{kin,0}/\textrm{erg})$ with mean 51.6 and standard deviation 0.5 \citep{Fong2015,RoucoEscorial2022}. Both of these methods obtain very similar energy distributions for short GRBs; the main difference being the low energy tail inferred by  \citet{Ghirlanda2016}, see also \citet{Wanderman2015,Salafia2023}.

\subsubsection{Afterglow detectability}
\label{sec:AGdet}

For every simulated BNS merger (\S \ref{sec:GWsims}) we perform a simple Monte Carlo simulation by sampling from the distribution of afterglow parameters $N$\,$=$\,$1,000$ times. This leads to $\sim$\,$2.4 \times 10^6$ and $3.0 \times 10^6$ simulated lightcurves for O5 and XG, respectively. 
We compute the afterglow lightcurves between observer frame times of $10^{-2}$ d and $6.5$ yr after merger in 100 log-spaced steps. 
The afterglow fluxes are computed at each frequency corresponding to every telescope we are considering for detectability. A table containing the list of telescopes, their effective frequencies, sensitivities, and whether they are expected to be available in O5 or next-generation is shown in Table \ref{tab:telescopes}. For each event that launches a jet, we determine the detectable fraction by comparing the afterglow lightcurve to these detector sensitivity limits at a given frequency. We perform this detectability simulation for seven different setups outlined in Table \ref{tab: simulationruns}. 

\begin{figure*}
    \centering
    \includegraphics[width=\textwidth]{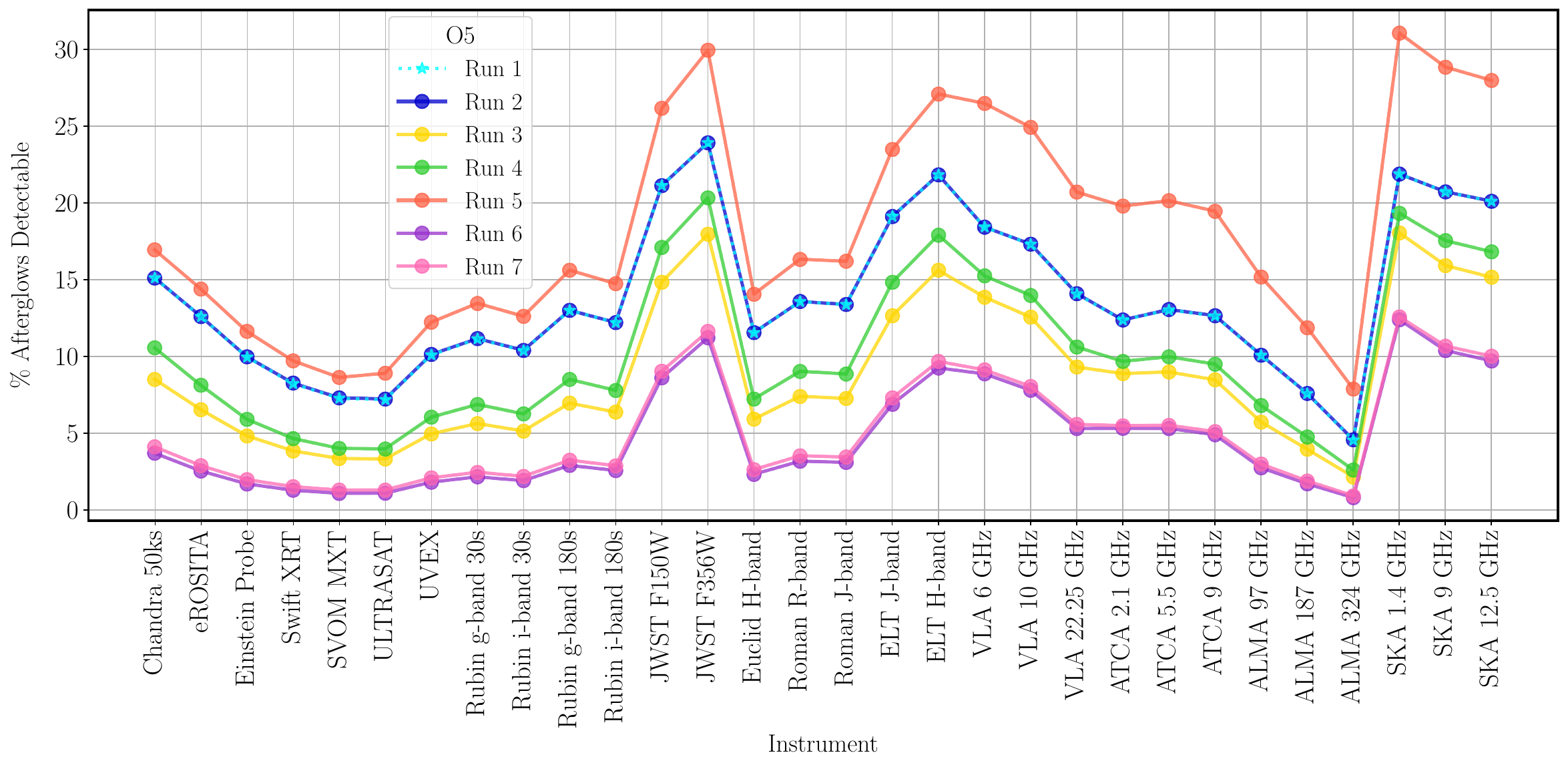}
    \includegraphics[width=\textwidth]{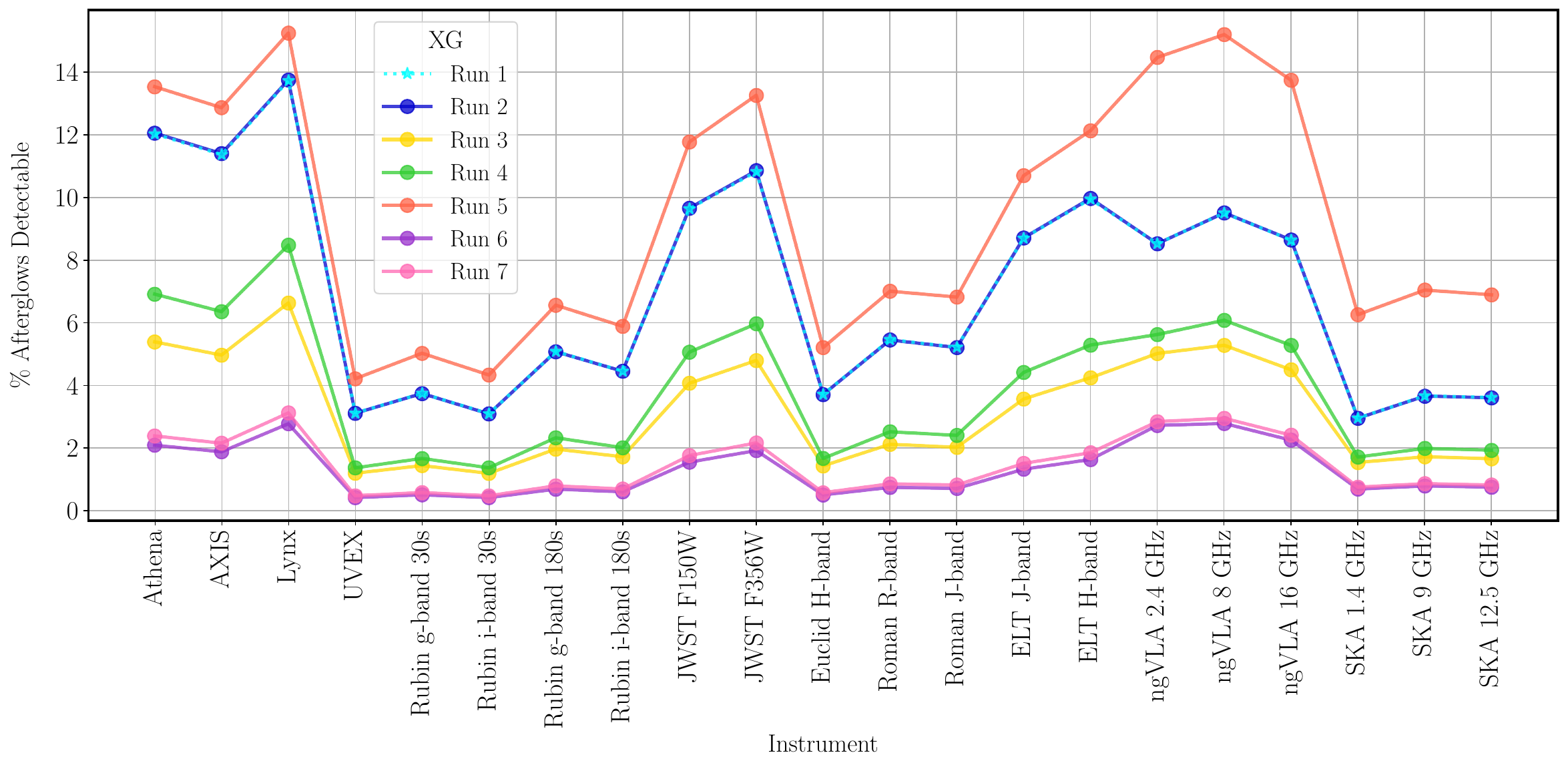}
    \caption{Total percent of afterglows detectable by each instrument for O5 (top) and XG (bottom) over all simulated times between $10^{-2}$ d and 6.5 yr after the merger. We show how these detections change based on each run (different color lines; shown in label). 
    }
    \label{fig:detfracsAG}
\end{figure*}

\begin{figure*}
    \centering
    \includegraphics[width=\textwidth]{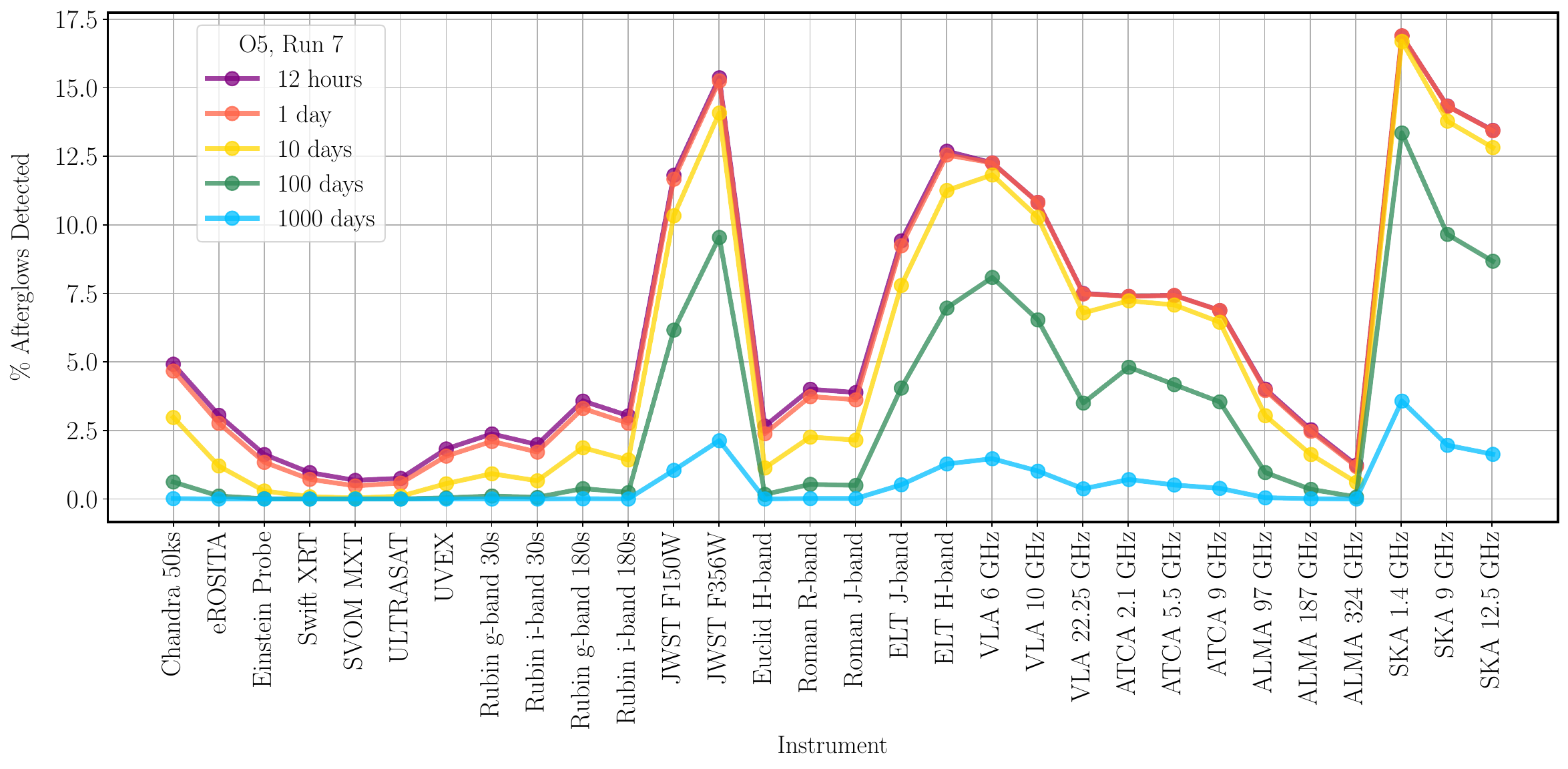}
    \includegraphics[width=\textwidth]{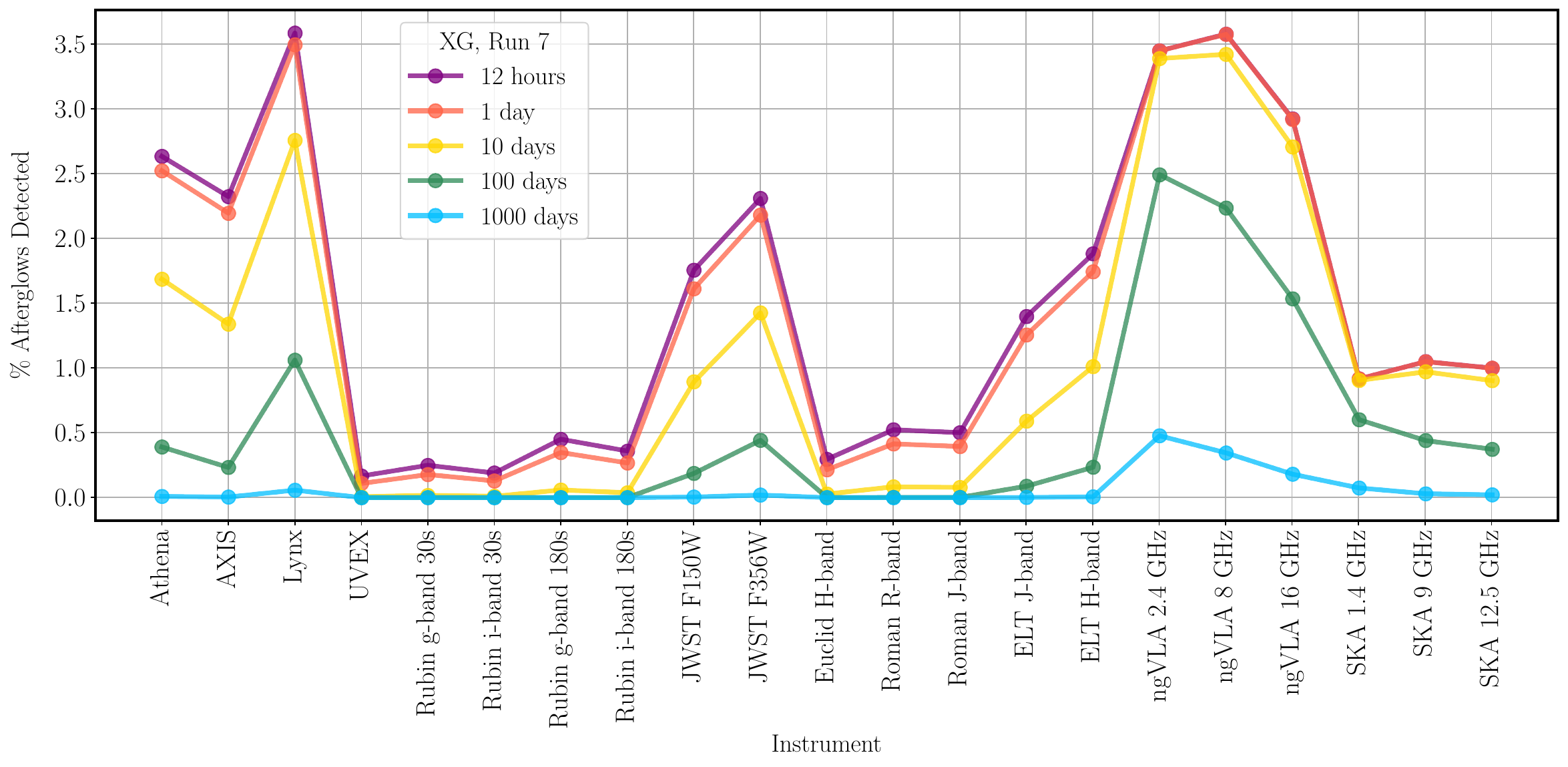}
    \caption{Percent of afterglows detectable by each instrument for O5 (top) and XG (bottom) as a function of time after the event (observer frame). We show how many afterglows are detectable after 12 hours, 1 day, 10 days, 100 days, and 1000 days after the event. Here we consider only our fiducial run: simulation Run 7. 
    }
    \label{fig:detfracsmaxtime}
\end{figure*}

An example of the different possible lightcurves that are generated for a single GW event (fixed viewing angle and distance) are displayed in Figure \ref{fig:170817lightcurves}.  In Figure \ref{fig:detfracsAG} we show the total afterglow detection percentage as a function of instrument (described in Table \ref{tab:telescopes}) and parameter assumptions in different simulation runs (described in Table \ref{tab: simulationruns}). While Figure \ref{fig:detfracsAG} does not account for the exact time of detection, we show how the time after merger impacts detectability in Figure \ref{fig:detfracsmaxtime}. This is discussed further in \S \ref{section:Results}. 

We do not account for the impact of dust extinction on optical and infrared sensitivity or hydrogen gas absorption to X-rays. 
Furthermore, we have not accounted for telescope observing constraints or the location of the simulated BNS on the sky with respect to the telescopes, Sun, or Moon. As such, the presented detection fractions and rates should be treated as upper limits to the true rate. We do however discuss the detectability timescales and peak times as a function of viewing angle in \S \ref{section:Results}. 
Our study is strictly based on the ability for a given telescope sensitivity to detect an afterglow, and we leave afterglow characterization to future work. 

\subsubsection{Gamma-ray detectability}
\label{sec: gammadet}

While we do not require that the prompt gamma-ray emission is detectable in order for the afterglow to be detectable, we also track this information in our Monte Carlo simulations and consider the probability of a joint detection (defined as both a GRB and afterglow detection; as all simulated events already have a GW detection based on \S \ref{sec:GWsims}) during LVK O5 and XG runs. In \S \ref{sec:AGparams} we outlined two methods for determining the core kinetic energy: \textit{i}) using the observed broken power-law gamma-ray luminosity function \citep{Ghirlanda2016} and applying a gamma-ray efficiency or \textit{ii}) adopting the inferred core kinetic energies from afterglow modeling efforts \citep{Fong2015,RoucoEscorial2022}. In order to convert the latter method to a core gamma-ray energy $E_{\gamma,0}$, we apply a gamma-ray efficiency in reverse such that $E_{\gamma,0}$\,$=$\,$E_\textrm{kin,0}\varepsilon_\gamma/(1-\varepsilon_\gamma)$. In this case we consider a 50\% gamma-ray efficiency as found by \citet{Fong2015}, which yields $E_{\gamma,0}$\,$=$\,$E_\textrm{kin,0}$.

As we adopt the same energy structure (Equations \ref{eqn: GJ} and \ref{eqn: PLJ}) for both the gamma-ray energy and kinetic energy, effectively we have assumed in this case that the gamma-ray efficiency is constant with angle. This provides an upper bound on the number of joint detections as the efficiency is expected to decrease with angle \citep{BeniaminiNakar2019,OConnor2024}. In addition, a decreasing Lorentz factor with angle can cause other issues for gamma-ray production, such as an increasing optical depth \citep[e.g.,][]{Gill2020} or  rapidly decreasing the dissipation radius \citep[e.g.,][]{LambKobayashi2016,BBPG2020}, both of which can potentially lead to a cutoff in gamma-ray production at large angles  \citep[$\theta_\textrm{obs}/\theta_\textrm{c}$\,$>$\,$2$;][]{BeniaminiNakar2019,BBPG2020,Gill2020}. For simplicity, we neglect the effects of these mechanisms in this work.

The observed line-of-sight (isotropic equivalent) gamma-ray energy $E_\gamma(\theta)$ depends on the viewing angle $\theta_\textrm{v}/\theta_\textrm{c}$ for both the GJ (Equation \ref{eqn: GJ}) and PLJ (Equation \ref{eqn: PLJ}). We consider only the jet's line-of-sight contribution to the prompt gamma-ray emission and do not account for de-beamed emission from the core \citep{Kasliwal2017,Ioka2018}. This choice has a negligible impact as the line-of-sight structured jet energy dominates over the de-beamed emission. 

However, we do consider the addition of a quasi-isotropic cocoon producing gamma-rays through shock breakout \citep{Lazzati2017,Duffell2018,Salafia2018,Bromberg2018,Gottlieb2018,Nakar2018,Matsumoto2019a}. We adopt the analytic model presented by \citet{Duffell2018}, as previously adopted for similar gamma-ray detectability studies of short GRBs \citep[e.g.,][]{Beniamini2019structuredjet,Dichiara2020,Bhattacharjee2024}. We adopt a cocoon breakout efficiency of $\eta_\textrm{br}$\,$=$\,$10^{-3}$ and a peak energy for the cocoon of 100 keV. This leads to typical isotropic-equivalent energies of $\approx$\,$10^{47}$ erg for the cocoon. 
We only apply the cocoon model for events simulated for LVK O5, as the further distances of events in XG do not allow for the detection of a cocoon.

The jet's core energy $E_{\gamma,0}$ is converted to the line-of-sight energy $E_\gamma(\theta_\textrm{v})$, which is taken to be in the rest frame $1$\,$-$\,$10,000$ keV energy range. This energy $E_\gamma(\theta_\textrm{v})$ is then transformed to a line-of-sight gamma-ray fluence $\phi_\gamma(\theta_\textrm{v})$ in different (observer frame) energy ranges $[e_\textrm{low},e_\textrm{high}]$ corresponding to the telescopes we consider here (see Table \ref{tab:telescopes}). For simplicity, and due to the short duration of the GRBs considered here, we consider equivalent fluence sensitivity limits to the flux limits displayed in Table \ref{tab:telescopes}. In order to determine the fluence in the observer frame bandpass we apply a bolometric correction:
\begin{align}
k_{\textrm{bol}}=\frac{\int^{10\,\textrm{MeV}}_{1\,\textrm{keV}}N(E)\,E\,dE}{\int^{(1+z)\,e_\textrm{high}}_{(1+z)\,e_\textrm{low}}N(E)\,E\,dE}, 
\label{eqn: kbol}
\end{align}
where $N(E)$ represents the prompt emission energy spectrum, defined by a Band function \citep{Band1993}. The prompt emission spectral parameters are taken to be the observed values for \textit{Fermi} short GRBs \citep{Nava2011}, where the low energy spectral slope is $\alpha$\,$=$\,$-0.5\pm0.4$ and (observer frame) peak energy at the jet's core is a normal distribution of $\log E_\textrm{p,0}$\,$=$\,$2.9\pm0.2$ keV. We fix the high energy spectral slope to $\beta$\,$=$\,$-2.25$. 

We also assume that the peak energy is constant in the comoving frame. The peak energy at a given viewing angle $E_\textrm{p}(\theta_\textrm{v})$ is then given by $E_\textrm{p}(\theta_\textrm{v})$\,$=$\,$E_\textrm{p,0}\Gamma(\theta_\textrm{v})/\Gamma_0$\,$=E_\textrm{p,0}E_\gamma(\theta_\textrm{v})/E_{\gamma,0}$ \citep{Bhattacharjee2024}, where the Lorentz factor profile $\Gamma(\theta)$ is taken to be the same as the energy profile (Equation \ref{eqn:GammaEvenMass}). As such the peak energy is independent of assumptions for the initial Lorentz factor $\Gamma_0$. 

\begin{figure*}
    \centering
    \includegraphics[width=\columnwidth]{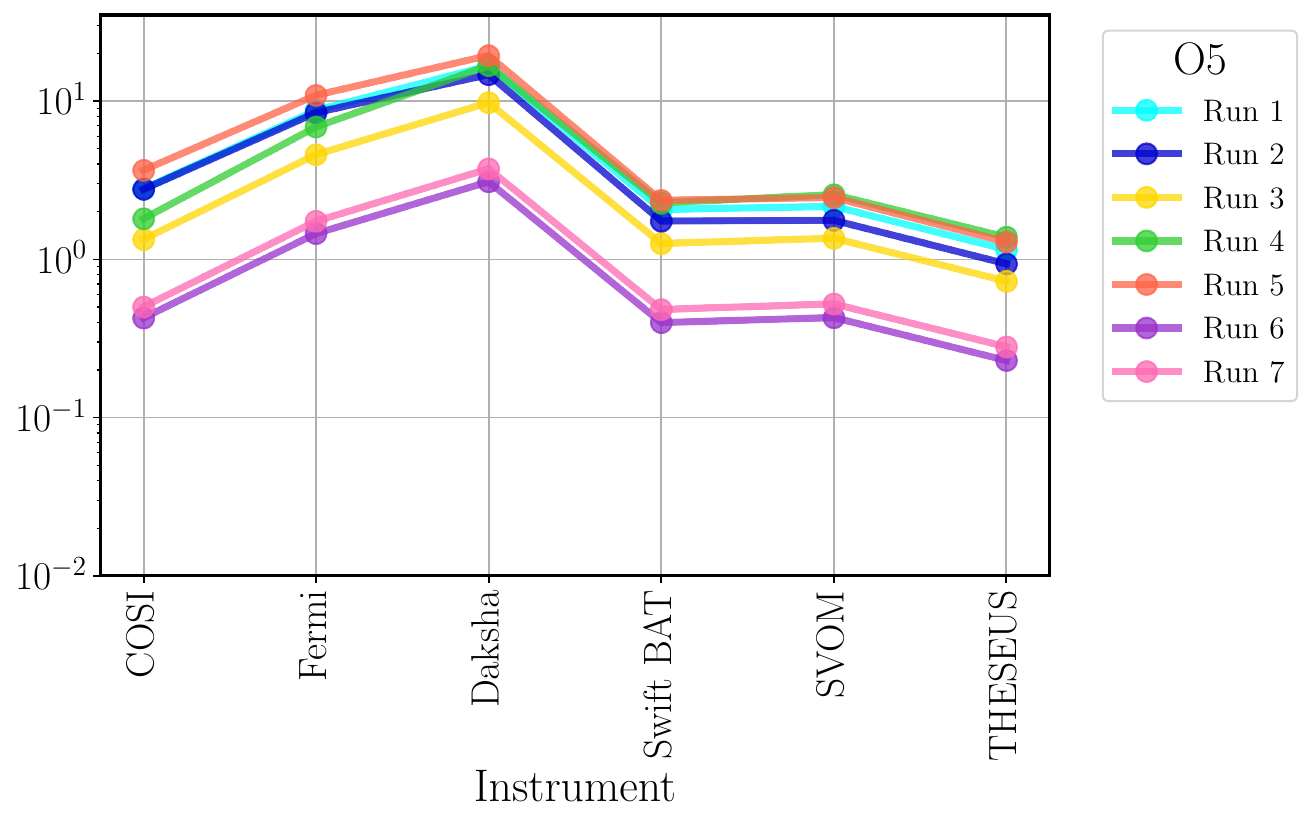}
    \includegraphics[width=\columnwidth]{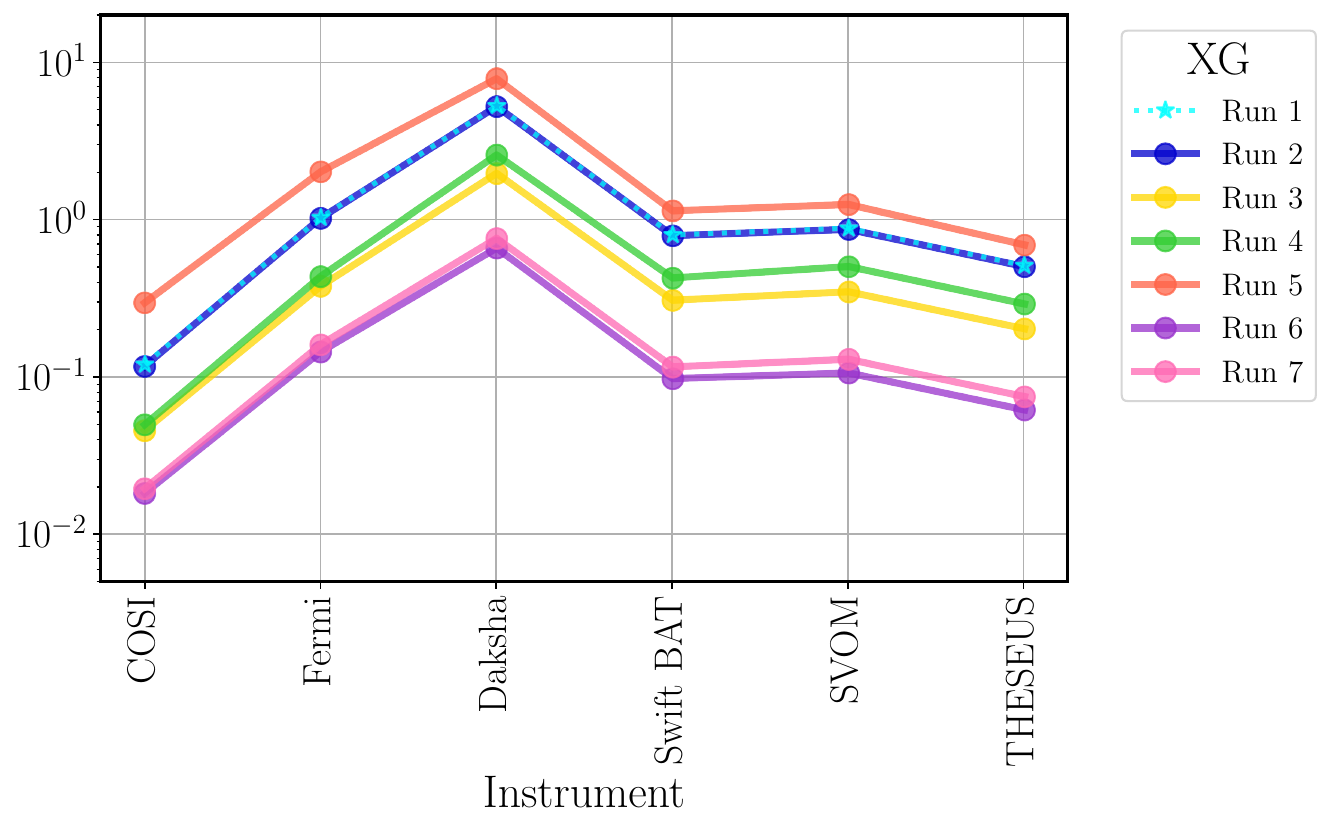}
    \caption{Percent of prompt GRB emission detectable in gamma-rays by each instrument for O5 (left) and XG (right). We show how these detections change based on each run and are corrected for the  field of view and duty cycle of each instrument.  
    }
    \label{fig:detfracsGamma}
\end{figure*}

The observed gamma-ray fluence can be computed as $\phi_\gamma(\theta_\textrm{v})$\,$=$\,$(1+z)E_\gamma(\theta_\textrm{v})/(4\pi d_\textrm{L}^2 k_\textrm{bol})$. If a given GRB has a fluence higher than the sensitivity $\phi_{\gamma,\textrm{lim}}$ of the considered telescopes (Table \ref{tab:telescopes}) in their observer frame bandpass $[e_\textrm{low},e_\textrm{high}]$ we consider it a detection. In Figure \ref{fig:detfracsGamma} we show the prompt gamma-ray detection percentage as a function of instrument (Table \ref{tab:gammasens}) and run (Table \ref{tab: simulationruns}). We correct the detection fraction for each gamma-ray facility based on its duty cycle, including the time spent in the South Atlantic Anomaly (SAA) and its instantaneous field of view (see \citealt{Bhattacharjee2024} for a discussion).

\begin{figure*}
    \centering
    \includegraphics[width=\columnwidth]{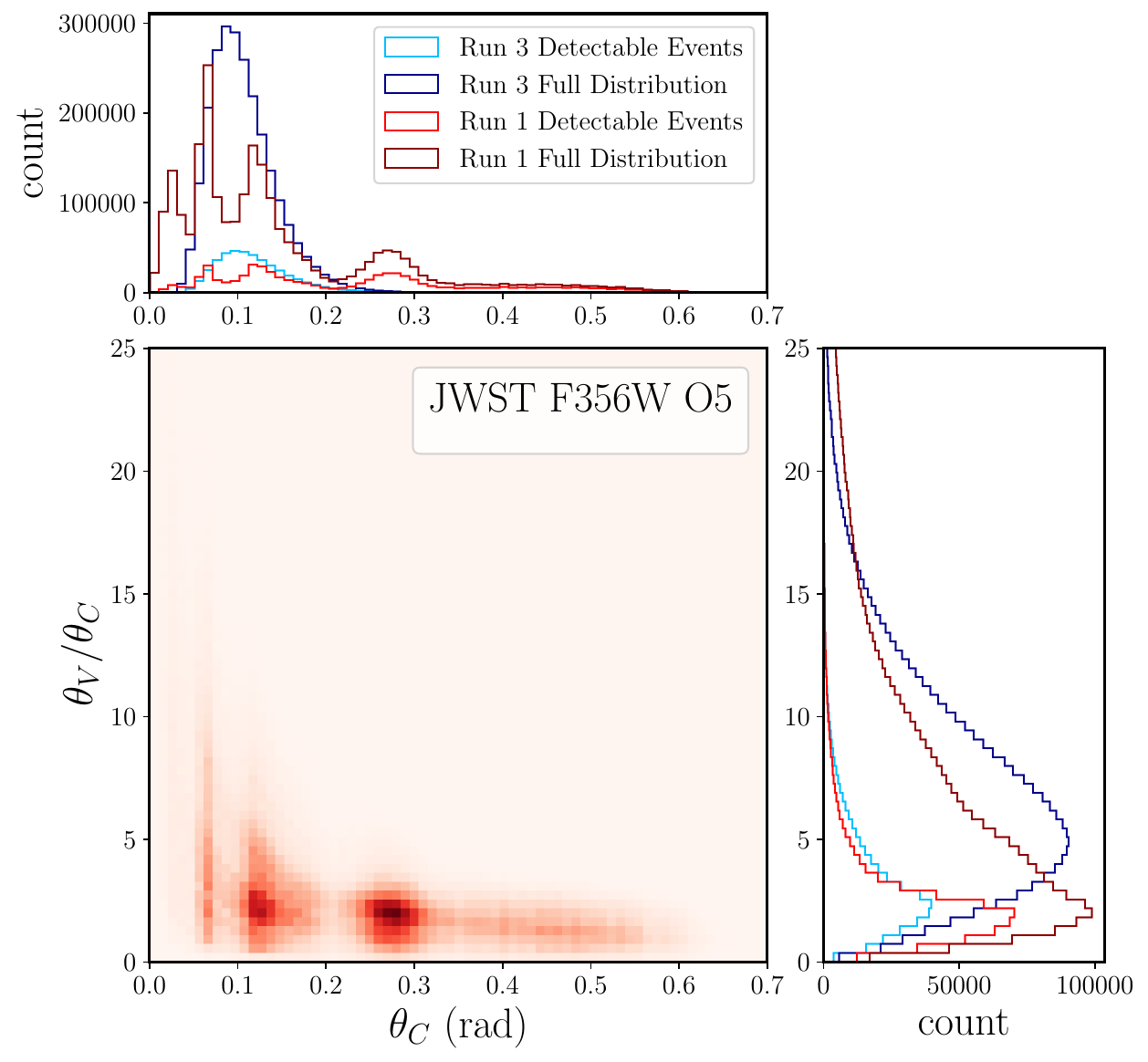}
    \includegraphics[width=\columnwidth]{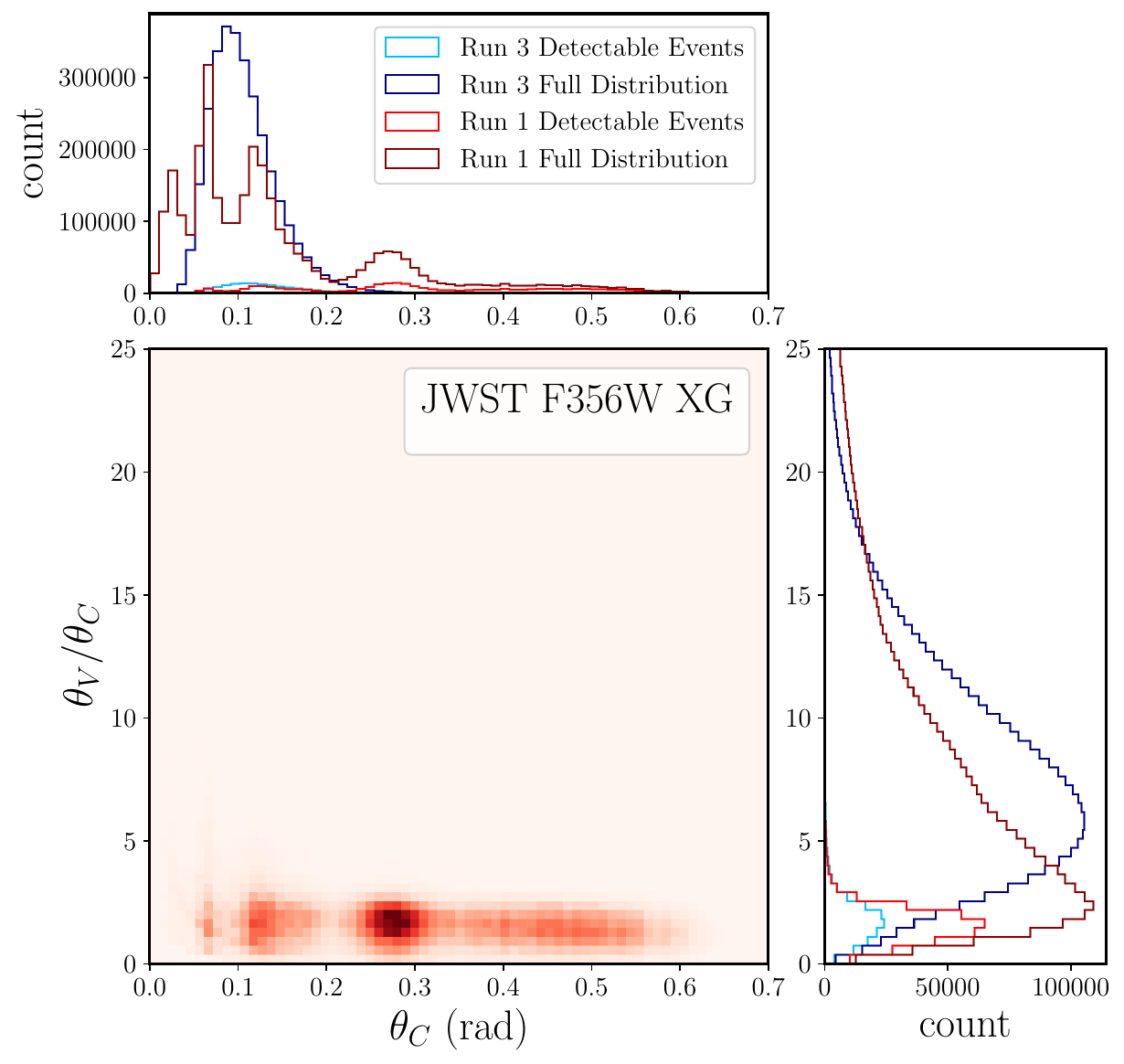}
    \caption{Distribution of jet opening angle and jet viewing angle for \textit{JWST/F356W} in O5 (left) and XG (right). In each panel we show the 2D distribution of the detected afterglows from Run 1, along with the corresponding 1D distributions for the two parameters, including both the  distribution of all simulated afterglows and the distribution of detected ones for both Run 1 and Run 3. }
    \label{thVthCplot}
\end{figure*}

\begin{figure*}
    \centering
    \includegraphics[width=\columnwidth]{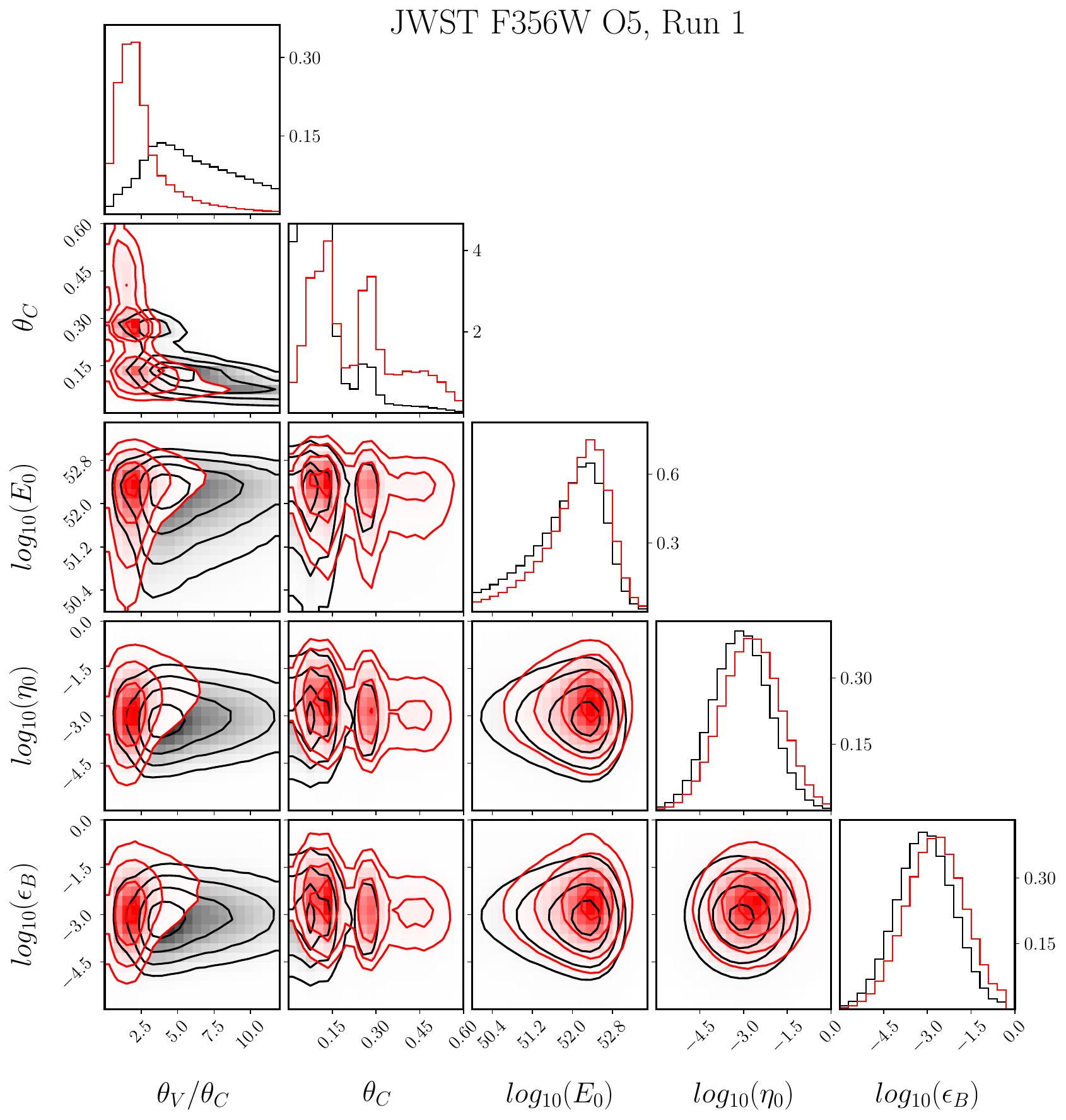}
    \includegraphics[width=\columnwidth]{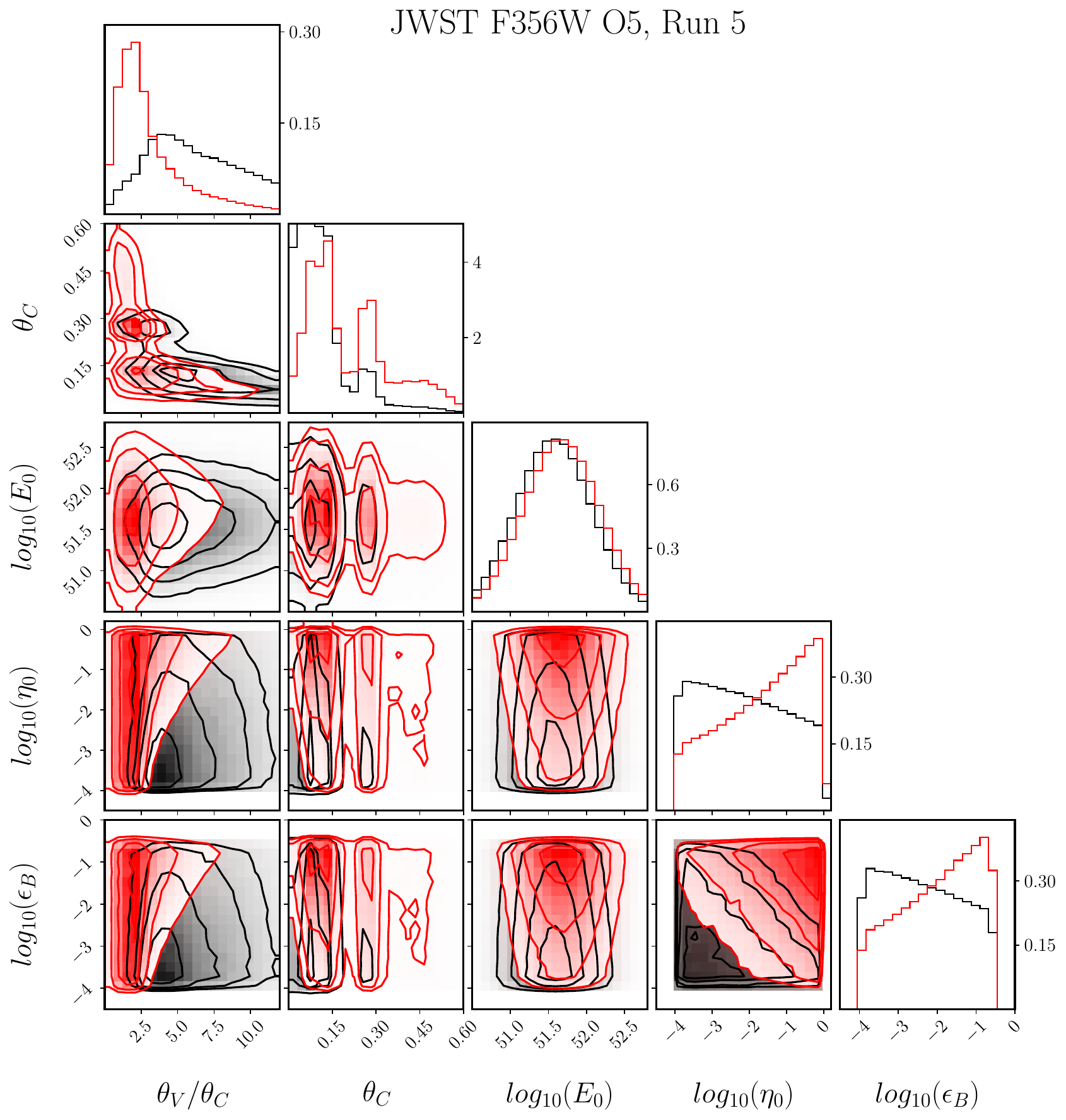}
    \caption{Distributions of O5 afterglow parameters for Run 1 shown on the left, and Run 5 shown on the right. Distributions of the events detectable by \textit{JWST/F356W} are shown in red, and the distributions of events that are not detectable are shown in black. Plots along the diagonal depict the 1D probability density function of each parameter, with the scale shown on the right.
    }
    \label{paramscornerO5}
\end{figure*}

\begin{figure*}
    \centering
    \includegraphics[width=\columnwidth]{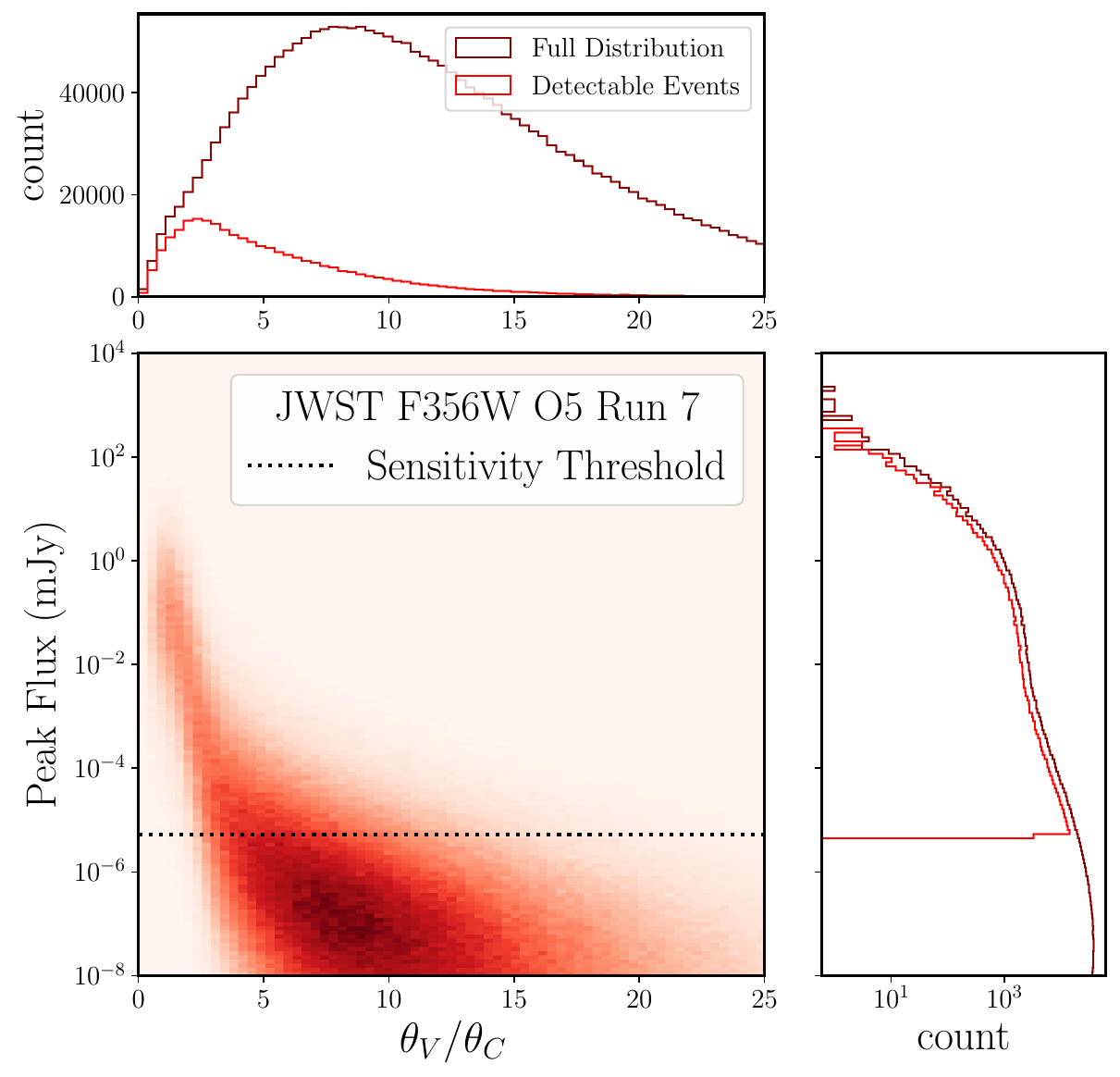}
    \includegraphics[width=\columnwidth]{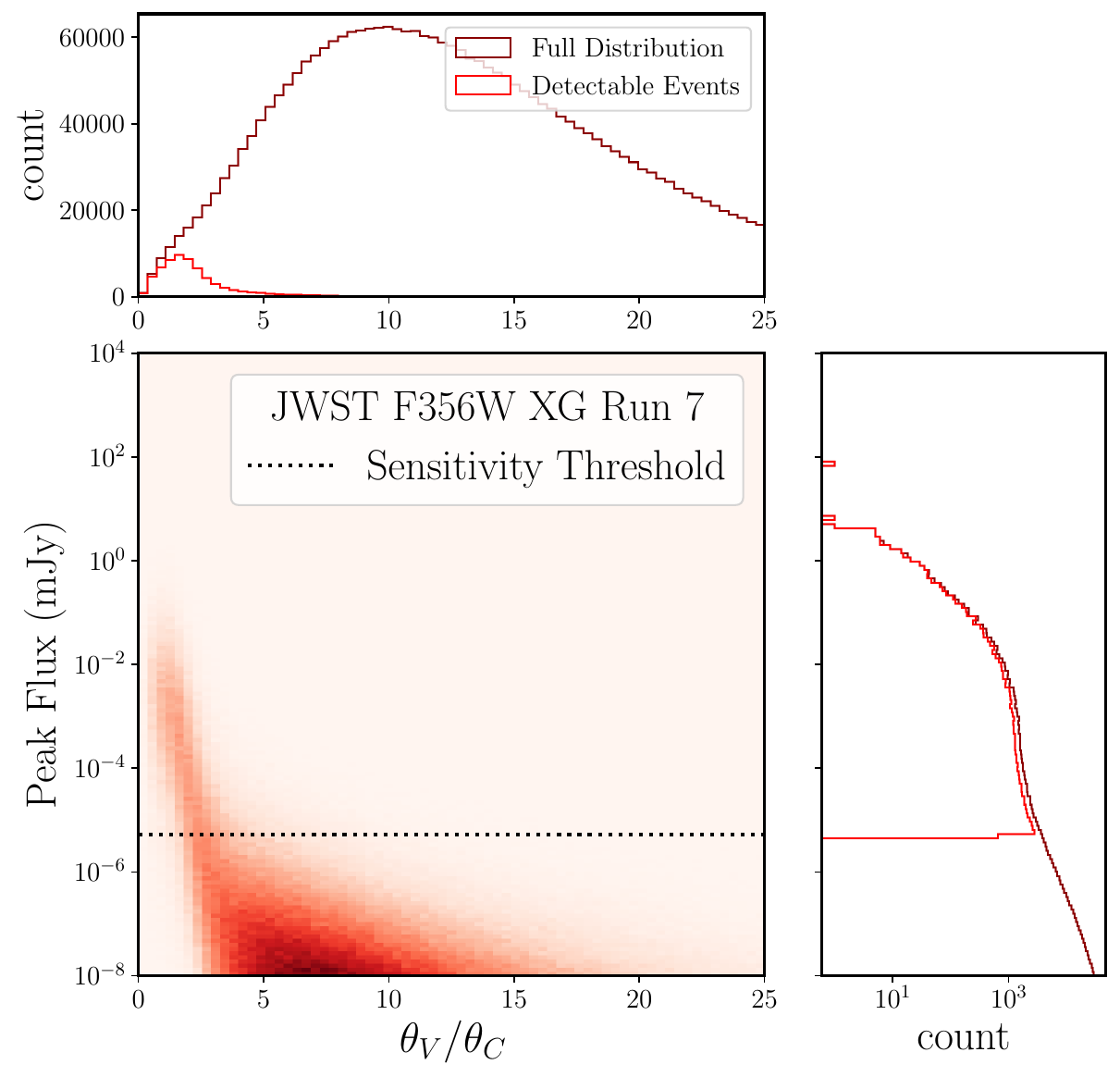}
    \caption{Peak flux versus the ratio of viewing angle $\theta_\textrm{v}$ to core angle $\theta_\textrm{c}$ for \textit{JWST/F356W} in O5 (left) and XG (right) for Run 7. The 2D distribution refers to all simulated afterglows, and the dotted line marks the sensitivity of the instrument; thus the portion of the 2D distribution above that line is detectable. The 1D distributions for the two parameters are shown, including both the full distribution of all simulated afterglows and the distribution of those detectable.  }
    \label{fluxthVthCplot}
\end{figure*}

\begin{figure*}
    \centering
    \includegraphics[width=0.92\columnwidth]{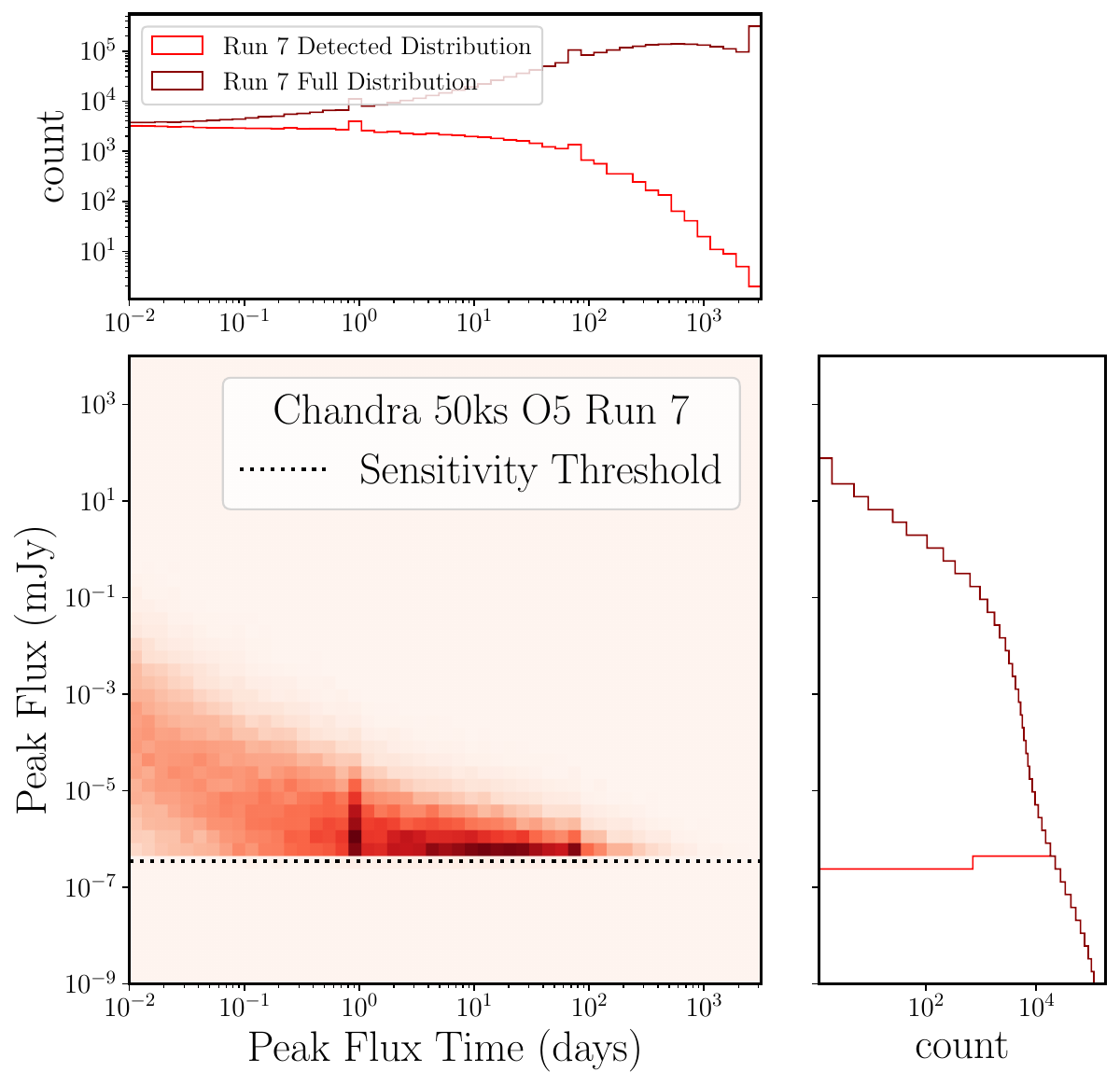}
    \includegraphics[width=0.92\columnwidth]{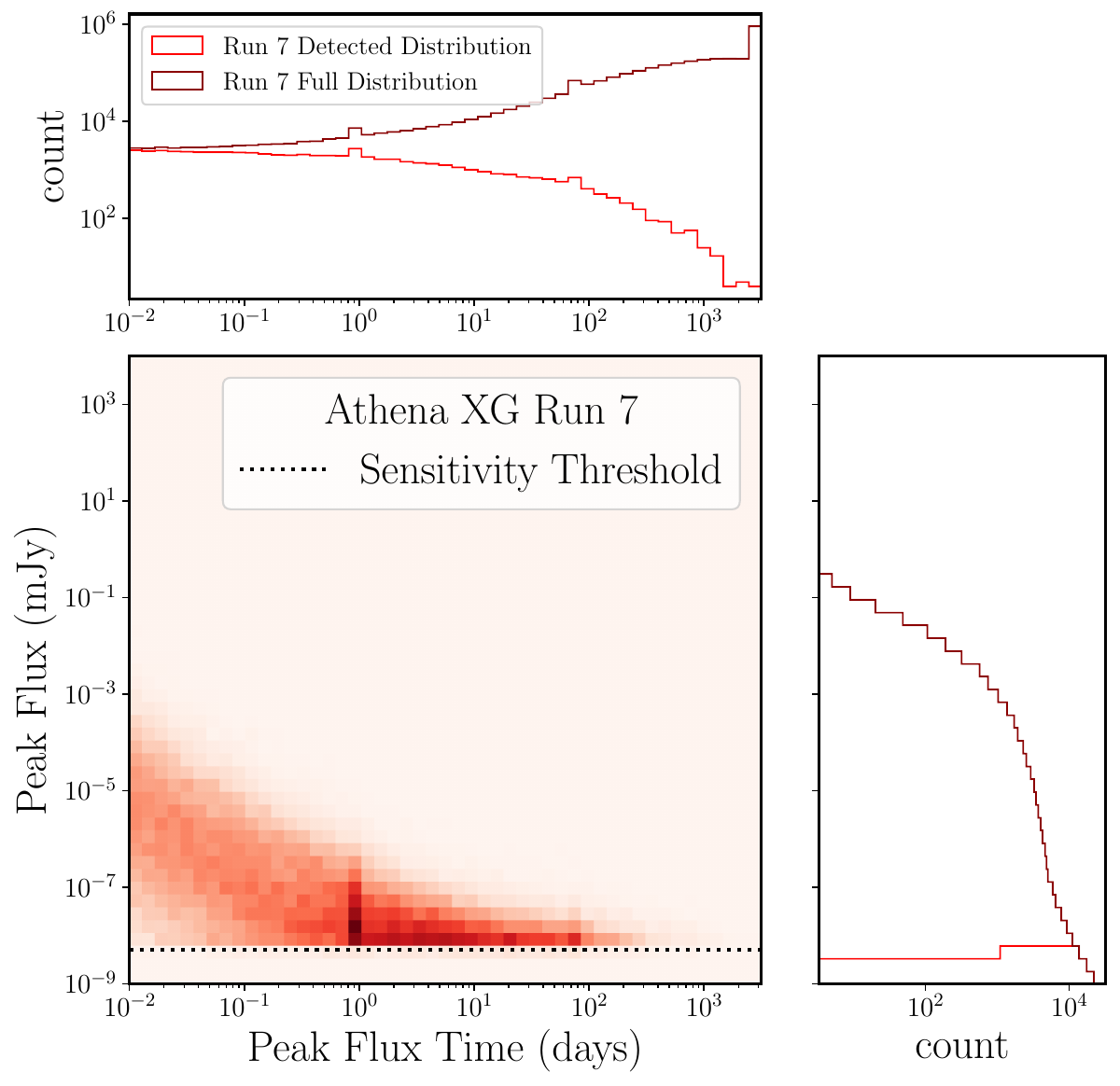}
    \includegraphics[width=0.92\columnwidth]{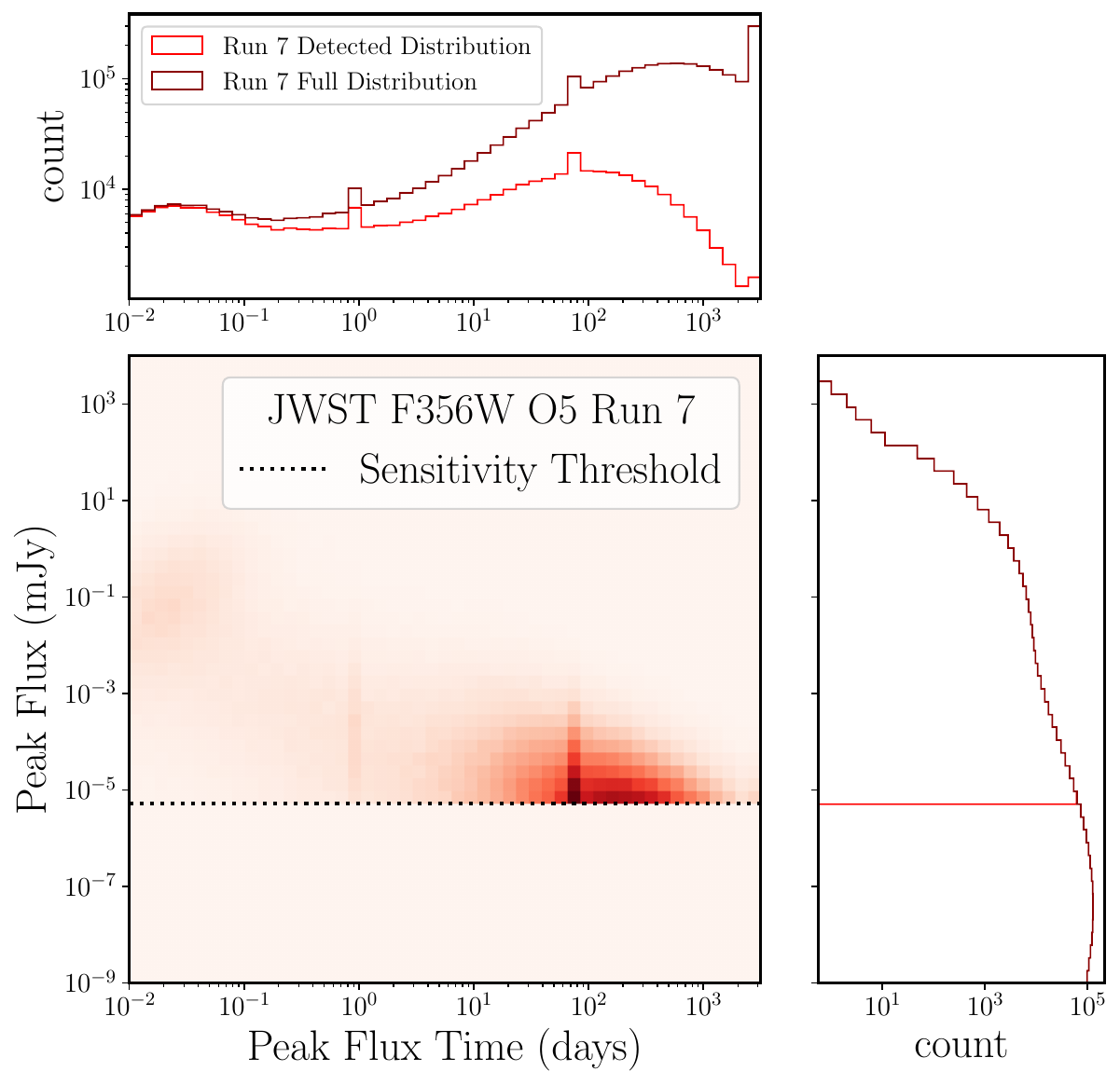}
    \includegraphics[width=0.92\columnwidth]{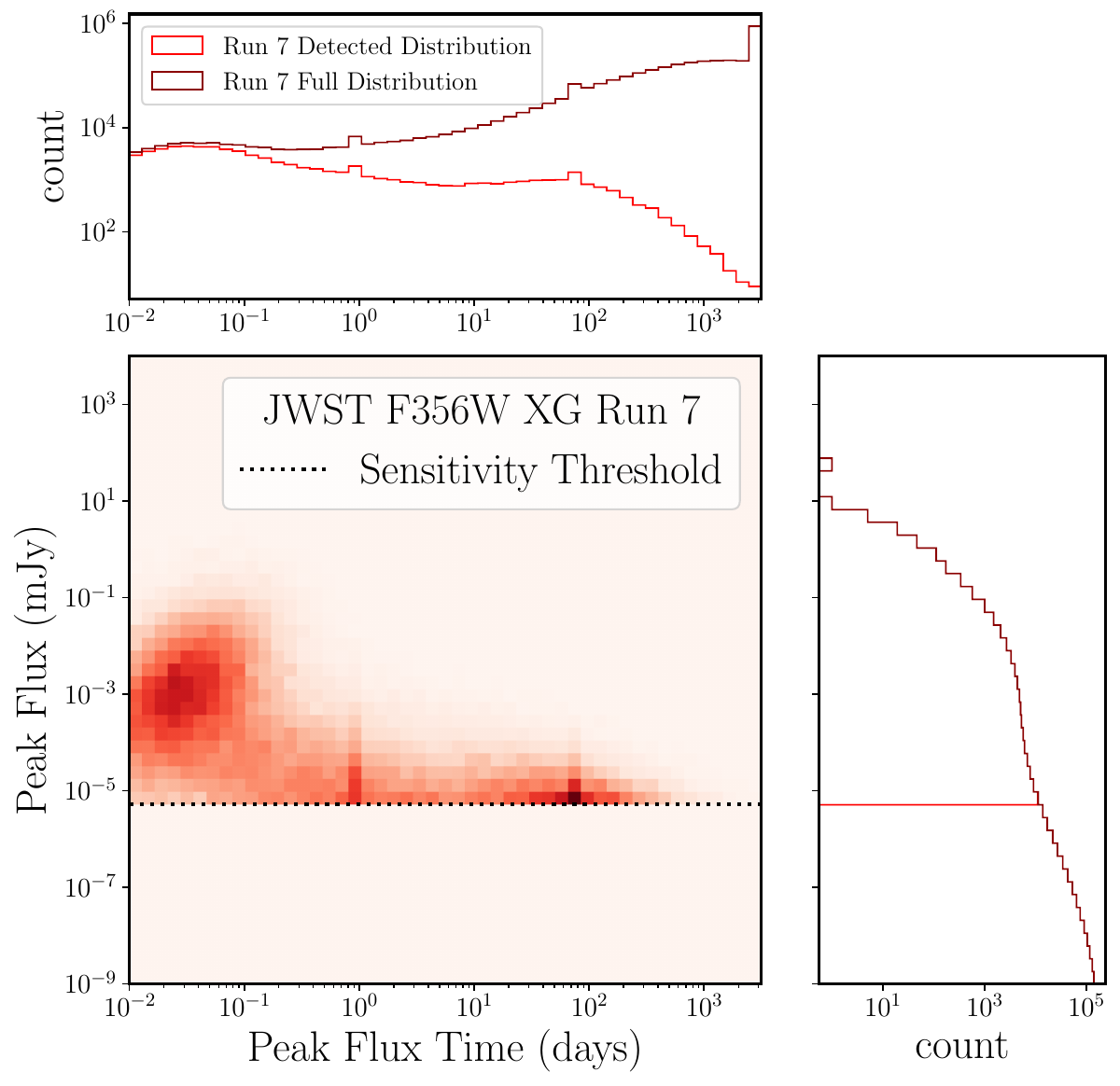}
    \includegraphics[width=0.92\columnwidth]{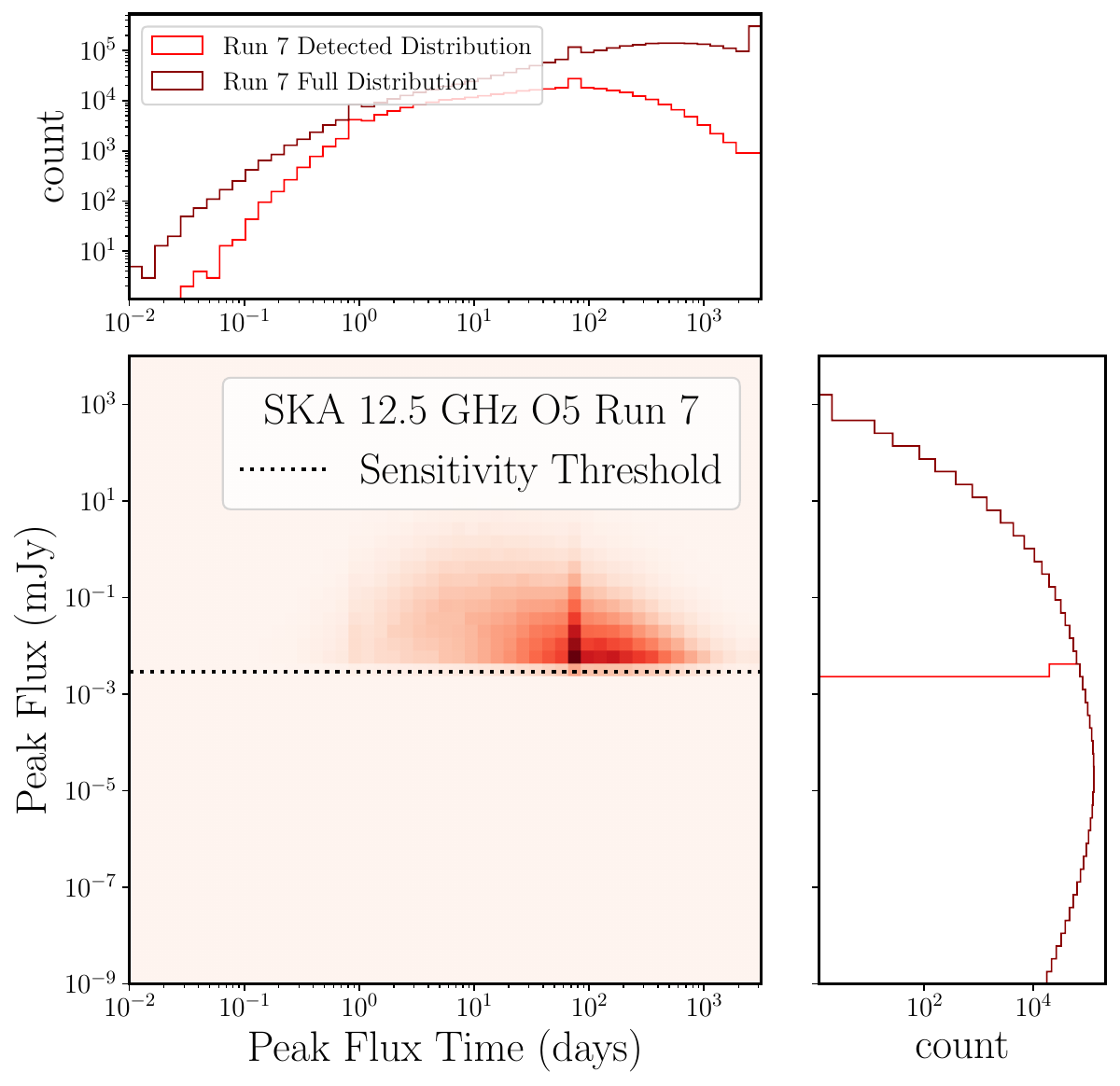}
    \includegraphics[width=0.92\columnwidth]{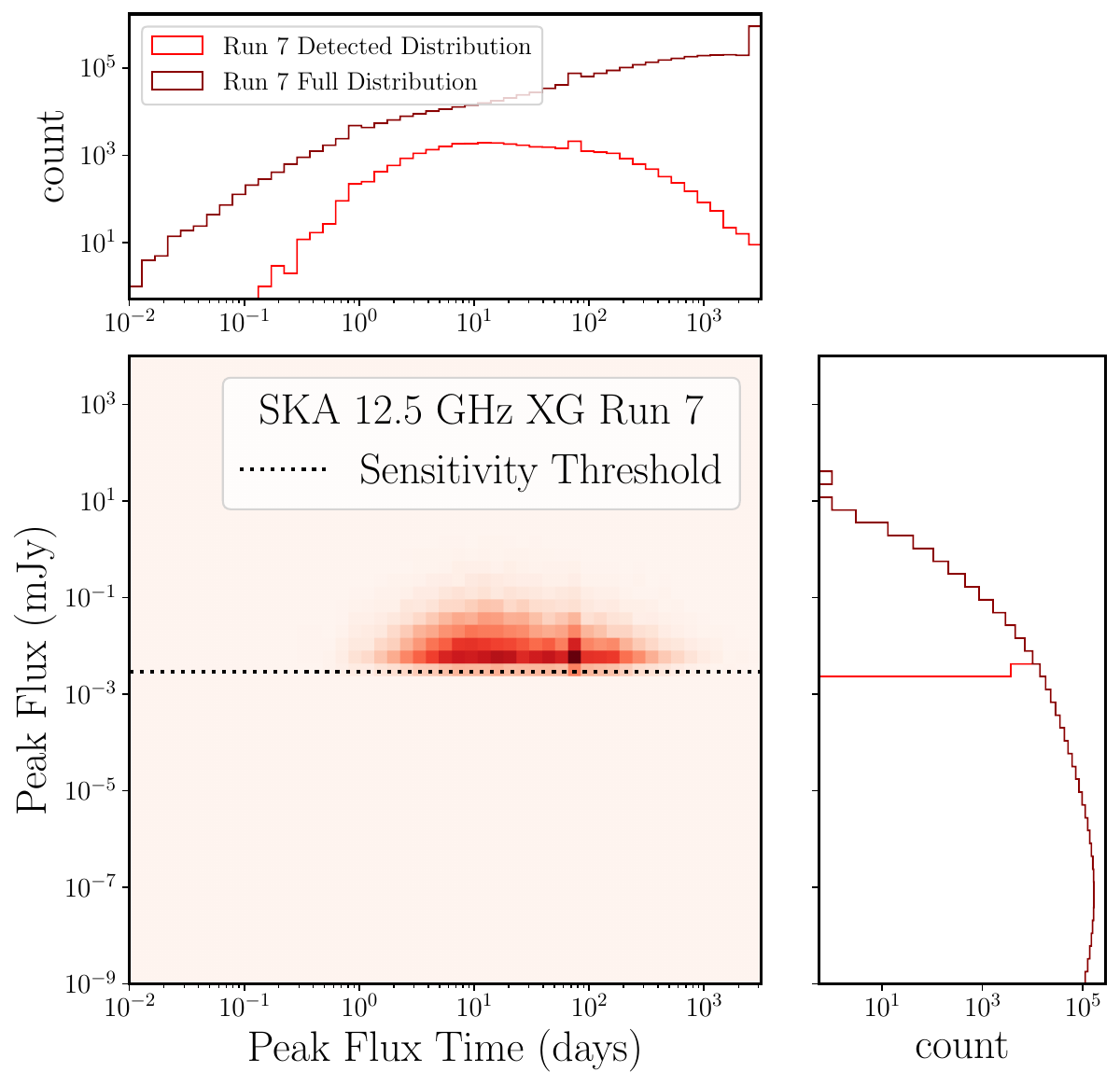}
    \caption{The peak flux of the simulated afterglow compared to the time at which the afterglow peaks for Run 7 at X-ray (top), near-infrared (middle), and radio (bottom) wavelengths. O5 is shown on the left, and XG on the right. The 2D histograms show only detected afterglows at each wavelength. The 1D histograms for the two parameters (top and right of each figure) are projected, including both the full simulated distribution and the distribution of only the detected afterglows. 
    }
    \label{fluxtimeplot}
\end{figure*}

\section{Results and discussion}
\label{section:Results}

\subsection{Impact of afterglow parameter and jet structure assumptions}
\label{sec: AGparamimpact}

Here we discuss the impact of our assumptions used in the seven simulations (Table \ref{tab: simulationruns}) on the afterglow detectability. There are a number of assumptions that have to be made to perform this study, such as: \textit{i}) the distribution of jet core half-opening angles $\theta_\textrm{c}$, which largely dictate detectability and are unknown a priori, \textit{ii}) the jet structure and truncation angle $\theta_\textrm{w}$ of the jet, and \textit{iii}) the distribution of jet kinetic energies, microphysical parameters, and surrounding environmental densities. Each of these are relatively unknown and display a broad diversity. We utilize available constraints from the observed population of cosmological short GRBs \citep[e.g.,][]{Fong2015,OConnor2020,RoucoEscorial2022} that are used to construct our simulation setup.

We find that the most critical assumption for detectability is the core half-opening angle, which is unconstrained for the vast majority of cosmological short GRBs and based on only around a dozen measurements \citep{Grupe2006,Burrows2006,Soderberg2006,Nicuesa2012,Berger2013,Fong2012,Fong2014,Fong2015,Jin2018,Troja2016jetbreak,Troja2019b,Lamb2019grb160821B,Becerra2019,OConnor2021kn,Fong2021kn,RE21,RoucoEscorial2022,Laskar2022}. We have tested four different distributions for the core angle, which we refer to as ``narrow'' (a log-normal distribution centered at 6 deg; 0.1 rad), ``very narrow" (a log-normal distribution centered at 3.4 deg; 0.06 rad), ``fixed" (the core angle at a fixed value of 3.4 deg), and the observed distribution from \citet{RoucoEscorial2022}, which we refer to as RE23 (see Table \ref{tab: simulationruns}). The RE23 distribution has a similar median value to the narrow (6 deg; \citealt{RoucoEscorial2022}), but has a small fraction (a tail) of wide jets with half-opening angles $>$\,$10$\,$-$\,$30$ deg ($0.17$\,$-$\,$0.5$ rad), while the very narrow distribution is centered at a GW170817-like core angle. In Figure \ref{fig:170817lightcurves} we compare the results between Run 1 and 7, and in Figure \ref{thVthCplot} we compare Run 1 and 3, all three of which differ only in the distribution adopted for $\theta_\textrm{c}$.

Figure \ref{thVthCplot} clearly shows that the detectability of afterglows (based on \textit{JWST} but this applies to all instruments) is significantly biased towards small values of the ratio $\theta_\textrm{v}/\theta_\textrm{c}$. As Run 1 produces a larger fraction of wider jets, with larger core angles, the ratio $\theta_\textrm{v}/\theta_\textrm{c}$ peaks at systematically smaller values where afterglows are easier to detect. 
This is reflected in the higher detection percentage of Run 1 versus Run 3 and 7 for all instruments and wavelengths (Figure \ref{fig:detfracsAG}). 
The detected fraction of afterglows decreases rapidly for $\theta_\textrm{v}/\theta_\textrm{c}$\,$>$\,$2$ (see also Figure \ref{paramscornerO5}) where for a GJ the energy along the line-of-sight has decreased by a factor of $\sim$\,$8$. This trend is the same for both Run 1 and Run 3, and thus the major indicator of the difference in detected fraction is the number of afterglows with $\theta_\textrm{v}/\theta_\textrm{c}$\,$<$\,$2$. This is also reflected in the observed population of cosmological GRBs which are limited to smaller viewing angles \citep[e.g., $\theta_\textrm{v}/\theta_\textrm{c}$\,$<$\,$2$;][]{BeniaminiNakar2019,OConnor2024}. Here we have shown that the small handful of short GRBs with wide jets \citep{Grupe2006,Berger2013,Laskar2022,RoucoEscorial2022} inferred based on their jet break times (or lack of a jet break to late-times) can have a large impact on the expected detection rate of the entire population. Thus, we avoid using Run 1 as the default, and use Run 7 as our fiducial case instead to avoid the small population of wide jets biasing our simulation results. 

In \texttt{afterglowpy} the other relevant parameter dictating the structure of a GJ is the jet's truncation angle $\theta_\textrm{w}$. We have tested the impact of our assumption by varying this value between Run 1 and Run 2 (which both adopt the RE23 distribution for $\theta_\textrm{c}$). In Run 1 we take $\theta_\textrm{w}$\,$=$\,$4.9\theta_\textrm{c}$ whereas in Run 2 we truncate the jet at smaller angles $\theta_\textrm{w}$\,$=$\,$3\theta_\textrm{c}$ outside of which there is no material along the line-of-sight (for $\theta_\textrm{v}$\,$>$\,$\theta_\textrm{w}$). Figure \ref{fig:detfracsAG} shows that this has no impact on afterglow detection. This is because far off-axis afterglow detection is generally dependent on the peak flux of the afterglow, and independent of the early afterglow behavior (and is only dependent on emission from the jet's core; Appendix \ref{sec: afterglowpyassumptions}), which is modified by this change in $\theta_\textrm{w}$. 

We also briefly tested the impact of jet structure. The difference between simulation Run 3 and Run 4 is the use of a GJ for Run 3 (and all other runs) and PLJ with $b$\,$=$\,$10$ \citep{Ryan2023} for Run 4. From Figure \ref{fig:detfracsAG} we see that the PLJ (Run 4) is slightly easier to detect, but overall very similar. This is expected as both structures are capable of fitting the full lightcurve of GW170817 \citep[e.g.,][]{Ryan2023}. It is clear that much shallower angular profiles would be significantly easier to detect, though an exhaustive exploration is beyond the scope of this work and challenging due to the unknown jet structure for all but one short GRB (i.e., GW170817).

Next we explore the impact of our afterglow parameter assumptions  which we refer to as Case 1 and Case 2 (see Table \ref{tab: afterglowparameters}). While Case 1 is our default parameter setup we explore the impact of broad uniform distributions (referred to as Case 2) in the microphysical parameter $\varepsilon_\textrm{B}$ and density $n$ in Run 5 (which can most directly be compared against Run 1). 
In general for short GRBs the emission at X-ray, optical, and infrared wavelengths, and sometimes radio (e.g., GW170817), is most commonly between the synchrotron injection frequency $\nu_\textrm{m}$ and the cooling frequency $\nu_\textrm{c}$ such that $\nu_\textrm{m}$\,$<$\,$\nu$\,$<$\,$\nu_\textrm{c}$. In this regime the flux is positively correlated with the kinetic energy, density, and the fraction of energy in magnetic fields such that $F_\nu$\,$\propto$\,$E_\textrm{kin}^\frac{3+p}{4}\varepsilon_\textrm{B}^\frac{1+p}{4}n^\frac{1}{2}$ \citep{Granot2002}. As such, afterglow detection is biased towards larger values of these parameters as observed in the corner plots in Figure \ref{paramscornerO5}. Thus, the unknown afterglow parameters can have a significant impact on the likelihood of detecting the afterglow of a GW event. 

Figure \ref{paramscornerO5} clearly demonstrates this trend, which is also seen in Figure \ref{fig:detfracsAG} where we observe that Run 5 has the highest detection fraction for all instruments. This is notably due to the allowance for larger fraction of events at higher values of $\varepsilon_\textrm{B}$ and $n$ (seen in Figure \ref{paramscornerO5}) which lead to brighter afterglows. 
The kinetic energies are not substantially different, though Run 1 does have a tail towards lower energies \citep{Ghirlanda2016}.

Figure \ref{fig:detfracsAG} shows that Run 5 has a particularly large increase in detectability at low frequency radio wavelengths. 
The large increase at radio wavelengths can be explained in two parts. First, the low frequency radio instruments are significantly more sensitive than those at higher frequencies. 
Secondly, the maximum flux density usually lies at radio wavelengths and is determined by flux at the injection frequency which is given by $F_{\nu_\textrm{m}}$\,$\propto$\,$E_\textrm{kin}\varepsilon_\textrm{B}^\frac{1}{2}n^\frac{1}{2}$ \citep{Granot2002}. As this maximum flux is constant in time in a uniform density environment, and the  injection frequency decreases swiftly in time $\nu_\textrm{m}$\,$\propto$\,$E_\textrm{kin}^\frac{1}{2}\varepsilon_\textrm{B}^\frac{1}{2} t^\frac{-3}{2}$, at late times the maximum flux will lie in the low frequency radio band, especially for events viewed off-axis such that the core emission comes into view only at later times.  

\subsection{Impact of viewing angle}
\label{sec:viewingimpact}

As discussed in \S \ref{sec: AGparamimpact}, the ratio $\theta_\textrm{v}/\theta_\textrm{c}$, which we refer to as the viewing angle interchangeably with $\theta_\textrm{v}$, largely dictates the detectability of an afterglow. This ratio also dictates the lightcurve behavior such as the rising and fading slopes, the time to peak, and the flux at the peak (see also Appendix \ref{sec: afterglowpyassumptions}). Figure \ref{fluxthVthCplot} shows the behavior of the peak flux of the afterglow lightcurve (here for \textit{JWST}) versus the viewing angle $\theta_\textrm{v}/\theta_\textrm{c}$. The inset 2D histogram displays all simulated afterglows, whereas the 1D histograms on the top and bottom present both all simulated afterglows and only those which are detected at the sensitivity of \textit{JWST}. 
We see that the peak flux declines rapidly with increasing $\theta_\textrm{v}/\theta_\textrm{c}$. This is due to the rapidly decreasing kinetic energy along the line-of-sight, which dominates the peak flux for $\theta_\textrm{v}/\theta_\textrm{c}$\,$<$\,$2$ (Appendix \ref{sec: afterglowpyassumptions}). At larger angles the material along the line-of-sight has such low energy and Lorentz factor that it becomes fainter than the rising emission from the jet's core becoming de-beamed to the observer. This produces a flattening in the peak flux distribution that is observed as a cloud in the 2D histogram between $\theta_\textrm{v}/\theta_\textrm{c}$\,$\approx$\,$5$\,$-$\,$10$. 

The lightcurve behavior is also modified. Figure \ref{fluxtimeplot} 
shows peak flux versus peak time for the O5 and XG instruments with the largest detection fraction at near-infrared (\textit{JWST}), radio (SKA), and X-ray (\textit{Chandra} and \textit{NewAthena})  wavelengths. Larger viewing angles $\theta_\textrm{v}/\theta_\textrm{c}$ push the time of the peak to later times (see also Figure \ref{fig:170817lightcurves}), similar to GW170817, which also produce lower fluxes as the shocked material has had a longer time to cool. In these figures the 2D histogram instead shows only the detectable lightcurves. 

The high density region between $10^{-2}$\,$-$\,$10^{-1}$ d in the 2D histogram for \textit{JWST/F356W} in Figure \ref{fluxtimeplot} is due to $\nu_\textrm{m}$ passing through the band, at later times ($\sim$\,$10$\,$-$\,$20$ d) $\nu_\textrm{m}$ shifts towards low frequency radio and produces an overdensity in the 2D histograms at the same flux levels at 12.5 GHz (Figure \ref{fluxtimeplot} for SKA). At later times ($\sim$\,$100$ d) the 12.5 GHz band lies between $\nu_\textrm{m}$\,$<$\,$\nu$\,$<$\,$\nu_\textrm{c}$ and the detectable emission is due instead to the emission from the core of the jet. In Figure \ref{fluxtimeplot} we see a different behavior for \textit{Chandra} and \textit{NewAthena} as the X-ray band lies above $\nu_\textrm{m}$\,$<$\,$1$ keV for all times considered here.

As the viewing angle $\theta_\textrm{v}/\theta_\textrm{c}$ increases, the injection frequency $\nu_\textrm{m}$ and the flux at the injection frequency $F_{\nu_\textrm{m}}$  decline rapidly due to their dependence on the line-of-sight kinetic energy. Thus for large angles ($\theta_\textrm{v}/\theta_\textrm{c}$\,$>$\,$3$) even the \textit{JWST/F356W} is above the injection frequency for all times considered here ($>$\,$10^{-2}$ d). Despite the slower temporal evolution of $\nu_\textrm{m}$ for off-axis viewing angles \citep{BGG2020,BGG2022} by $\sim$\,$20$ d the injection frequency is $<$\,$10$ GHz for our default parameter setup (and most parameters in general) at all viewing angles. For the time of peak for far off-axis events ($>$\,$100$ d) all frequencies considered in this work lie above $\nu_\textrm{m}$\,$<$\,$\nu$, where only the X-ray band is sometimes above the cooling frequency $\nu_\textrm{c}$\,$<$\,$1$ keV. 

Figures \ref{fig:O5detectionsJWST} (O5) and \ref{fig:XGdetectionsJWST} (XG) show the detected fraction as a function of viewing angle and distance where each point at fixed viewing angle and distance represents a simulated GW event. For each event, 1,000 lightcurves are simulated based on the afterglow parameter assumptions (Tables \ref{tab: afterglowparameters} and \ref{tab: simulationruns}) and the color of the datapoint represents the fraction of those 1,000 lightcurves that are detected by \textit{JWST}. GW events observed closer to the axis of the jet are significantly more likely to be detected with an afterglow, though this is a function of both distance and the assumed afterglow parameters (in particular $\theta_\textrm{c})$.

\subsection{LVK O5}
\label{sec:O5results}

\subsubsection{Afterglow detection}

Here we discuss the afterglow detection percentage for simulated GW events during LVK O5. We focus only on our fiducial setup, referred to as Run 7 (Table \ref{tab: afterglowparameters} and \ref{tab: simulationruns}). As this uses the smallest half-opening angles, it is the least optimistic for future detections (Figure \ref{fig:detfracsAG}). The detection percentage for each instrument is listed in Table \ref{tab:telescopes}. This is also displayed in Figure \ref{fig:detfracsAG} for each simulation run. For this fiducial case, we further present the fraction of afterglows detectable after a given time (observer frame), see Figure \ref{fig:detfracsmaxtime}. The highest detection fractions are achieved by \textit{SKA} at 1.4 GHz with a detection percentage of 12.6\%. In the range of UVOIR wavelengths, \textit{JWST} in the \textit{F356W} filter has the highest detectable percentage of 11.6\%. Lastly, at X-ray wavelengths, we find that \textit{Chandra} is capable of detecting 4.1\% of all simulated afterglows.

\textit{JWST} requires 14 d of activation time for a non-disruptive Target of Opportunity (ToO) observation, though faster responses are possible, thus we also consider the fraction of afterglows detectable after a given time from merger (Figure \ref{fig:detfracsmaxtime}). In this discussion, we consider afterglows for which the peak of the lightcurve is detectable after a given time, as detecting the peak is useful for characterizing the afterglow. Of the 11.6\% of afterglows from our full simulation that are detectable in the \textit{JWST/F356W} filter, $55 \%$ peak after 10 days. Thus, \textit{JWST/F356W} would be able to detect 6.4\% of afterglows from our full simulation suite after 10 days post-merger, with the number dropping to 3.2\% after 100 days. 

At radio wavelengths, 98\% of afterglows detectable by SKA at 12.5 GHz peak at $>$\,$1$ days, as shown in Figure \ref{fluxtimeplot}. As such, at 12.5 GHz, SKA is capable of detecting 9.8\% of all simulated afterglows after 1 day. That number drops slightly to 7.5\% at 10 days and further to 3.2\% at 100 days. Therefore, SKA detection fractions are comparable to \textit{JWST} over these time periods. 

At X-ray wavelengths, we observe higher peak fluxes at early times (as discussed in \S \ref{sec:viewingimpact}) with the peak flux from the afterglow decreasing as the time of the peaking (and viewing angle) increases (Figure \ref{fluxtimeplot}). Of all simulated X-ray afterglows that Chandra is able to detect, $26\%$ peak at $>$\,$1$ d.  
\textit{Chandra} is able to detect 1\% of all simulated afterglows after 1 day, and 0.4\% after 10 days. 

Because fast-turnaround ToOs on instruments like \textit{JWST} are limited, it is important to consider the advantages of using less sensitive wide-field instruments to capture afterglows at $<$\,$1$ day that would otherwise not be probed by either \textit{JWST} or \textit{Chandra}. At UVOIR wavelengths, we consider Rubin in $g$-band with a 180 s exposure. Rubin has a detection percentage of 3.3\% across all simulated afterglows with 79\% of the detectable afterglows peaking before 1 day. This emphasizes the need for a rapid ToO response from Rubin to GW events \citep[e.g.,][]{Andreoni2022a}. 

We perform the same checks for \textit{Roman} and \textit{ULTRASAT}. Of all afterglows detectable in \textit{Roman} ($R$-band), 76\% peak within 1 day of the merger, and for afterglows detectable by \textit{ULTRASAT}, 98\% peak within 1 day. We see this same trend at X-ray frequencies as well; 97\% of the afterglows detectable by \textit{Einstein Probe} peak within 1 day.

While deep field instruments are still crucial for detecting afterglows at later times, these wide-field instruments are essential for the detection and localization of near on-axis events at short timescales and a rapid response by these facilities is still strongly encouraged to aid in model constraints and increase the detection likelihood.

Based on this analysis, we conclude that radio telescopes have the best chance of detecting an afterglow counterpart to a BNS merger during LVK O5. It is important to note that all afterglow detections in this paper do not take into account constraints regarding visibility, Sun positioning, and day/night observing times, all of which will realistically lower the rate of afterglow detection. Thus the detected fractions and rates should be treated as upper limits\footnote{Assuming a total $A_V$ extinction value of $\sim$1 mag from dust and gas absorption and assuming a sky observability of 50\% for any given afterglow, for the optical Rubin g-band we would have to scale our current detectable fractions and rates to 19\% of their current value. Doing a similar calculation in the IR, for \textit{JWST F356W} we would have to scale our current numbers to 47\% of their current value. This calculation for minimum detection fractions would yield a detectable fraction of 0.6\% for Rubin g-band (180s) and 5.5\% for \textit{JWST F356W} in O5, lowered from 3.3\% and 11.6\% respectively. We offer this back of the envelope calculation for a reference here, and leave the full treatment to future work.}. Given that radio telescopes have a larger range of visibility, and are not as impacted by day/night constraints, this only improves the likelihood of detecting an afterglow in radio as compared to other wavelengths. However, it must be noted that \texttt{afterglowpy v0.8.0} does not currently include synchrotron self-absorption which could impact the lower frequency radio data considered here (e.g., SKA at 1.4 GHz).

\begin{figure*}
    \centering
    \includegraphics[width=\columnwidth]{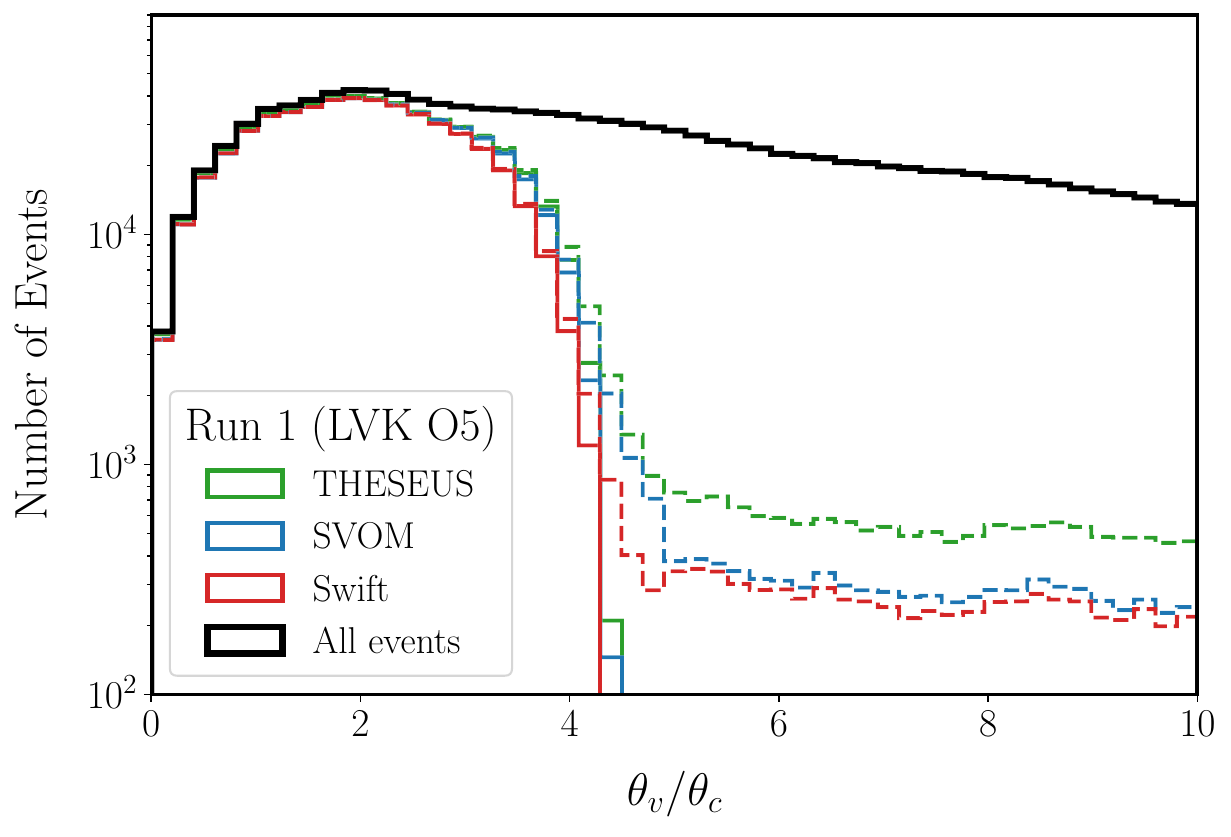}
    \includegraphics[width=\columnwidth]{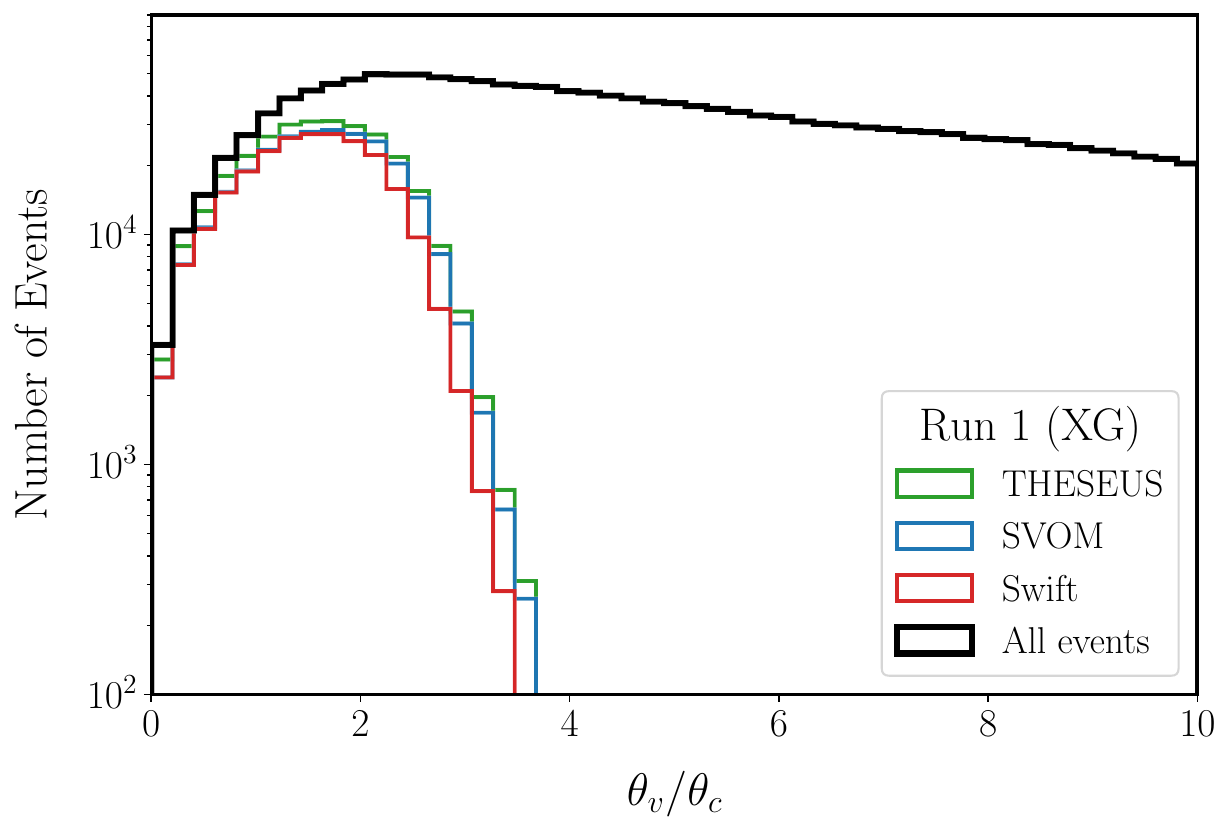}
    \includegraphics[width=\columnwidth]{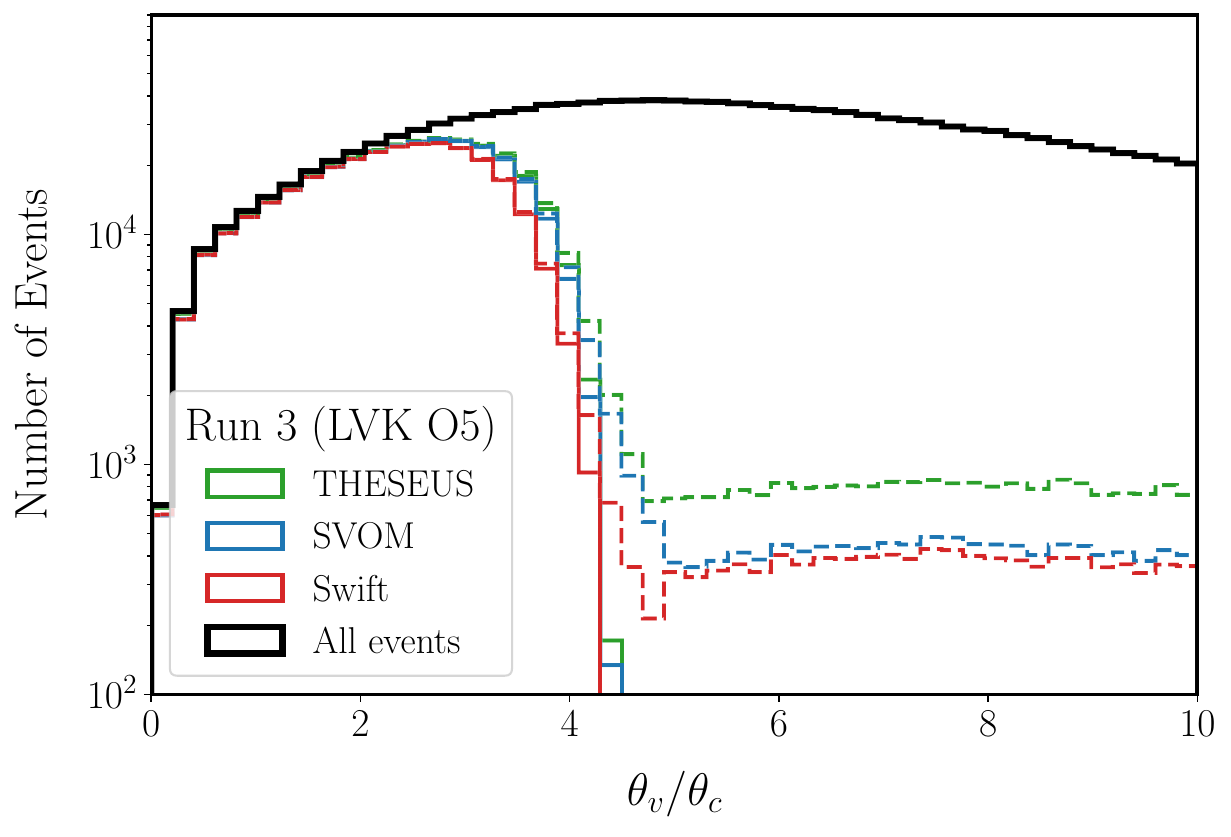}
    \includegraphics[width=\columnwidth]{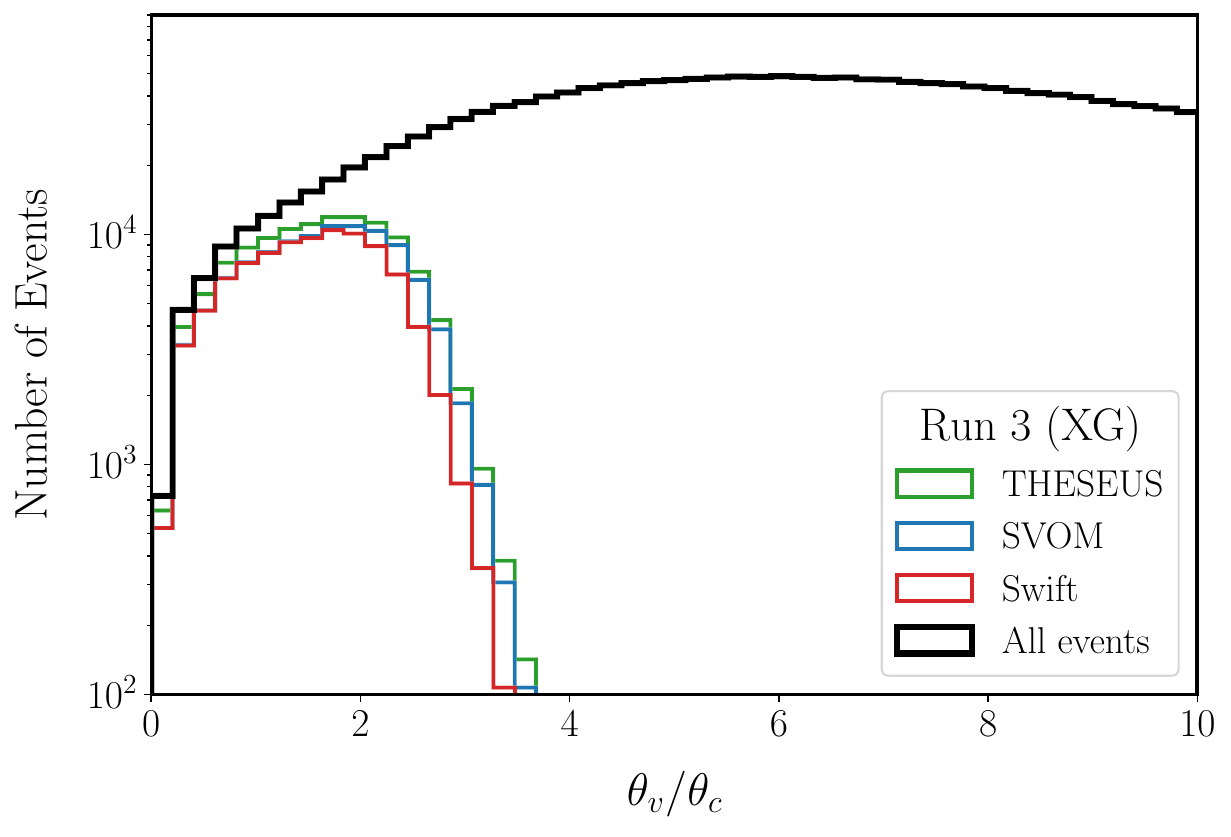}
    \includegraphics[width=\columnwidth]{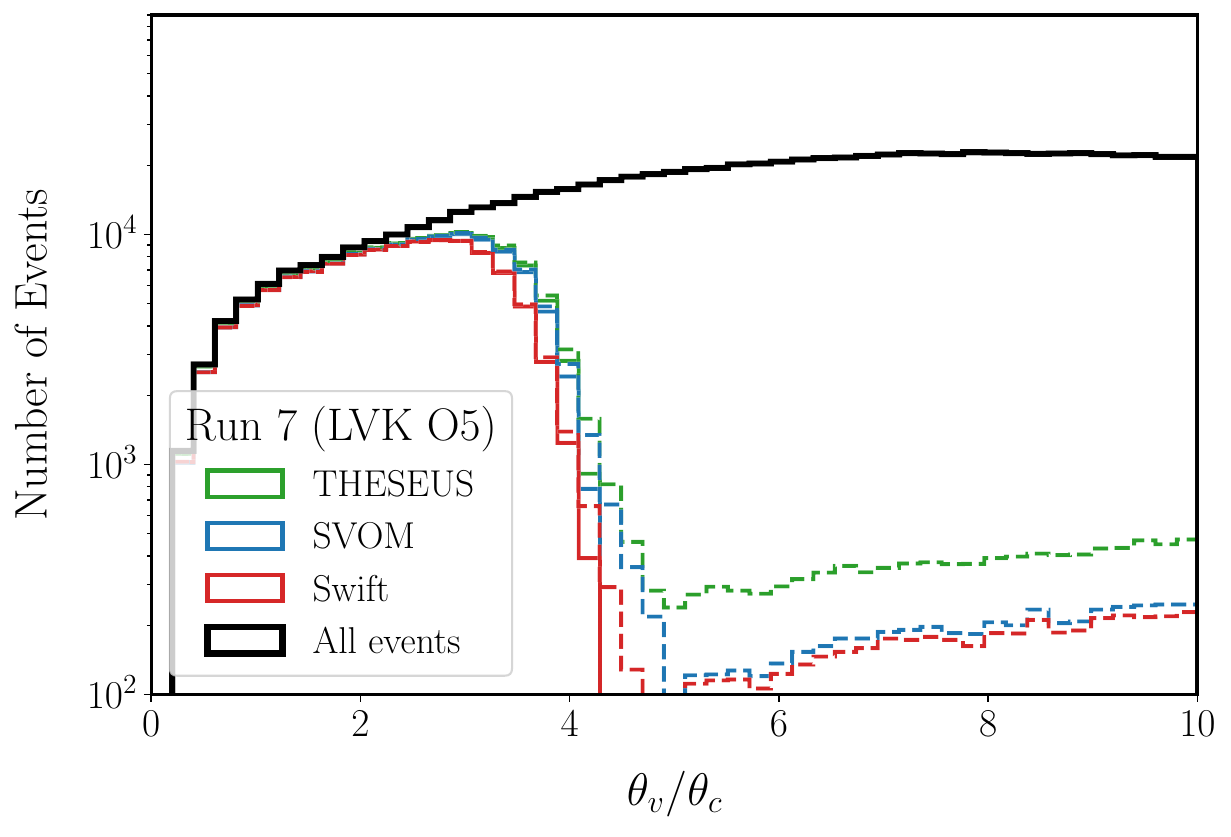}
    \includegraphics[width=\columnwidth]{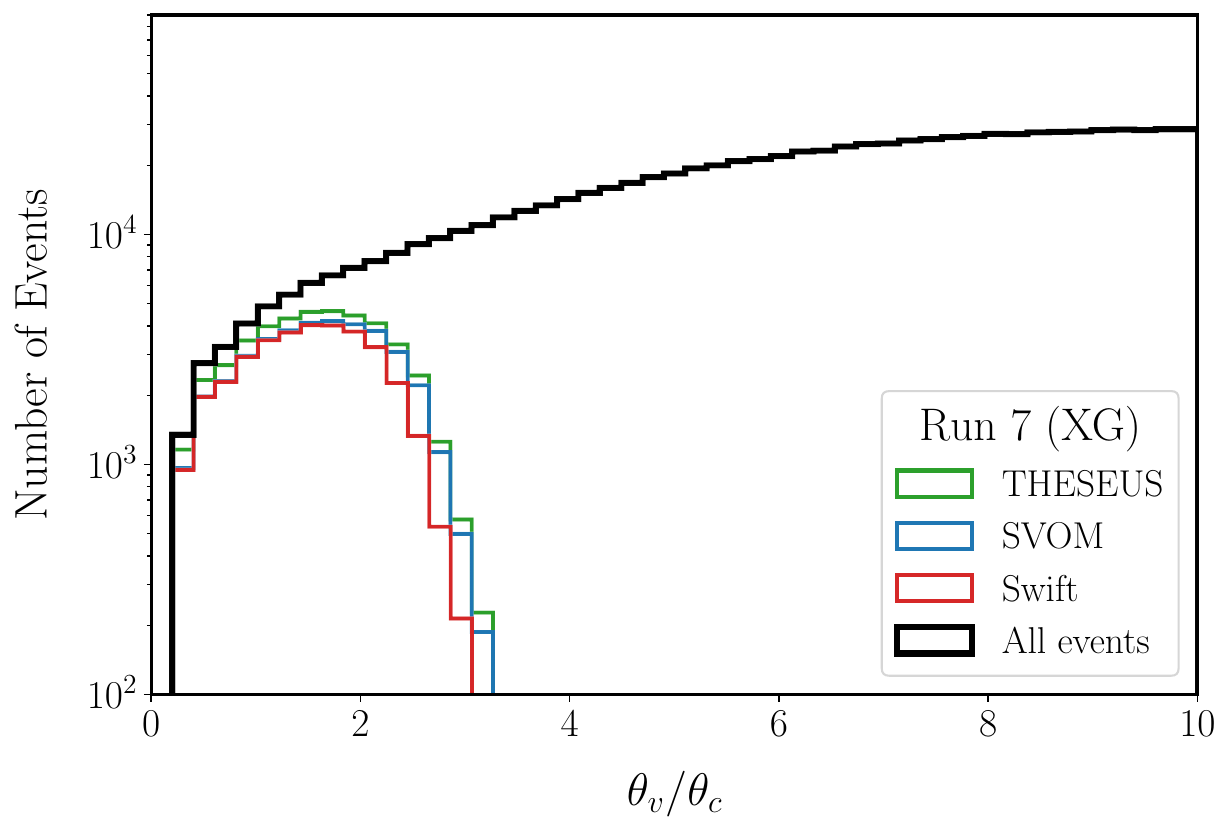}
    \vspace{-0.3cm}
    \caption{Distribution of viewing angles $\theta_\textrm{v}/\theta_\textrm{c}$ for events detected in gamma-rays. Left panels are for LVK O5 and right panels are for XG. We compare Run 1 (RE23), Run 3 (Narrow), and Run 7 (Very narrow) for both O5 and XG. Solid lines represent the GJ model, whereas dashed lines are for the GJ model with the addition of a quasi-isotropic cocoon (only for LVK O5). 
    }
    \label{fig:gammadet-vs-angle}
\end{figure*}

\subsubsection{Prompt gamma-ray detection}

Here we quantify the detection prospects for prompt gamma-rays associated to our simulated BNS mergers. The detection percentages for each instrument (Table \ref{tab:gammasens} for Run 7) are shown in Figure \ref{fig:detfracsGamma} for each run. We show the detection percentage considering both emission from the jet, and a simple quasi-isotropic cocoon model \citep{Duffell2018} with breakout efficiency $\eta_\textrm{br}$\,$=$\,$10^{-3}$. Our results favor the prompt emission detection of $\sim$\,$0.5$\,$-$\,$10\%$ of BNS mergers during LVK O5 considering current instruments (e.g., \textit{Swift}, \textit{SVOM}). We have accounted for both the sky coverage or duty cycle of the available instruments (see also, e.g., \citealt{Bhattacharjee2024}).

While Figure \ref{fig:detfracsGamma} shows the fraction of simulated events detected in gamma-rays, it does not include information related to  the viewing angle. Therefore, in Figure \ref{fig:gammadet-vs-angle} we show the distribution of viewing angles $\theta_\textrm{v}/\theta_\textrm{c}$ of events detected in gamma-rays versus the total simulated population for multiple assumptions for the jet's core half-opening angle. 
In both panels we have adopted a GJ with truncation angle $\theta_\textrm{w}$\,$=$\,$5\theta_\textrm{c}$, and we observe that the steep drop in detections occurs before this truncation angle. All detections at angles larger than $\theta_\textrm{v}$\,$>$\,$\theta_\textrm{w}$ (dashed lines) are due to emission from the quasi-isotropic cocoon \citep{Duffell2018}. Due to the low assumed breakout efficiency $\eta_\textrm{br}$\,$=$\,$10^{-3}$ the cocoon has a very small impact on detectability. However higher values, e.g., $\eta_\textrm{br}$\,$=$\,$10^{-2}$, lead to a significant fraction of detected events at large angles.

Notably our range of gamma-ray detection percentages is quite large presented in Figure \ref{fig:detfracsGamma}, owing to the uncertainty in the distribution of jet opening angles. This is highlighted most clearly in Figure \ref{fig:gammadet-vs-angle}, where we compare the distribution of viewing angles $\theta_\textrm{v}/\theta_\textrm{c}$ generated for the different assumptions for the distribution of core half-opening angles. The detections are concentrated towards small viewing angles where the jet's energy is highest. As such, the detected fractions depend very strongly on the assumptions for the distribution of core opening angles, which dictates the ratio $\theta_\textrm{v}/\theta_\textrm{c}$. Only a small fraction of these distribution (RE23 for Run 1 or Narrow for Run 3; see Table \ref{tab: simulationruns}) are viewed within the core of their jet ($\theta_\textrm{v}/\theta_\textrm{c}$\,$<$\,$1$) with only 1.7\% for the narrow distribution of core angles and 4\% for the RE23 distribution \citep{RoucoEscorial2022}. The RE23 distribution also has a significantly larger fraction within $\theta_\textrm{v}/\theta_\textrm{c}$\,$<$\,$3$ ($5$) with 25\% (40\%) within these angles, compared to 14\% (33\%) for the narrow core distribution. This difference favors the detection of a higher fraction of EM signals when using the RE23 distribution, due to the larger fraction of wide jets, and leads to drastically lower (by a factor of $\sim$\,$10\times$) detection rates when using core angles like that seen in GW170817 \citep{Mooley2018,Ghirlanda2019,Ryan2023}. The very narrow distribution of core angles used in Run 7 lead to a viewing angle distribution peaking at $\theta_\textrm{v}/\theta_\textrm{c}$\,$\approx$\,$15$. The significantly smaller fraction of simulated events at lower viewing angles is the cause of the lower detection fractions in Run 7, which is based on GW170817's jet \citep{Ryan2023}.

The cocoon model begins to take over all detections outside of $\theta_\textrm{v}$\,$\gtrsim$\,$4$\,$-$\,$5\,\theta_\textrm{c}$. The transition occurs when the line-of-sight gamma-ray energy $E_\gamma(\theta_\textrm{v})$ of the jet drops below $\approx$\,$10^{47}$ erg which is the approximate energy of the cocoon in our setup. The quasi-isotropic cocoon \citep{Duffell2018} is assumed to be independent of viewing angle, unlike the jet, and is detectable to a distance of $\approx$\,$50$ to $300$ Mpc for \textit{Fermi} and  \textit{SVOM}, respectively. 
This distance is significantly smaller than the expected BNS distances for XG detectors (Figure \ref{fig:XGdetectionsJWST}), and for this reason we do not consider the cocoon for our XG simulations. 

A major caveat of the gamma-ray detections computed here is the application of a constant gamma-ray efficiency with angle, though this has been adopted in previous works on the topic \citep[e.g.,][]{Ronchini2022}. There is growing evidence that the efficiency of gamma-ray production is likely to decrease with angle \citep[e.g.,][]{Gill2019,BeniaminiNakar2019,OConnor2024}, which would lead to an even smaller fraction of off-axis detections at $\theta_\textrm{v}/\theta_\textrm{c}$\,$\gtrsim$\,$2$. In addition, the distribution of gamma-ray energies or luminosities of cosmological short GRBs is still uncertain and other prescriptions \citep[e.g.,][]{Wanderman2015,Salafia2023} are likely to lead to different results. In particular, the luminosity distribution derived by \citet{Wanderman2015}, utilized in past works \citep{BeniaminiNakar2019,Bhattacharjee2024}, would substantially decrease the detection rate due to its lower average gamma-ray luminosity $L_\gamma$.

\subsection{XG detectors}
\label{sec:XGresults}

\subsubsection{Afterglow detection}

The major critical difference between O5 and XG is the distance of the detected BNS mergers. The distances of BNS mergers detectable in GWs during O5 range up through 20 Gpc (see Figure \ref{fig:redshift} for redshift or Figures \ref{fig:O5detectionsJWST} and \ref{fig:XGdetectionsJWST} for luminosity distance), but are highly concentrated within 2.5 Gpc, 
while the distances for the GW detectable BNS mergers in the next-generation simulations range up through 20 Gpc, and are more evenly distributed throughout the entire range.
This is reflected in Figures \ref{fluxthVthCplot} and \ref{fluxtimeplot} 
where the distribution of peak fluxes for XG afterglows is shifted to smaller values compared to O5; this is a distance dependent shift and independent of viewing angle. Therefore, the detected afterglows are concentrated to smaller viewing angles, as shown in Figure \ref{fig:XGdetectionsJWST} where the majority of detections are obviously concentrated to within $14$ deg ($0.25$ rad). 
The further distance of XG events means that the currently active EM telescopes we are considering in this study will not be able to pick up the fainter signals from the most distant events. 
As the next-generation instruments (e.g., ngVLA, SKA, \textit{NewAthena}, \textit{Lynx}) are more sensitive to fainter signals (signals from more distant events), we are still capable of detecting this distribution. However, this is one reason we see lower detection rates overall, across all telescopes, for the next-generation simulations. The detection of the afterglows of BNS mergers in this era will require rapid ToO response (e.g., \textit{NewAthena}) and this should be a key goal for facilities over the next decade. 

We find that the top facilities in this era at each wavelength are \textit{NewAthena} and \textit{Lynx} at X-ray wavelengths with detectable percentages of 2.4\% and 3.1\% respectively, \textit{JWST} at near-infrared wavelengths with a detectable percentage of 2.2\% in the \textit{F356W} filter, and ngVLA at radio wavelengths with a detectable percentage of 3.0\% at 8 GHz. 

At UVOIR wavelengths, we find that \textit{JWST/F356W} is able to detect 0.5\% of all simulated afterglows after 1 day and 0.3\% after 10 days. 
At X-ray wavelengths, \textit{NewAthena} has a similar detectable fraction, with 0.6\% after 1 day and 0.2\% after 10 days. Lastly, at radio wavelengths, ngVLA (8 GHz) is capable of detecting 2.9\% of all simulated afterglows after 1 day, which is 99\% of all possible detectable afterglows at that frequency, and the number decreases to 2.7\% after 10 days, and to 1\% after 100 days. For XG era merger events, where signals are fainter and afterglow peaks less likely to be detected than in O5, radio will be the optimal wavelength range to search in, outperforming \textit{JWST}.

\subsubsection{Prompt gamma-ray detection}

The detection fraction of GRBs from simulated XG-detectable BNS is in the range $\sim$\,$0.1$\,$-$\,$1\%$, lower than the $\sim$\,$0.5$\,$-$\,$10\%$ expected for the distances of events during LVK O5. 
Due to the larger distance of XG events (Figures \ref{fig:redshift} and \ref{fig:XGdetectionsJWST}) the gamma-ray detections are likewise concentrated closer to the core of the jet (see Figure \ref{fig:gammadet-vs-angle}). In fact 
50\% (90\%) of the detected distribution is viewed at $\theta_\textrm{v}/\theta_\textrm{c}$\,$<$\,$1.5$ ($2.4$) if using the RE23 (Run 1) distribution of core half-opening angles (see Figure \ref{fig:gammadet-vs-angle}). In terms of the inclination angle this translates to $\theta_\textrm{v}$\,$<$\,$21^{\circ}$ ($40^{\circ}$). As only a small fraction of events have $\theta_\textrm{v}/\theta_\textrm{c}$\,$>$\,$3$ (which is $\theta_\textrm{w}$ for Run 2) there is no large deviation between the gamma-ray detections of Run 1 and Run 2 in XG (Figure \ref{fig:detfracsGamma}). In any case, the lack of far off-axis detections during XG is in agreement with inferences from cosmological short GRBs that they are not detected far off-axis at such distances \citep{OConnor2024}. 

We note that the predicted rate of joint GW and afterglow detections in the XG era depends strongly on the number of BNS mergers that are expected to be detectable. This number depends sensitively on the assumed BNS merger rate, which is notably uncertain and has continued to decrease in every GW observing run since the discovery of GW170817. Therefore, we caution that the XG rate of events may be overestimated if the BNS rate itself is overestimated. This rate will continue to be refined over the next decade through the O5 and A\# observing runs and beyond, and as we continue to approach the XG era the true rate will continue to become more clear. 

\section{Conclusions}
\label{section:Conclusion}

In this work we have simulated the gamma-ray prompt emission and the afterglows of BNS mergers detected in GWs during both LVK O5 and by next-generation detectors, such as the Cosmic Explorer and Einstein Telescope, in order to quantify the prospects for multi-messenger astronomy in the next decade. We consider a range of instruments across the electromagnetic spectrum from low frequency radio wavelengths to gamma-rays. We aim to quantify the best instruments, both current and future, for GW observations focused on the afterglow.  We simulated a population of BNS mergers and quantified their detectability by future GW instrumentation. The jet afterglows were simulated based on observations of cosmological short GRBs. 

We have not accounted for telescope visibility and therefore our detected fractions and rates should be treated as upper limits. Future work will explore a more realistic accounting of the observability as a function of time (which is important given the potential for late peaking afterglows) depending on a specific sky position with relation to the Sun, Moon, and a given telescope. 

As expected, we find that viewing angle and jet's core half-opening opening angle have a significant impact on the detectability of the afterglow, with lower viewing angles $\theta_\textrm{v}/\theta_\textrm{c}$ corresponding to increased detectability. 
The best chances of detecting the jet afterglow of a BNS merger are at radio (e.g., SKA, ngVLA) or near-infrared (e.g., \textit{JWST}) wavelengths at later times (Figures \ref{fig:detfracsAG} and \ref{fig:detfracsmaxtime}). This is partially due to the difficulty in localizing an afterglow (or kilonova) on rapid timescales ($<$\,$1$ d) to an area small enough for the sensitive wide field instruments (e.g., \textit{Rubin}, \textit{Roman}, \textit{ULTRASAT}). 
Instruments such as \textit{Einstein Probe} with a 3600 deg$^2$ field of view are prime for rapid observations of the full sky localization of a GW event, especially during O5 and XG, but these instruments lack the sensitivity to detect the majority of afterglows that are likely to be produced by these GW events. Therefore, if the afterglow (or kilonova) is not rapidly localized and targeted at $<$\,$1$ d the prospects improve at $\gtrsim$\,$20$ d at radio wavelengths and $\sim$\,$100$ d in the near-infrared. Over these later timescales it is easy to re-point deep field instruments, and improve their likelihood of detecting the emission from the jet, though the issue of their field of regard remains.

We find that for LVK O5 there are several telescopes that are useful for detecting the late-time afterglow of a BNS merger. For example, on timescales of $>$\,$10$ days \textit{JWST} can detect $\sim$\,$6.4\%$ of all simulated afterglows, SKA can detect $\sim$\,$7.5\%$, and \textit{Chandra} can detect $\sim$\,$0.4\%$. Similar detection likelihoods are expected during the A\# era (McIver et al. in prep.), which will follow the O5 run prior to the implementation of XG detectors.

The detection likelihood is lower for our Cosmic Explorer and Einstein Telescope simulations, owing to the larger distance of the detectable XG mergers, which is $\sim$\,$10\times$ further than the median distance of the BNS mergers detectable during O5. The best instrument is the \textit{Lynx} mission concept at X-ray wavelengths, having the highest detectable fraction at $3.1\%$ with 1.1\% detected at $>$\,$1$ d (Figures \ref{fig:detfracsAG} and \ref{fig:detfracsmaxtime}). Events viewed within $<$\,$15$ deg of the jet's axis have the highest detection fraction, which remains somewhat constant out to $z$\,$\sim$\,$1$\,$-$\,$2$ (Figure \ref{fig:XGdetectionsJWST}). The ngVLA and SKA are incredibly useful at low-frequency radio wavelengths and are complementary given their locations in the Northern and Southern Hemispheres, respectively. ngVLA has an improved detectable fraction over SKA after $\sim$\,$1$ d, with $\sim$\,$2.9\%$ as compared to SKA's $\sim$\,$0.8\%$. Finally, at UVOIR wavelengths, \textit{JWST} is the optimal telescope, with an overall detection percentage of 0.53\% after $\sim$\,$1$ d.

In the future it may be useful to further explore afterglows as a tool to identify the EM counterpart to GW events, without the need of GRB or KN detection. We have approached this here by considering the detectability of afterglows by both wide FoV instruments and sensitive, deep instruments with small FOVs. However, identification and characterization may not be straightforward. Although orphan afterglows are rare \citep{Ho2020,Freeburn2024} and therefore the probability of chance coincidence between a GW event and an afterglow at a redshift consistent with the merger's distance will be low, it will be useful to properly quantify the probability of association for afterglows peaking months after the GW event using statistical tools such as those in \citet{Ashton_2018,Palmese:2021wcv} and astrophysical arguments based on multimessenger  observations. Moreover, detectability of jet afterglows from binary black hole mergers should also be explored, given the possibility that these objects also launch jets when embedded in the disks of Active Galactic Nuclei (e.g.,  \citealt{2023ApJ...950...13T}). These jets can be detected as nuclear transients in coincidence with GW binary black hole mergers \citep{Graham_2020,2023ApJ...942...99G,2024arXiv240710698C}, and show great promise for cosmological measurements with standard sirens \citep{PhysRevD.110.083005}.

\section*{Acknowledgments}
\begin{acknowledgments}
The authors thank Alicia Rouco Escorial and Wen-fai Fong for sharing their distribution of short GRB core angles. 
R.K. acknowledges helpful comments from, and discussions with, Raffaella Margutti on her Undergraduate Honors Thesis at the University of California Berkeley, which served as the basis for this work. 
B.O. acknowledges useful discussions with Geoff Ryan and Paz Beniamini. B.O. thanks Gabriele Bruni and Alexander van der Horst for assistance with radio telescope sensitivities, and Eliza Neights and John Tomsick for assistance with gamma-ray telescope sensitivities. A.P. thanks Amanda Farah and Viviane Alfradique for help with the simulations, and Marica Branchesi and Om Salafia for useful discussion.

B.O. is supported by the McWilliams Postdoctoral Fellowship at Carnegie Mellon University. A. P. acknowledges support for this work was provided by NASA through the NASA Hubble Fellowship grant HST-HF2-51488.001-A awarded by the Space Telescope Science Institute, which is operated by Association of Universities for Research in Astronomy, Inc., for NASA, under contract NAS5-26555.

This research used resources of the National Energy Research
Scientific Computing Center, a DOE Office of Science User Facility
supported by the Office of Science of the U.S. Department of Energy
under Contract No. DE-AC02-05CH11231 using NERSC award
HEP-ERCAP0029208 and HEP-ERCAP0022871. This work used resources on the Vera Cluster at the Pittsburgh Supercomputing Center.
\end{acknowledgments}

\appendix

\section{Simulated BNS merger distances and viewing angles}
\label{sec:appendixsimulatedevents}

\begin{figure*}
    \centering
    \includegraphics[width=\textwidth]{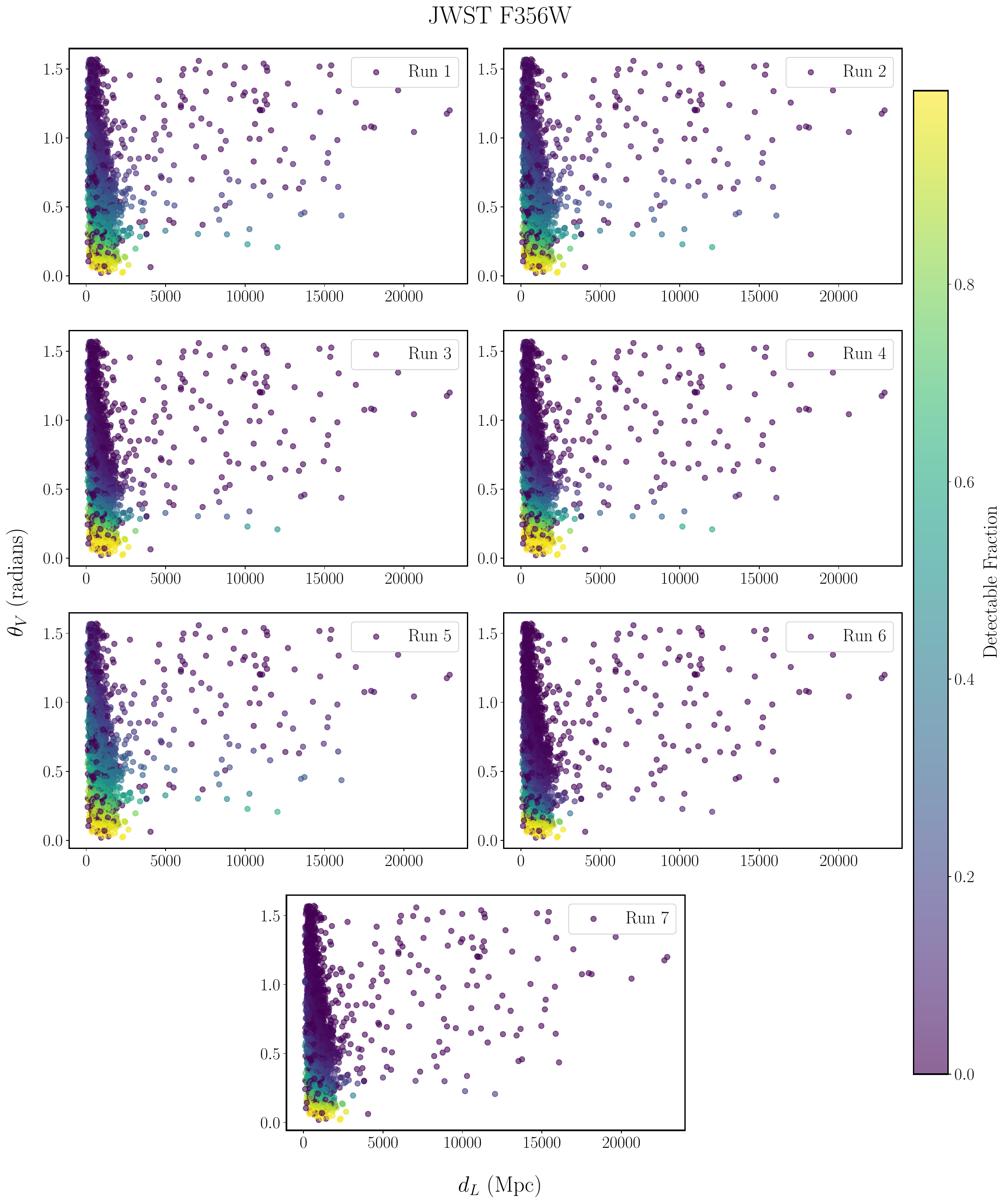}
    \caption{The distance and viewing angle of all simulated BNS mergers detected in GWs during LVK O5 for all seven simulation runs. The colorbar shows the fraction of the 1,000 simulated afterglows which are detected by \textit{JWST} in the \textit{F356W} filter for each of the simulated mergers.
    }
    \label{fig:O5detectionsJWST}
\end{figure*}

\begin{figure*}
    \centering
    \includegraphics[width=\textwidth]{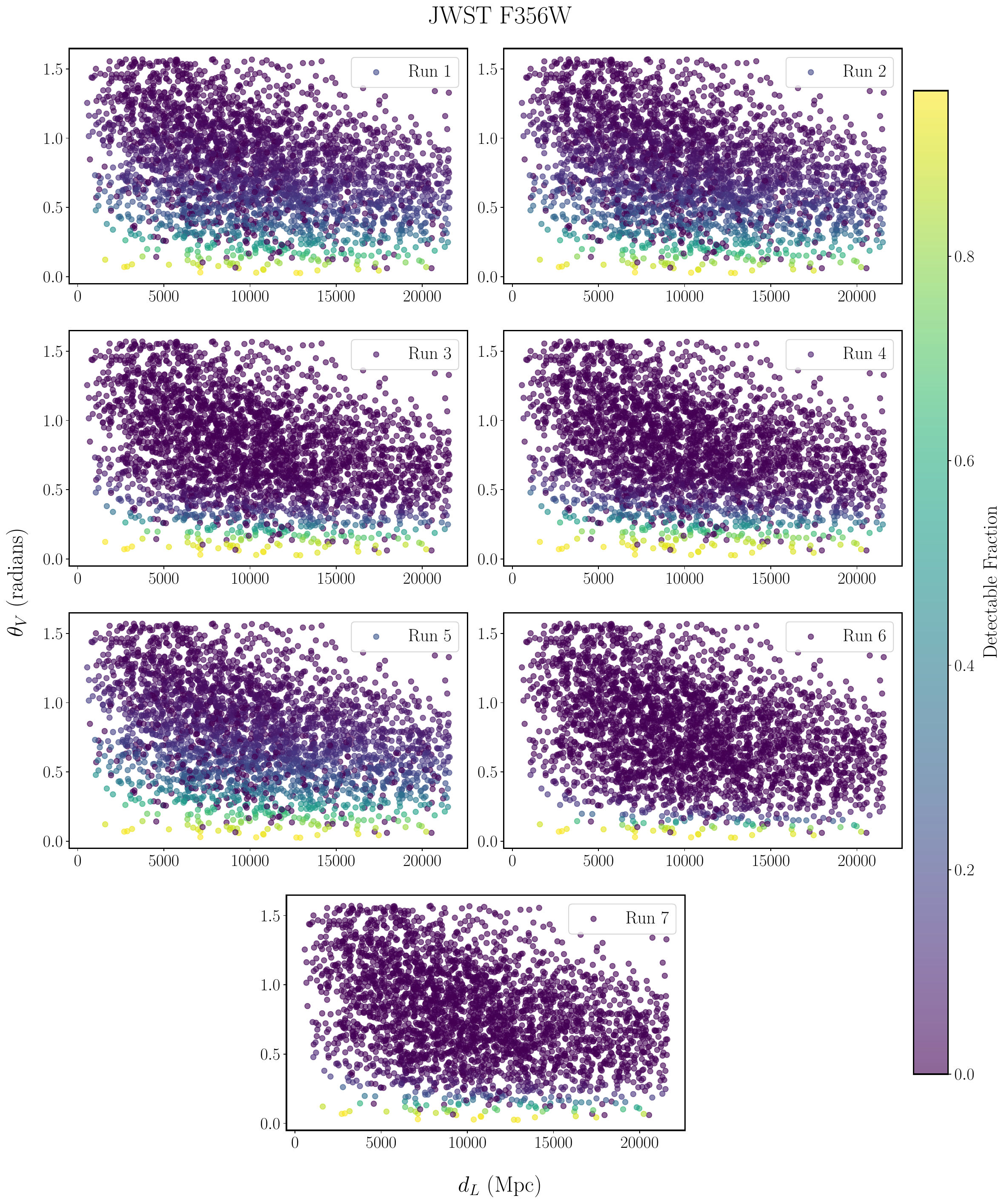}
    \caption{Same as Figure \ref{fig:O5detectionsJWST} for BNS mergers detectable by our XG detector setup.
    }
    \label{fig:XGdetectionsJWST}
\end{figure*}

Here we show the distance and viewing angle of all BNS mergers detected by LVK at O5 sensitivity (Figure \ref{fig:O5detectionsJWST}) and our XG detector setup (Figure \ref{fig:XGdetectionsJWST}). Each of the points in Figures \ref{fig:O5detectionsJWST} and \ref{fig:XGdetectionsJWST} represent a single GW event for which we perform our afterglow simulations. The afterglow detection fraction at each distance and viewing angle is shown based on  the colorbar.

\section{Impact of Lorentz factor and jet spreading}
\label{sec: afterglowpyassumptions}

By default \texttt{afterglowpy} \citep{Ryan2020,Ryan2023} does not include an angular profile for the bulk Lorentz factor $\Gamma(\theta)$. However, we found that this assumption, and in particular the choice of an infinite Lorentz factor (defined by default as \texttt{GammaInf}) leads to a larger fraction of detected afterglows, especially at early times. Therefore, we apply the \texttt{GammaEvenMass} flag (Equation \ref{eqn:GammaEvenMass}) and an initial bulk Lorentz factor at the jet's core of $\Gamma_0$\,$=$\,$300$ \citep{Ghirlanda2018}. This flag, and the relaxation of the default \texttt{GammaInf} setting, is only accurate within \texttt{afterglowpy} when lateral jet spreading is  neglected. We find that our afterglow detectability inferences are not significantly impacted by lateral spreading, which has the largest effect after the afterglow peak (see also the comparison in \citealt{jetsimpy}). As such we do not believe that our inferences for the afterglow detection percentages would significantly change if spreading were included in our calculations.

The detected fraction of events depends most strongly on the peak flux of the afterglow. Here we discuss the analytic expectations for the peak afterglow flux and how different assumptions within \texttt{afterglowpy} match these expectations. 
The angular Lorentz factor profile $\Gamma(\theta)$ modifies the time of the lightcurve peak for smaller viewing angles $\theta_\textrm{v}/\theta_\textrm{c}$\,$\lesssim$\,$2$. For these smaller angles the time of the afterglow peak and the peak flux can be computed by using the line-of-sight energy and Lorentz factor (see also \citealt{OConnor2024} for a discussion). As we adopt the same angular profile for energy and Lorentz factor (Equation \ref{eqn:GammaEvenMass}) this can be easily derived. The (on-axis) time of deceleration is given by $t_\textrm{dec,0}$\,$\propto$\,$(E_\textrm{kin,0}/n)^{1/3}\Gamma_0^{-8/3}$ for a uniform density environment \citep{Sari1999,Molinari2007,Ghisellini2010,Ghirlanda2012,Nava2013,Nappo2014,Ghirlanda2018}. 
Using the same scalings in energy and Lorentz factor, the deceleration of line of sight material produces an afterglow peak at: 
\begin{align}
\label{eqn:structdec}
   &t_\textrm{dec}(\theta_\textrm{v}) = t_\textrm{dec,0}\, E_\textrm{kin}(\theta)^{-7/3}. 
\end{align}

A similar scaling can be derived for the peak flux due to deceleration:
\begin{align}
\label{eqn:structdecflux}
   &F_\textrm{dec}(\theta_\textrm{v}) = F_\textrm{dec,0}\,E_\textrm{kin}(\theta)^{17/5}. 
\end{align}
For a GJ this can be written as:
\begin{align}
    &t_\textrm{dec}(\theta_\textrm{v}) = t_\textrm{dec,0}\, \exp\Bigg(-\frac{1}{2}\frac{\theta_\textrm{v}^2}{\theta_\textrm{c}^2}\Bigg)^{-7/3}, \\
   &F_\textrm{dec}(\theta_\textrm{v}) = F_\textrm{dec,0}\, \exp\Bigg(-\frac{1}{2}\frac{\theta_\textrm{v}^2}{\theta_\textrm{c}^2}\Bigg)^{17/5}.
\end{align}
These equations hold for smaller viewing angles $\theta_\textrm{v}/\theta_\textrm{c}$\,$\lesssim$\,$2$ and $\theta_\textrm{v}$\,$\lesssim$\,$\theta_\textrm{w}$ for a GJ and $\theta_\textrm{v}/\theta_\textrm{c}$\,$\lesssim$\,$3$ for a PLJ, whereas for larger viewing angles the time of the afterglow peak begins to be dictated by the jet's core becoming de-beamed to the observer \citep[see, e.g.,][for a discussion]{BGG2020,BGG2022,OConnor2024}. 

For larger viewing angles and no lateral jet spreading  the lightcurve will peak when the jet's core becomes de-beamed to the observer which occurs when $\Gamma$\,$\approx$\,$\theta_\textrm{v}^{-1}$, such that the peak time $t_{\rm p}$ is \citep{Nakar2002,BGG2020,OConnor2024} 
\begin{align}
&  t_{\rm p} \propto t_{\rm j} \left( \frac{\Delta\theta}{\theta_c} \right)^{8/3}, 
\end{align}
where $t_{\rm j}$ is the on-axis time of the jet break  \citep{SariPiranHalpern1999,Rhoads1999,Frail2001} and $\Delta\theta$\,$=$\,$\theta_\textrm{v}-\theta_\textrm{c}$. The far off-axis scaling for the peak flux $F_{\rm p}$ is independent of jet spreading and provided by \citep{Nakar2002,Lamb2021rev}
\begin{align}
&  F_{\rm p} \propto F_{\rm j} \left( \frac{\Delta\theta}{\theta_c} \right)^{-2p}, 
\end{align}
where $F_{\rm j}$ is the flux at on-axis jet break time. As the peak flux dictates detectability and the fact the dependence is not modified by jet spreading we are confident the impact of our assumptions are negligible. 
These equations hold for a tophat jet, but are also accurate for \texttt{afterglowpy} at large angles or for $\theta_\textrm{v}$\,$\gtrsim$\,$\theta_\textrm{w}$.

As an example we compute lightcurves for a GJ over 200 log-spaced viewing angles between $\theta_\textrm{v}/\theta_\textrm{c}$\,$=$\,$\{0,9\}$ for a core half-opening angle $\theta_\textrm{c}$\,$=$\,$0.1$ rad and a truncation angle $\theta_\textrm{w}$\,$=$\,$0.5$ rad. 

\begin{figure*}
    \centering
    \includegraphics[width=0.48\columnwidth]{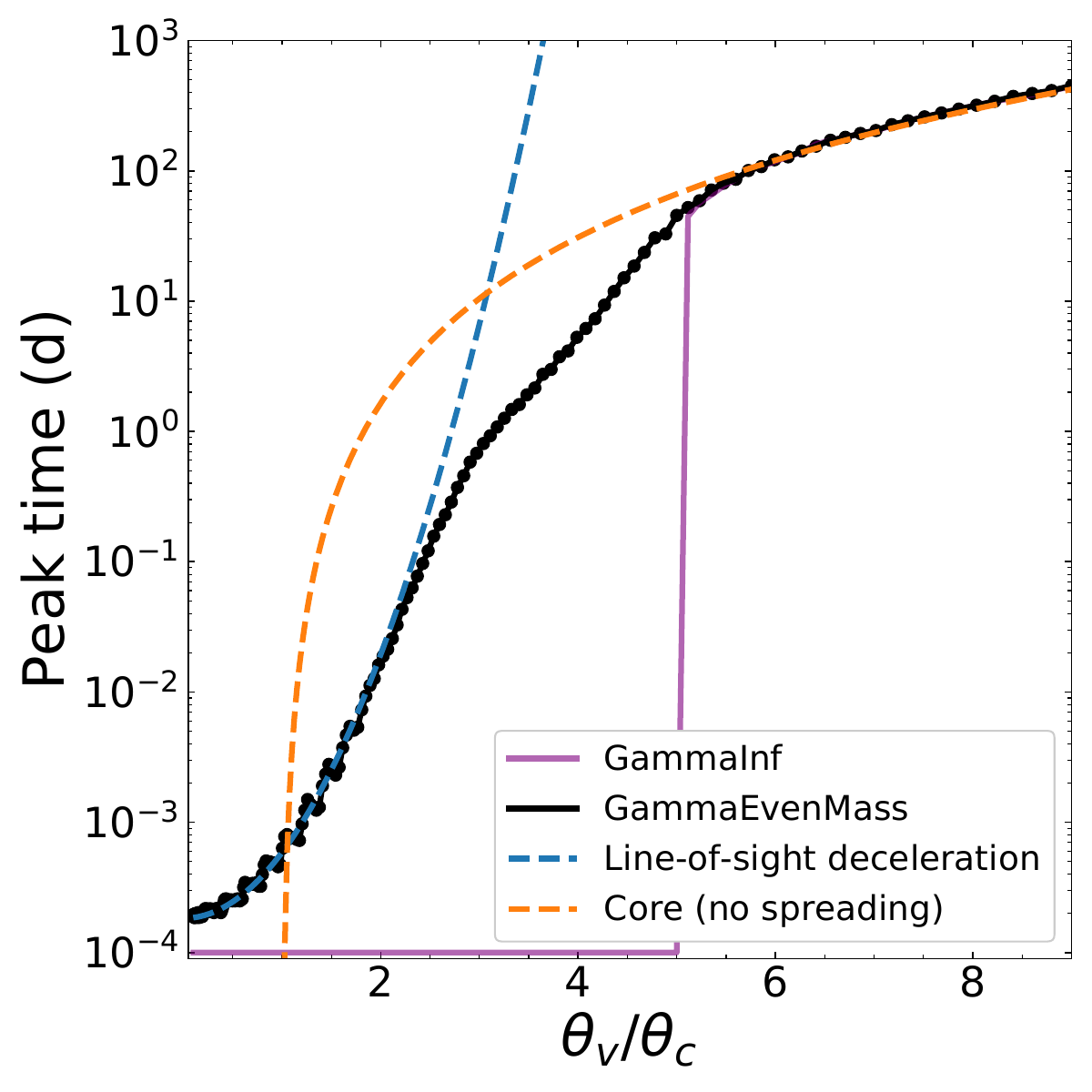}
    \includegraphics[width=0.48\columnwidth]{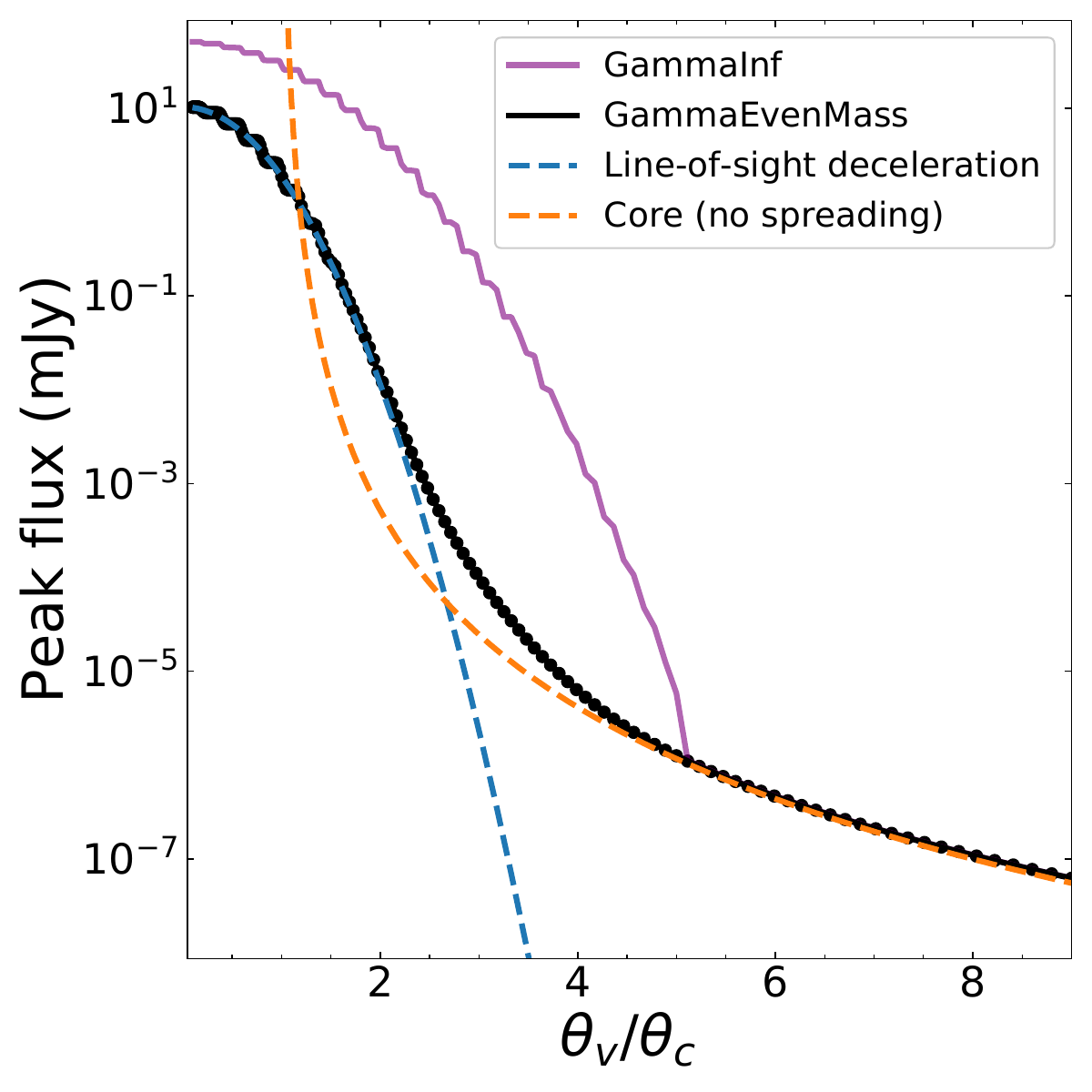}
    \vspace{-0.3cm}
    \caption{Peak time $t_\textrm{p}$ and flux $F_\textrm{p}$ versus viewing angle $\theta_\textrm{v}/\theta_\textrm{c}$ using \texttt{afterglowpy}. We show the results for a GJ for both the default setting (\texttt{GammaInf}) and an angular Lorentz factor profile (\texttt{GammaEvenMass}). The peak time and flux match the expected analytic scalings. The afterglow parameters are: $\Gamma_0$\,$=$\,$300$, $E_\textrm{kin,0}$\,$=$\,$10^{52}$ erg, $n$\,$=$\,$10^{-2}$ cm$^{-3}$, $\varepsilon_e$\,$=$\,$10^{-1}$, $\varepsilon_B$\,$=$\,$10^{-2}$, $p$\,$=$\,$2.2$, $\theta_\textrm{c}$\,$=$\,$0.1$ rad, and $\theta_\textrm{w}$\,$=$\,$0.5$ rad computed at a distance of 150 Mpc at $\nu$\,$=$\,$1$ keV.  
    }
    \label{fig:lorentz}
\end{figure*}

We compute the peak time and flux of these lightcurves for both the default \texttt{GammaInf} setting and for \texttt{GammaEvenMass} with an initial core Lorentz factor $\Gamma_0$\,$=$\,$300$. The lightcurves are computed at an observer frame frequency of 1 keV between $10^{-4}$ and $1500$ d (observer frame). The results are displayed in Figure \ref{fig:lorentz}. We verified these scaling relations also hold for a PLJ. 

We see good agreement between the analytic scalings described above and the \texttt{afterglowpy} lightcurves for \texttt{GammaEvenMass}. There is a natural transition regime between the analytic expectations that is due to the assumed value of $\theta_\textrm{w}$\,$=$\,$5\theta_\textrm{c}$. For viewing angles $\theta_\textrm{v}$\,$\gtrsim$\,$\theta_\textrm{w}$ the tophat jet scalings exactly match the lightcurve behavior. For smaller values of $\theta_\textrm{w}$ these equations will apply for smaller angles from the core. 

It is worth pointing out that for the \texttt{GammaInf} assumption that the peak time is always the first value in the lightcurve grid when $\theta_\textrm{v}$\,$<$\,$\theta_\textrm{w}$. Hence the purple line showing a steep vertical rise from our smallest time value to the tophat expectation at precisely $\theta_\textrm{w}$\,$=$\,$5\theta_\textrm{c}$. 

If we consider detectability only for times greater than $1$ day after the merger then these issues no longer greatly impact detectability and the peak flux is very similar for both Lorentz factor assumptions. However, for times less than $1$ day, using $\texttt{GammaInf}$ can lead to higher inferred peak fluxes by a factor of up to $1,000\times$ for a GJ (and more for a PLJ) and thus significantly increase the fraction of detected afterglows. This deviation would be more significant if earlier timescales ($<$\,$10^{-4}$ d) are considered as for $\texttt{GammaInf}$ the flux is always higher at early times (unless $\theta_\textrm{v}$\,$>$\,$\theta_\textrm{w}$). 

Another mechanism to consider is potential variations to the initial Lorentz factor of the jet's core. We adopted a canonical value of $\Gamma_0$\,$=$\,$300$ which is the median value inferred for long GRBs in a uniform environment by \citet{Ghirlanda2018}. It is reasonable to expect a range of Lorentz factors for short GRBs, and we comment that using a smaller Lorentz factor, e.g., $\Gamma_0$\,$=$\,$100$, leads to a later peak time ($t_\textrm{dec,0}$\,$\propto$\,$\Gamma_0^{-8/3}$) and therefore a lower peak flux. For the parameters considered here (Figure \ref{fig:lorentz}) this variation in Lorentz factor leads to a higher peak flux by a factor of $\lesssim10$ (a factor of $\sim$\,$3$ for $\theta_\textrm{v}/\theta_\textrm{c}$\,$=$\,$2$) and no variation for times $>$\,$1$ day. 

We further note that these comparisons apply only to frequencies above the synchrotron injection frequency $\nu_\textrm{m}$, typically the optical and X-ray bands. For radio wavelengths, which are likely below $\nu_\textrm{m}$ (especially at early times), the peak time and flux of the afterglow do not depend at all on deceleration of line-of-sight material and instead occur either when $\nu_\textrm{m}$ passes through the observed band or when the jet break $t_\textrm{j}(\theta_\textrm{v})$ occurs. As such, the radio afterglow detectability is not impacted by any of these assumptions for the initial Lorentz factor, Lorentz factor angular structure, or the jet truncation angle. We verified this using \texttt{afterglowpy} with a similar methodology to that outlined above for Figure \ref{fig:lorentz}. 

\bibliography{apssamp}
\bibliographystyle{aasjournal}

\end{document}